\documentclass{pasa}
\usepackage{multirow}
\usepackage{graphicx}

\title{The central spectra of massive star-forming galaxies}

\author[Jaimie R. Sheil et al.]{
Jaimie R. Sheil,$^{1}$\thanks{E-mail: jaimiersheil@gmail.com}
Michael J. I .Brown$^1$,
Virginia A. Kilborn$^2$,
Michelle E. Cluver$^2$,
~and Thomas Jarrett$^{3,4}$
\\
\affil{$^1$ School of Physics \& Astronomy, Monash University, Clayton, Victoria 3800, Australia}
\affil{$^2$ Department of Physics \& Astronomy, Swinburne University of Technology, Melbourne, Victoria, Australia}
\affil{$^3$ Institute for Astronomy, University of Hawaii at Hilo, 640 N Aohoku Pl 209, Hilo, HI 96720, USA}
\affil{$^4$ Department of Astronomy, University of Cape Town, Rondebosch, Cape Town, 7700, South Africa}
}

\jid{PASA}
\doi{10.1017/pas.\the\year.xxx}
\jyear{\the\year}

\usepackage{aas_macros}
\usepackage{hyperref} 
\hypersetup{colorlinks,citecolor=blue,linkcolor=blue,urlcolor=blue}

\hypersetup{draft}

\begin{document}

\begin{frontmatter}
\maketitle

\begin{abstract}
We have examined the nuclear spectra of very massive star-forming galaxies at $z \sim 0$ to understand how they differ from other galaxies with comparable masses, which are typically passive.  We selected a sample of 126 nearby massive star-forming galaxies  ($<100~{\rm Mpc}$, $10^{11.3}~\rm{M_\odot} \leq M_{\rm stellar} \leq 10^{11.7}~\rm{M_\odot}$, $1 ~{\rm M_\odot~yr^{-1}}< {\rm SFR} <13 ~{\rm M_\odot~yr^{-1}}$) from the 2MRS-Bright WXSC
catalogue. LEDA morphologies indicate at least 63\% of our galaxies are spirals, while visual inspection of Dark Energy Survey images reveals 75\% of our galaxies to be spirals with the remainder being lenticular. Of our sample 59 have archival nuclear spectra, which we have modelled and subsequently measured emission lines ([NII]$\rm{\lambda 6583}$, H$\alpha\rm{\lambda 6563}$,  [OIII]$\rm{\lambda 5008}$, and H$\beta\rm{\lambda 4863}$), classifying galaxies as star-forming, LINERS or AGNs. Using a BPT diagram we find $83 \pm 6$ \% of our galaxies, with sufficient signal-to-noise to measure all 4 emission lines, to be LINERs. Using the [NII]$\rm{\lambda 6583}$/H$\alpha\rm{\lambda 6563}$ emission line ratio alone we find that $79 \pm 6$ \% of the galaxies (46 galaxies) with archival spectra are LINERs, whereas just $\sim 30\%$ of the overall massive galaxy population are LINERs \citep{Belfiore2016}. Our sample can be considered a local analogue of the Ogle et al. (2016, 2019) sample of $z \sim 0.22$ massive star-forming galaxies in terms of selection criteria, and we find 64\% of their galaxies are LINERs using SDSS spectra. The high frequency of LINER emission in these massive star-forming galaxies indicates that LINER emission in massive galaxies may be linked to the presence of gas that fuels star formation. 
\end{abstract}



\end{frontmatter}



\section{Introduction}
\label{sec:intro}



    The bimodal distribution of galaxies is well established observationally and most galaxies can be described as either blue, star-forming, spiral galaxies, or as red, passive, elliptical galaxies \citep[e.g.][]{Strateva2001, Bell2003, Blanton2003, Cattaneo2006}. The colour of a galaxy is a key observable parameter, being dominated by light from the most luminous stellar population in a galaxy and therefore reflecting a galaxy's star formation history, including starbursts and quenching \citep[e.g.][]{Tinsley1968,BC2003, Buta2011}. Spiral galaxies are typically bluer in colour, indicating active star formation and an abundance of young, high mass stars \citep[e.g.][]{Strateva2001, Buta2011}, while elliptical galaxies are typically redder as they have older stellar populations of $\approx 10$ \rm{Gyr} \citep[e.g.][]{Tinsley1968,Greene_2013}. Galaxy bimodality is also a strong function of galaxy mass, with the majority of galaxies with stellar masses $\geq 10^{10.5}~\rm{M_\odot}$  being elliptical galaxies without star formation \citep[e.g.][]{Baldry2004, Cattaneo2006,Kauffmann2003, bluck2020}. This bimodality is also seen in the spectra of elliptical and spiral galaxies, with elliptical galaxies lacking strong emission lines and having spectra dominated by the stellar continuum of the red, old stellar population \citep[e.g.][]{Masters2010}, while star-forming, spiral galaxies typically have strong emission lines, in particular the star formation tracer H$\alpha\rm{\lambda 6563}$ emitted by the gas ionised by young stars \citep[e.g.][]{kennicutt1998}. 

    Galaxy bimodality is reproduced by hierarchical galaxy formation models. Spiral galaxies may be transformed into elliptical galaxies through major mergers that destroy galaxy disks and produce quenched, elliptical galaxies \citep[e.g.][]{Toomre1972, Mihos1996, Zucker2016, Nelson2017}.  There are several plausible mechanisms that could be responsible for truncating star-formation, including galaxy mergers, starbursts, virial accretion shocks and AGN feedback \citep[e.g.][]{Mihos1996,Birnboim2003,Hopkins2006,Cattaneo2006}. The specific mechanism responsible for the truncation of star-formation and morphological transformation is not essential for this paper, as there are clear observational correlations between star formation, morphology and mass. Regardless of the specific astrophysics involved, star-forming galaxies (typically spirals) that populate one peak of the bimodal distribution can become passive galaxies (typically ellipticals) that populate the other peak of the bimodal distribution \citep[e.g.][]{Bell2004}.
    
 

     Whilst most galaxies follow galaxy bimodality, there are exceptions that could be illustrative of galaxies moving between the two peaks of the bimodal distribution or that highlight when quenching can occur without morphological transformation or vice versa. These exceptions may provide insights into the conditions required for quenching and morphological transformation. Green valley galaxies fall between the two peaks of the bimodal distribution, and have lower specific star formation rates than most star-forming galaxies but are not passive \citep{Martin2005, Martin2007}. These green valley galaxies may represent galaxies that are in the process of being quenched and transforming from blue, star-forming galaxies to red, passive galaxies \citep[e.g.][]{Wyder_2007,Goncalves2009, Nogueira2018}.  
     Another exception are passive, spiral galaxies, which show that while morphology is correlated with star-formation, an elliptical morphology is not required to form a passive galaxy \citep[e.g.][]{McKelvie2016, Shimakawa2022}. These galaxies may also reveal the influence of bars or environment on star formation rates although no one mechanism obviously accounts for the quenching seen in these galaxies \citep{McKelvie2017}. 

     
     Massive star-forming galaxies are another example of exceptions to bimodality and are the topic of this paper. These galaxies are exceptional as the quenching of star formation is correlated with galaxy mass and very massive galaxies (M$\rm_{{stellar}}> 10^{11.3}$ \rm{M$_{\odot}$}) are typically passive \citep[e.g.][]{Bell2004,Schawinski2014, Ogle2016}. These galaxies include massive flocculent spiral galaxies \citep[e.g.][]{Elmegreen1981, Elmegreen1982, Sheth2010, Sergey2020} and super spiral and super lenticular galaxies \citep{Ogle2016, Ogle2019}. A recent study by \citet{Lisenfeld2023} examined the molecular gas in super spirals and found that these galaxies have comparable molecular gas to stellar mass ratios to lower mass galaxies. They suggest that high mass itself is not sufficient to quench a galaxy and that these super spirals have enough molecular gas present to continue to form stars and not quench. Understanding why these galaxies have spiral morphologies and continue to form stars despite their high masses may provide insights into quenching, and is a motivation for this work. 
     
\begin{figure}
        \centering
        \includegraphics[width=0.46\textwidth,angle=0]{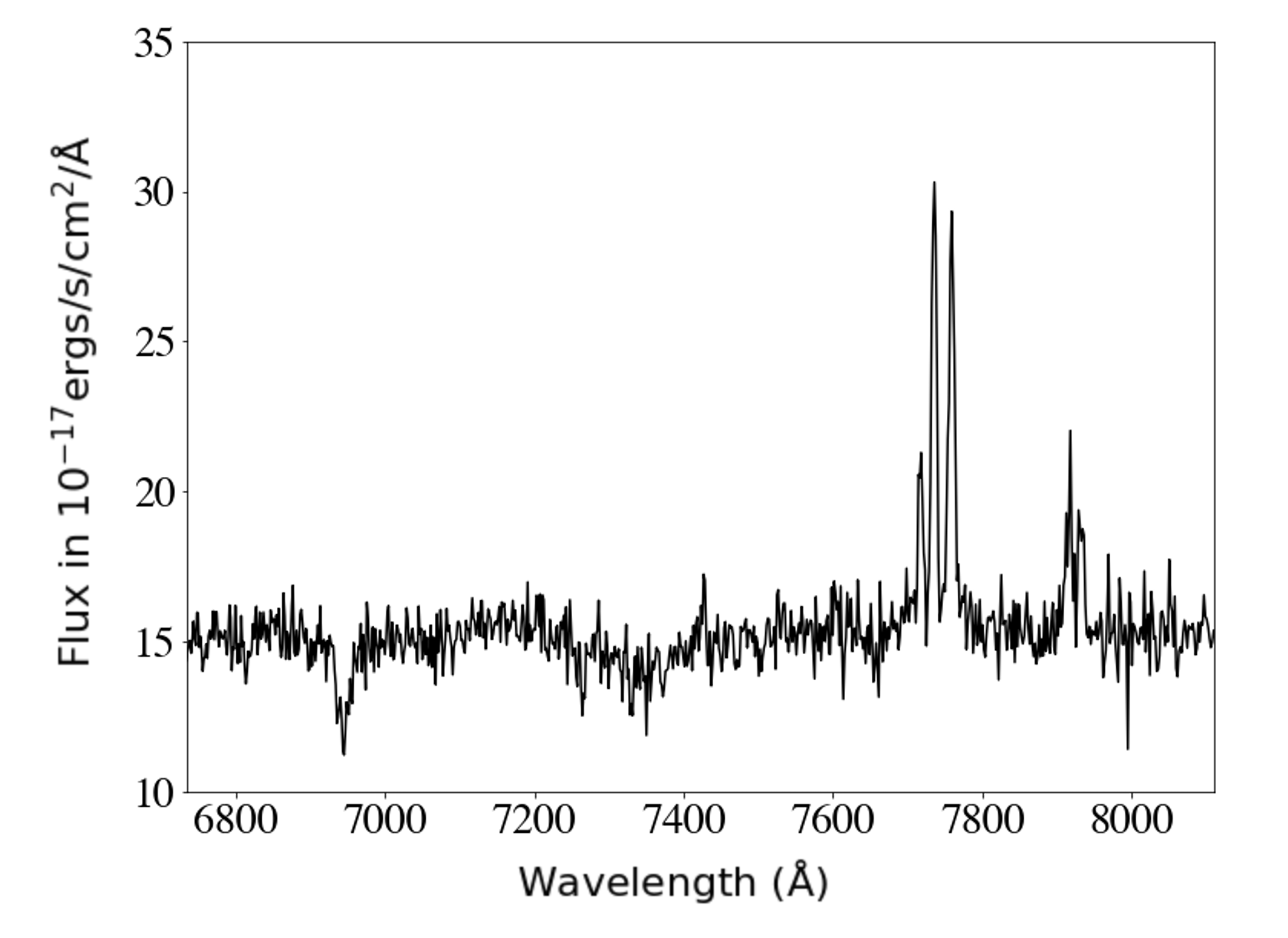}
        \caption{The SDSS spectrum of the super spiral 2MASX J07404205+4332412, which was identified by \citet{Ogle2016, Ogle2019}. The redshifted H$\alpha\rm{\lambda 6563}$ and [\rm{NII}]$\rm{\lambda 6583}$ are evident and are comparable in strength, which identifies this galaxy as a LINER.}
        \label{fig:Ogle}
    \end{figure}

     The \citet{Ogle2016, Ogle2019} super spiral galaxies have higher star formation rates (SFR), ranging from $1-30$ \rm{M$_{\odot}$ $\rm{yr}^{-1}$}, corresponding to specific star formation rates (sSFR) of $0.02-1.5\times 10^{-10}$ $\rm{yr}^{-1}$ and different morphologies than most massive galaxies. They often have different nuclear spectra too, as Figure~\ref{fig:Ogle} illustrates, the $z\sim 0.3$ super spiral galaxies of \cite{Ogle2016, Ogle2019} have strong nebular emission lines with comparable [\rm{NII}]$\rm{\lambda 6583}$/H$\alpha\rm{\lambda 6563}$ ratios to low ionisation nuclear emission region (LINER) galaxies. 
     
    While most very massive galaxies have little cool gas \citep[e.g.][]{Forman1985}, minimal star formation and consequently have absorption line spectra \citep[e.g.][]{Faber1992, Worthey1992}, the central spectra of the \citet{Ogle2016, Ogle2019} galaxies are an exception. These galaxies show strong nebular emission lines with emission line ratios that are inconsistent with ionisation by high mass main sequence stars \citep{Heckman1980}. As only 30\% of typical massive galaxies are LINERs \citep{Belfiore2016}, the abundance of [\rm{NII}]$\rm{\lambda 6583}$ and the high ratios of [\rm{NII}]$\rm{\lambda 6583}$/H$\alpha\rm{\lambda 6563}$ in the \citet{Ogle2016, Ogle2019} sample is unusual and warrants further investigation.
    

    \begin{figure}
        \centering
        \includegraphics[width=0.47\textwidth,angle=0]{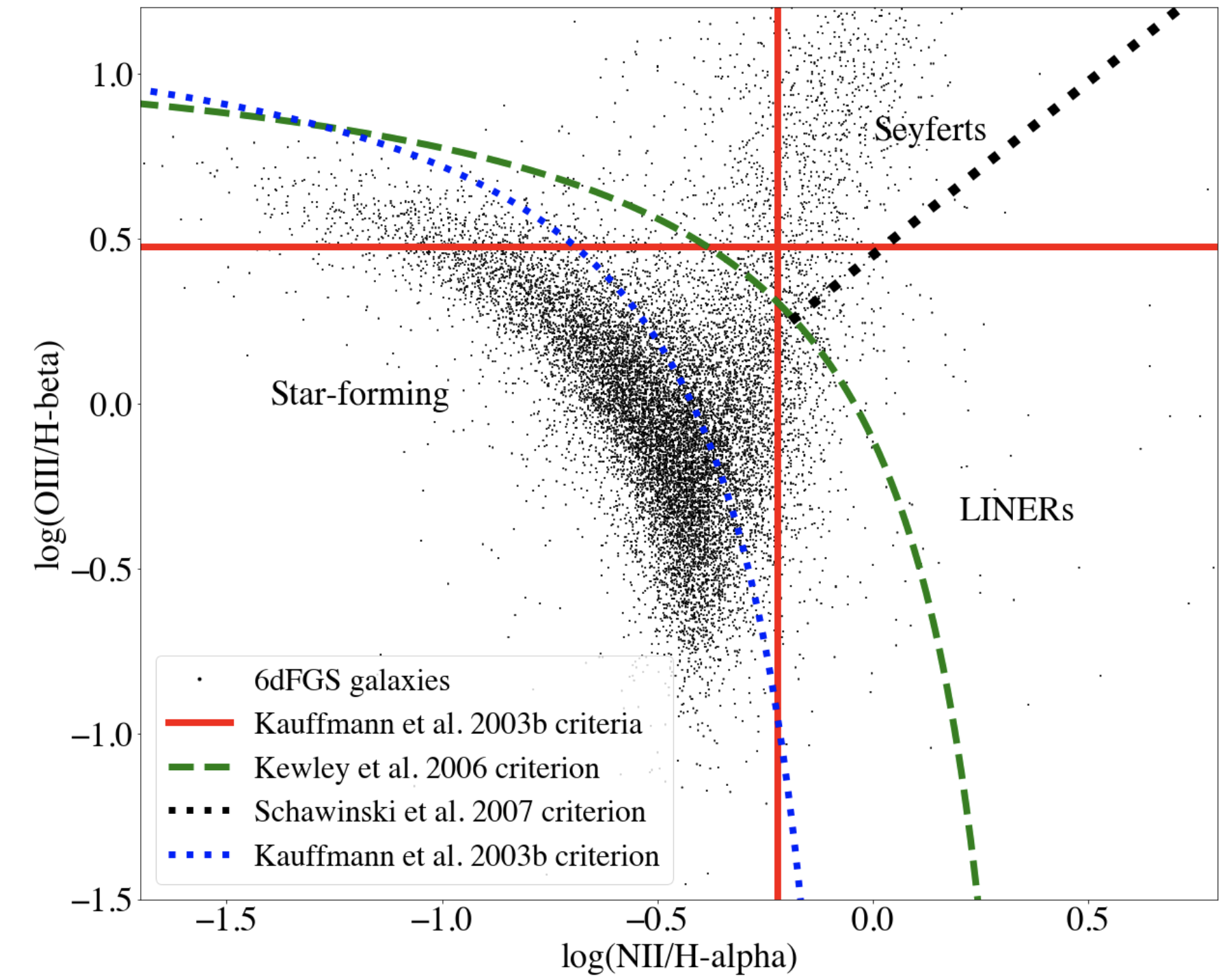}
        
        \caption{Our BPT diagram of 6dF galaxies along with commonly used galaxy classification criteria from the literature \citep{BPT1981}. The green line from \citet{Kewley2006} separates star-forming galaxies from Seyferts and LINERs, while the black dotted line from \citet{Schawinski2007} separates Seyferts and LINERs. The Kauffmann criteria shows alternative cuts to classify galaxies with and the emission line ratios required for LINER classification in red, [\rm{NII}]$\rm{\lambda 6583}$/H$\alpha\rm{\lambda 6563} > 0.6$ and [\rm{OIII}]$\rm{\lambda 5008}$/H$\beta\rm{\lambda 4863} <3$ \citep{Kauffmann2003LINER}. The different criteria are motivated by theory and observation, producing samples with varying completeness and contamination.}
        \label{fig:Example_BPT}
    \end{figure}
    
    Low Ionisation Nuclear Emission Regions (LINERs) have strong low-ionisation emission lines at [OI]$\rm{\lambda 6300}$, [SII]$\rm{\lambda 6731}$, [NII]$\rm{\lambda 6583}$, and [OIII]$\rm{\lambda 5008}$ \citep{Heckman1980}. The criteria for classifying LINERs generally includes BPT diagrams \citep{BPT1981} with specific classification criteria that have been updated with improvements in both data and theory \citep{Kewley2006, Schawinski2007, Kewley2013}. An example of such a diagram is shown in Figure~\ref{fig:Example_BPT}. Emission-line ratios alone can also be used to classify LINERs which are consistent with the current BPT diagrams commonly used, typically these ratios are [\rm{NII}]$\rm{\lambda 6583}$/H$\alpha\rm{\lambda 6563} \geq 0.6$ and [\rm{OIII}]$\rm{\lambda 5008}$/H$\beta\rm{\lambda 4863} \leq3$ \citep{Kauffmann2003LINER}.  


    If LINERs are particularly common in massive star-forming galaxies, it may be due to the presence of relatively cold gas that can be ionised by various sources. The ionisation source (or sources) that produces the observed spectra is an area of ongoing research with the two most prominent candidates being low luminosity AGN (LLAGN) and stars off the main sequence such as post-asymptotic giant branch (pAGB) stars \citep[e.g.][]{Binette1994, Kauffmann2003LINER, Groves2008, Ho2008, Coldwell2017, Marques2017, Kewley2019, Percival2020}. 

    Several previous studies have investigated LINERs and their association with atomic hydrogen and molecular gas \citep[e.g.][]{Gallimore1999, Olsson2010, Parkash2018}. Galaxies with detectable cool gas will typically have accompanying star formation, but it is not clear that the presence of cool gas with accompanying star formation is a necessary condition for LINER spectra. 


    \citet{Kauffmann2003LINER} and \citet{Kewley2006} studied the host galaxy properties of low luminosity emission-line AGNs (including LINERs, which are not always distinguished from AGNs in their studies), while our study takes the contrasting approach of studying the nuclear spectra of the most massive star-forming galaxies, which may or may not host AGNs. That said, \citet{Kauffmann2003LINER} do find that $\sim 30\%$ of galaxies with young stellar populations (${\rm D4000} <1.5$) and high stellar masses (M $\sim10^{11.3}~M_{\odot}$) host powerful AGNs (defined by L[OIII] > $10^{7} L_{\odot}$). \citet{Kauffmann2003LINER} has a large sample size drawn from SDSS, but is limited by a large projected aperture which may limit the detection of weaker LINERs. The \citet{Kauffmann2003LINER} sample has a projected aperture diameter of $\sim5.5$ kpc at the mean redshift of z $\sim 0.1$, while our sample has a projected diameter ranging from 0.48-3.18~kpc at our maximum distance of 100~Mpc, hence we may detect more weak AGNs in our sample than \citet{Kauffmann2003LINER} do in their sample. For this work we focus on the nuclear spectra of very massive, local star-forming galaxies, to characterise the central spectra, be they passive, star-forming, AGNs or LINERs. 

    The abundance of galaxies with large [\rm{NII}]$\rm{\lambda 6583}$ peaks in the spectra of \citet{Ogle2016, Ogle2019}'s population of massive, star-forming galaxies may indicate a link between what is producing star-formation in these galaxies and the ionisation that is producing LINER emission. The existing literature suggests a strong connection between star formation in massive galaxies and the presence of LINERs, but this has yet to be confirmed systematically \citep[e.g.][]{Kauffmann2003LINER, Graves2007, Belfiore2016}. Previous studies have large apertures such as, SDSS spectra at z $\sim 0.1$ and z $\sim 0.22$ \citet{Kauffmann2003LINER, Ogle2019} respectively, these samples correspond to aperture radii of $\sim 2.8$ kpc and $\sim 5.3$ kpc respectively. Therefore, aperture bias may result in the fraction of LINERs in massive star-forming galaxies being underestimated. To address this we have selected a local sample of galaxies that meet similar criteria to those of \citet{Ogle2019} so that we can compare our sample to one at higher redshift. The \citet{Ogle2019} selection criteria have stellar masses greater than $10^{11.3}~{\rm M_\odot}$ and star formation rate (SFR) greater than or equal to $1$ \rm{M$_{\odot}$ $\rm{yr}^{-1}$}, but unlike \citet{Ogle2019} our sample are all within $100~{\rm Mpc}$ of Earth. The masses in \citet{Ogle2019} are calculated using WISE $W_{1}$ photometry, assuming $M/L = 0.6$, while we are using the \citet{Cluver2014} relation with Log$(M/L) = -2.54 (W_1-W_2) - 0.17$ and since most of our galaxies have $W_1-W_2\sim0$, we get $M/L \sim 0.676$. Therefore, our masses are approximately 0.076 dex or 19\% larger than those of \citet{Ogle2019}. This is relatively small in absolute terms and small compared to the mass range of our sample, so we don’t expect this to have a major impact on our conclusions. Our sample's star formation rates range from $1-13$ \rm{M$_{\odot}$ $\rm{yr}^{-1}$} with a median of 2.6\rm{M$_{\odot}$ $\rm{yr}^{-1}$} while \citet{Ogle2019}'s sample ranges from $1.2-79$ \rm{M$_{\odot}$ $\rm{yr}^{-1}$} with a median of 7.7~\rm{M$_{\odot}$ $\rm{yr}^{-1}$}. The star formation rates in our sample are slightly lower than in \citet{Ogle2019}, however they are comprised using the same star-formation rate criteria and our sample contains the highest star-formation rates in local, high mass galaxies. As can be seen in Figure \ref{fig:Archival_spectra_LINER} the galaxies in our sample and the \citet{Ogle2019} sample are comparable although at different distances. 
    
    This paper is structured as follows. In Section 2 we describe our sample selection process and criteria. In Section 3 we detail the archival spectra we have used and the way in which we have modelled these spectra and measured emission line ratios. In Section 4, we classify the galaxies in our local sample and calculate the percentage of LINERs in this population and in Section 5, discuss the implications of these classifications. Throughout this paper we use the Vega magnitude system, a Kroupa initial mass function (IMF) \citep{Kroupa}, and distances from the WXSC 2MRS-Bright Catalogue (Jarrett et al. 2019, Jarrett et al, in prep) which contains both redshift dependent and independent distances.

    \section{Sample selection and characterisation}
     \footnote{If AGNs are present they will increase the infrared emission leading to overestimates of the stellar masses and star formation rates. However, the WISE colour-colour diagram in Figure 3 shows the $W_{1}-W_{2}$ colours of the bulk of our sample have not been increased by infrared emission from hot dust, with the one exception being MRK1239. This is not unexpected, given that low luminosity AGNs associated with LINERs have lower X-ray luminosities and Eddington ratios than local Seyferts \citep{Goncalves2009}.}
    To study these massive, star-forming galaxies we utilised archival photometry and redshifts, primarily from the 6dFGS \citep{6df2006, Jones2009} and 2MASS redshift survey \citep[2MRS; ][]{Huchra2012}. The 2MRS includes original and archival spectra (including 6dFGS), including redshifts for 44,599 galaxies with K-band magnitudes less than 11.75 and covers 91\% of the sky. Using both archival and new spectra, 2MRS provides redshifts for these galaxies and 20,860 of these galaxies with K<11.25 have visually determined or literature morphological types \citep{Huchra2012}. 
    
    To measure galaxy star formation rates and stellar masses we have utilised the Wide-field Infrared Survey Explorer (WISE) all sky survey, which covers four wavelength bands, $W_{1}$, $W_{2}$, $W_{3}$, and $W_{4}$ at 3.4, 4.6, 12, and, 23 $\mu$m respectively. The $W_{1}$ and $W_{2}$ bands trace the continuum emission from evolved stellar populations, typically K and M type stars, and hence are good indicators of the stellar mass in a galaxy \citep[e.g.][]{Cluver2014}, as well as AGNs creating a hot dust signature \citep[e.g.][]{Jarrett2011, Stern2012, Jarrett2013}. The $W_{3}$ and $W_{4}$ bands are sensitive to star formation history within the galaxy, due to them detecting warm dust \citep{Jarrett2011, Cluver2017}. 
    
\begin{figure}
        \centering
        \includegraphics[width=0.48\textwidth,angle=0]{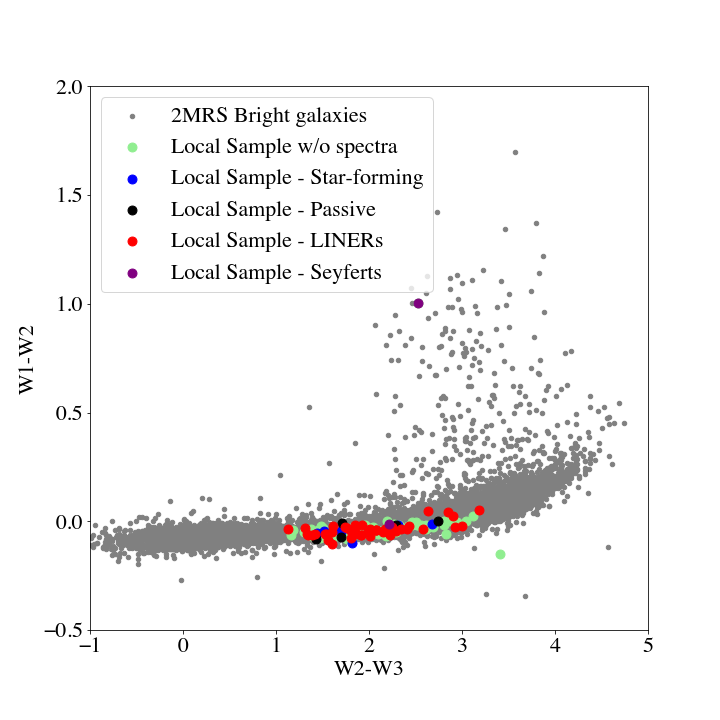}
        \caption{The WISE colour-colour diagram for all 2MRS-Bright galaxies (grey) and for our local, massive, star-forming sample of galaxies (coloured). Our sample lies in the spiral region of the $W_{1}$-$W_{2}$ against $W_{2}$-$W_{3}$ plot, which reflects that these galaxies have some star-formation present, as we selected for. We do not have an AGN locus for this plot currently as it is part of the ongoing work by Jarrett et al. in prep. There is one Seyfert shown in this plot in purple, classified by emission line width, MRK1239, and one galaxy, 2MASX J22444577+3327381, which lies below the rest of our sample due to a star contaminating the $W_{1}$-$W_{2}$ colour. However, with these two exceptions, the local sample of massive, star-forming galaxies have very uniform colours and fall just below the star-forming main sequence (SFMS) \citep{Chang2015}. }
        \label{fig:Colour}
    \end{figure}

    The brightest sources in the 2MRS ($K_{\rm{s(total)}}<10.5$) had their WISE photometry accurately measured for the WISE Extended Source Catalogue \citep[WXSC; ][]{Jarrett2019} and as part of the 2MRS-Bright WXSC catalogue (Jarrett et al, in prep.). The 2MRS-Bright WXSC catalogue contains WISE and 2MRS data for $\sim10,000$ galaxies as well as derived quantities such as the $W_{1}$-$W_{2}$ and $W_{2}$-$W_{3}$ colours (shown in Figure \ref{fig:Colour}), $W_{1}$ magnitude, k-corrected fluxes, and morphologies, which can be used to study the properties of these galaxies \citep{Jarrett2013, Cluver2017, Jarrett2019, Parkash2019}.
    
    The global stellar mass and star formation rate are estimated from the WISE mid-infrared photometry, where the short wavelength bands of WISE are sensitive to the stellar component and the longer wavelengths to star formation \citep{Jarrett2013}. We determined the stellar masses using the calibration of \citet{Cluver2014}, where the stellar mass-to-light ratio is a function of the $W_{1}$-$W_{2}$ colour (Equation 1). The stellar mass hence follows, with uncertainties (10-20\%) propagating from the stellar mass calibration and WISE photometry. For the star formation rate, we use the calibration from \citet{Cluver2017}, where the mid-infrared luminosity ($W_{3}$ and $W_{4}$ bands) is tightly correlated to the total-infrared luminosity (Equation 2). Typical uncertainties, from the star formation rate calibration and photometric measurements, are from 20 to 40\%.

    We select a sample of massive star-forming galaxies that have stellar masses greater than $\rm{M}_{\rm{stellar}}>10^{11.3}$ $\rm{M_\odot}$ and star formation rates of SFR $>1$ \rm{M$_{\odot}$ $\rm{yr}^{-1}$} from the 2MRS-Bright WXSC catalogue (Jarrett et al, in prep.). Our selection criteria is comparable to the $z \sim{0.22}$ super-spiral selection criteria used by \citet{Ogle2019} so that our sample is consistent with the literature on super spirals. The exact mass criteria for super spiral and lenticular populations in \citet{Ogle2019} are $10^{11.3}< \rm{M}_{\rm{stellar}} <10^{12.3}$ \rm{M$_\odot$} and $1<$ SFR $<30$ \rm{M$_{\odot}$ $\rm{yr}^{-1}$}, but as there are few massive galaxies with SFRs greater than $30$ \rm{M$_{\odot}$ $\rm{yr}^{-1}$} or stellar masses greater than $10^{12.3}$ \rm{M$_\odot$}, we take our criteria from the lower bounds of mass and star-formation rates. The range of SFRs in our sample is $1<$ SFR $\leq 13$ \rm{M$_{\odot}$ $\rm{yr}^{-1}$} (Mrk~1239 has a SFR$_{12\mu m}$ of $37$~\rm{M$_{\odot}$ $\rm{yr}^{-1}$} but we believe this is due to AGN dust contamination) and our masses range from $10^{11.3}< \rm{M}_{\rm{stellar}} \leq10^{11.7}$ \rm{M$_\odot$}. To calculate the stellar mass we use the following equations for low redshift sources from \citet{Cluver2014}:
    \begin{equation}
       \rm{log_{10}}\left(\frac{M_{stellar}}{L_{W_1}}\right) = -2.54\times(W_1-W_2) - 0.17
    \end{equation}
    where $\rm{L_{W_1}}$ is the $\rm{W_1}$ luminosity in Solar units, $\rm{M_{Sun}} = 3.24 $, $\rm{M_{W_{1_{abs}}}}$ is the absolute magnitude of the galaxy in W$_1$ and W$_1 -$ W$_2$ is the rest frame colour of the galaxy. Similarly we use the following relation from \citet{Cluver2017} to calculate SFR and subsequently sSFR: 
    \begin{equation}
    \begin{aligned}
        \rm{log_{10}}(SFR (\rm{M_{\odot} \rm{yr}^{-1}})) & = (0.889 \pm 0.018)\times
        \rm{log_{10}}(L_{12\rm{\mu m}}(L_\odot))  \\ & - (7.76 \pm 0.15)
    \end{aligned}
    \end{equation}
    From these values of SFR and mass we can select galaxies that match our selection criteria of stellar mass $\geq10^{11.3}$ \rm{M$_\odot$}, SFR $>1$ \rm{M$_{\odot}$ $\rm{yr}^{-1}$}, and distance $<100$ \rm{Mpc}. This provides a sample of 126 galaxies, which is shown in Figure~\ref{fig:Archival_spectra_samplespace}. As there is no infrared excess in Figure 3 there is no reason to think that hot dust is substantially inflating either the mass or the SFR. In Figure 15 of \citet{Yao2020} the AGNs are well separated in W1-W2 with values over 0.6. While in our sample the maximum value of W1-W2 is 0.15 with the vast majority being near 0. Given this and that the global wise colours are not contaminated, we don’t believe AGNs are significantly contaminating our sample. These galaxies are local and hence are easier to study in detail compared to similar high redshift samples e.g. \citet{Ogle2016, Ogle2019}. One galaxy, 2MASX J03370047+3240081, has been excluded from the sample as its photometry is contaminated with Galactic Cirrus. The sample of local galaxies, along with their observed and derived properties, are summarised in Table \ref{results}. 
        
    
\section{Spectra}
     To determine the abundance of LINERs in the most massive nearby star-forming galaxies, we have collated as many archival optical spectra of our sample galaxies as possible. We have used archival spectra from SDSS \citep{SDSSDR12}, 2MRS FAST spectra \citep{Huchra2012}, 6dFGS \citep{6df2006, Jones2009}, and the \citet{Ho1995} sample of nearby active galaxies. 
    The Sloan Digital Sky Survey (SDSS) covers over one third of the sky using fibre fed spectrographs with fibre diameters of 3$^{\prime\prime}$ and 2$^{\prime\prime}$  with the SDSS and BOSS spectrographs respectively. Both spectrographs have sufficient wavelength coverage to measure the Balmer series, [\rm{OIII}]$\rm{\lambda 5008}$ and [\rm{NII}]$\rm{\lambda 6583}$ for nearby galaxies, which are required to classify LINERs with the \citet{Kauffmann2003LINER, Kewley2006, Schawinski2007} criterion. The Six-degree Field Galaxy Survey (6dFGS) also uses a fibre fed multi-object spectrograph, it has fibre size of 6.7$^{\prime\prime}$ and has taken 136,304 spectra of 125,071 galaxies with $\rm{K_s} <12.65$. The 6dFGS sample spans over 83\% of the southern sky, $|b|>10^{\circ}$ and use two gratings, V and R to cover a wavelength range of 4000-7500 \AA\, making them ideal for our measurements and providing us with the majority of our spectra. 

\begin{center}
\begin{tabular}{ |p{2.4cm}|p{1.7cm}|p{3cm}| } 
    \hline
    \multicolumn{3}{|c|}{Aperture radii at 100 Mpc} \\
    \hline
    Spectra source & Fibre size & Aperture size (kpc) \\
     \hline
     SDSS   & $3^{\prime\prime}$    & 0.71\\
     SDSS (BOSS)&   $2^{\prime\prime}$  & 0.47\\
     6dFGS & $6.7^{\prime\prime}$ & 1.59\\
     2MRS FAST    &$1.5^{\prime\prime}$ & 0.36\\
     \citet{Ho1995}    &$1-2^{\prime\prime}$ & 0.24-0.47\\
     \hline
    \end{tabular}
    \captionof{table}{Table of aperture radii at 100Mpc for our sample which is sourced from SDSS, 6dFGS, 2MRS FAST and \citet{Ho1995}. As 100Mpc is the upper limit of distance for our sample these are also the upper limit of aperture radii. We may be more likely to detect faint AGN nuclei with our small aperture sizes compared to \citet{Ogle2016}.}
    \label{aperture}
\end{center}
    
    The sample of spectra in \citet{Ho1995} is largely sourced from the 1995 Palomar spectroscopic survey of nearby galaxies which was designed to gather spectra for bright, local galaxies that were thought to host AGNs. Given this goal, the wavelength range, 6210-6860 \AA\ and 4230-5110 \AA, is narrower than the previously discussed spectra. This survey collected long slit spectra for 486 northern galaxies with magnitudes $B_{T}\leq12.5$ and while it contains many bright AGN sources, it can also be used to search for LINERs. Finally, the 2MRS FAST survey collected optical spectra for 11,600 galaxies most of which have $\rm{K_s}\leq 11.75$. This survey has a wavelength range of 3500-7400 \AA\ which again is suitable for this work. 
    
\begin{figure}
        \centering
        \includegraphics[width=0.47\textwidth,angle=0]{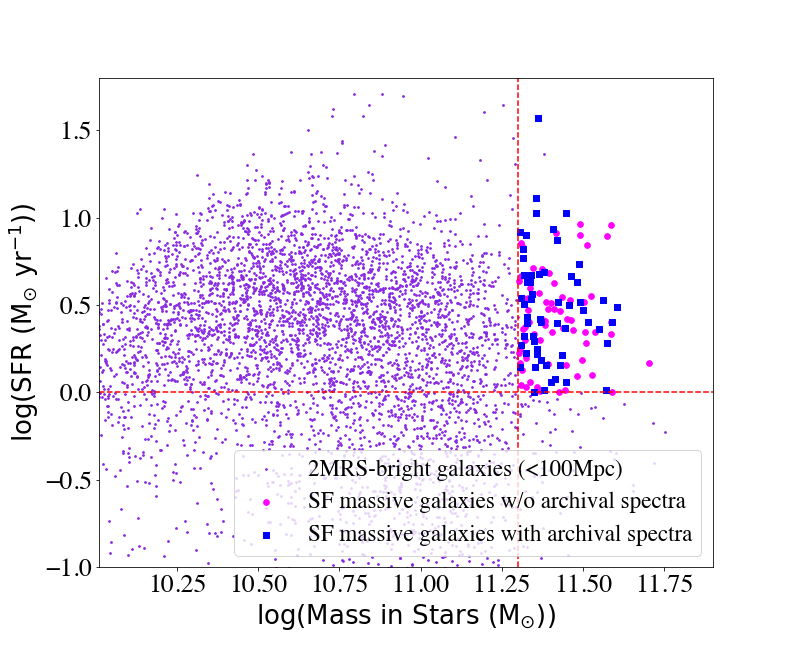}
        \caption{The sample for this paper consists of 126 galaxies with M$_{\rm{stellar}}$ $\geq10^{11.3}$ \rm{M$_\odot$}, SFR $>1$ \rm{M$_{\odot}$ $\rm{yr}^{-1}$}, and distance $ <100$ \rm{Mpc} (red), selected from the 2MRS-Bright < 100 Mpc sample (purple). The region the sample space populates against the wider selection of galaxies can be seen, with our sample being both massive and star-forming, which distinguishes them from typical galaxies and makes them the local analogue to the \citet{Ogle2016, Ogle2019} sample. We also show the $\sim 47\%$ of galaxies that have archival spectra (blue) and the remaining galaxies in the sample for which there is no available spectra.}
        \label{fig:Archival_spectra_samplespace}
    \end{figure}

    Together these surveys provide 59 unique heterogeneous slit and fibre spectra for our sample. The range of aperture sizes are shown in Table \ref{aperture} and span $0.24-1.62$ kpc for our furthest galaxies at \rm{100~Mpc}, this may result in aperture bias increasing our chance of detecting faint AGNs compared to samples like \citet{Kauffmann2003LINER} and \citet{Ogle2016, Ogle2019}.The sources of each of the spectra used to study our sample are shown in Table \ref{specdetails}. The spectra available enable us to measure [\rm{NII}]$\rm{\lambda 6583}$, H$\alpha\rm{\lambda 6563}$, H$\beta\rm{\lambda 4863}$, and [\rm{OIII}]$\rm{\lambda 5008}$ emission lines for 47\% of our sample, shown in Figure~\ref{fig:Archival_spectra_samplespace}. The sub-sample of galaxies in our selection criteria which have available spectra is representative of the larger sample of 126 galaxies in mass, SFR as can be seen in Figure \ref{fig:Archival_spectra_samplespace}. The sub-sample is also representative of the larger sample in z with the sub-sample and full sample having a median redshift of 0.018, and morphology with the sub-sample being 70\% spiral while the full sample is 63\% spiral, as such we expect the results from the sub-sample to be true for the whole sample as well.

\begin{center}
\begin{tabular}{ |p{2.4cm}|p{2.4cm}|} 
    \hline
    \multicolumn{2}{|c|}{Details of sample's spectra} \\
    \hline
    Spectra source & No. of spectra  \\
     \hline
     SDSS   & 8   \\
     6dFGS & 41  \\
     2MASS FAST    & 4  \\
     \citet{Ho1995}    & 6  \\
     \hline
    \end{tabular}
    \captionof{table}{The sources of each of the spectra used to study our sample of local galaxies. Together, they provide an inhomogeneous sample of nearby, massive, star-forming galaxies for which we can measure emission lines.}
    \label{specdetails}
\end{center}

\subsection{Modelling spectra}
 
    As the archival spectra are inhomogeneous, have flux calibrations of varying quality, and only some have accompanying emission line measurements, we perform our own continuum subtraction and emission line measurements to produce homogenous line measurements, which we can later use to classify galaxies. We must also account for absorption lines arising from the host stellar populations, such as K-giants \citep{Kennicutt1992}, which can be significant for H$\alpha\rm{\lambda 6563}$.

\begin{figure}
        \centering
        \includegraphics[width=0.49\textwidth,angle=0]{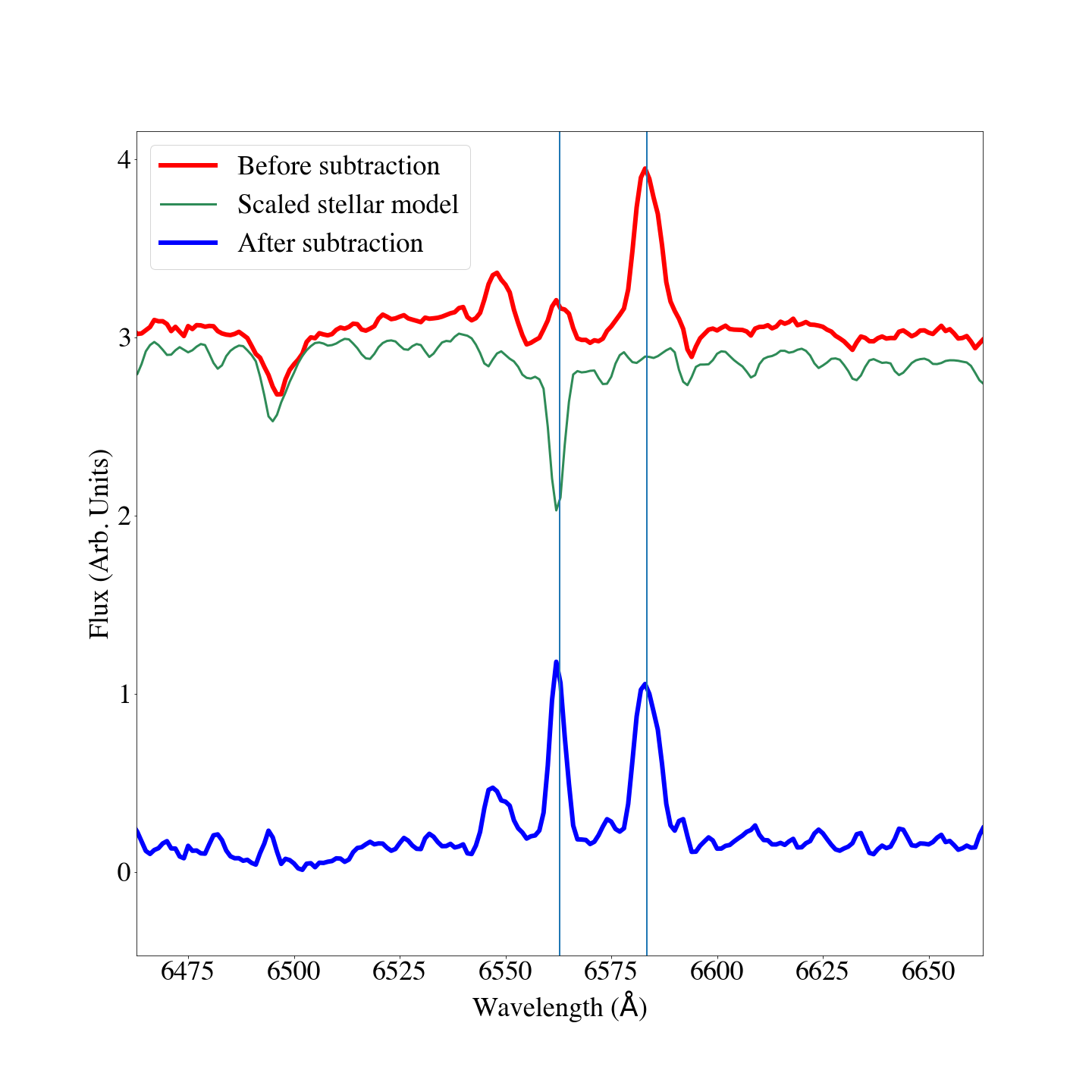}
        \includegraphics[width=0.49\textwidth,angle=0]{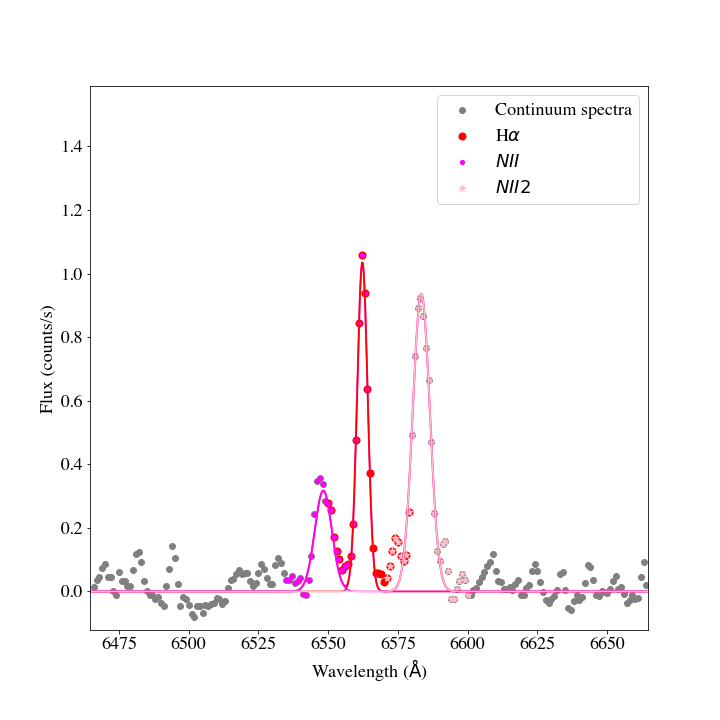}
        \caption{An example of the corrections and fits applied to archival spectra of galaxies in our local sample. The top panel, shows the original spectra in red, at wavelengths surrounding the redshifted emission lines H$\alpha\rm{\lambda 6563}$ and [\rm{NII}]$\rm{\lambda 6583}$. The \citet{BC2003} model is shown in green and the scaled model is then subtracted from the original spectra, resulting in the spectra shown in blue. In the bottom panel the corrected spectra is modelled using three Gaussians and hence the emission line ratios can be measured.}
        \label{fig:Example_spectra_tilting}
    \end{figure}

    If we don't fit any stellar continuum model and instead simply subtract a linear correction we find that we underestimate H$\alpha\rm{\lambda 6563}$ in-fill, which can be seen in the corrected spectra in Figure \ref{fig:Example_spectra_tilting}. The underestimation of H$\alpha\rm{\lambda 6563}$ results in an overestimation of the number of LINERs in the sample and hence it is important to appropriately fit and subtract stellar continuum models. 
    
    We have modelled the continuum using a \citet{BC2003} simple stellar population model, with an age of 11 Gyr and Z $= 0.008$. We have chosen a single model that is representative for our galaxy population rather than fitting for age and/or metallicity due to the varying quality of our spectra (particularly the flux calibration). In order to choose an appropriate model we considered that our galaxies are very massive and while they have some star-formation, the most prominent stellar populations in their (generally red) stellar bulges will be older stars. \citet{Greene_2013} found that the average age of the stellar populations in their massive, elliptical galaxies was $10$ Gyr and that these were relatively metal poor with $[Fe/H] \approx -0.5$. Further literature such as \citet{Mehlert2003} is in agreement with this range of ages, giving a mean age of elliptical galaxies in the Coma cluster of $10.5 \pm 3$ Gyr. Given these studies, we adopt a model of $11$ Gyr to represent an older stellar population and a metallicity of $Z = 0.008$, which is close to, but lower than Solar metallicity to represent the observed lower metallicity in \citet{Greene_2013}. 
    
    We tested several different \citet{BC2003} simple stellar population continuum models with different ages (ranging from 1.4Gyr to 11Gyr) and metallicities ranging from (0.008 to 0.05), to confirm our chosen \citet{BC2003} model is appropriate and that our sample selection is not highly sensitive to the continuum model.  When we applied a continuum model with a higher metallicity, $Z = 0.02$, and the same age as our preferred model, $11$ Gyr, we found that the resulting change was insignificant around the emission lines of interest. Further increasing the metallicity to $Z = 0.05$, while keeping the age consistent, made the absorption lines in the model deeper, producing stronger H$\alpha\rm{\lambda 6563}$ in-fill. This pushed the galaxies towards the star-forming region of the BPT diagram and produced large residuals for individual spectra. Visually inspecting these models, we saw that they did not fit our galaxy's stellar continuum spectra well and hence subtracting this model resulted in non-physical emission-line ratios. We also tried fitting models with the same metallicity as our preferred model, but with a younger stellar age of $5$ Gyr and found that all galaxies that were classified as LINERs with our preferred model were also classified as LINERs with this stellar age. This younger model produced similar results and as literature such as \citet{Greene_2013} predicts that the older stellar age is closer to the age of these galaxies and we see comparable fits with either model we continue to use our preferred model with $Z = 0.008$ and an age of $11$ Gyr. 

    
\begin{figure}
        \centering
        \includegraphics[width=0.49\textwidth,angle=0]{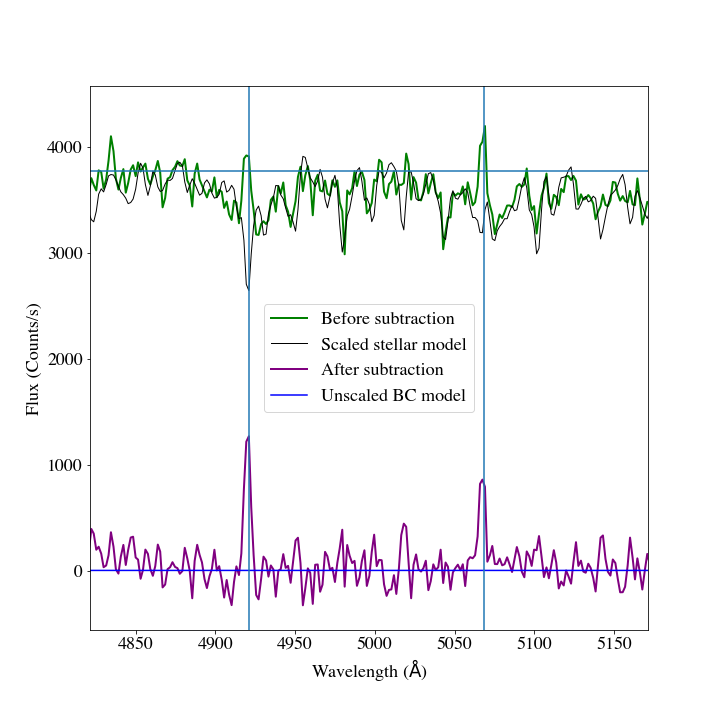}
        \includegraphics[width=0.49\textwidth,angle=0]{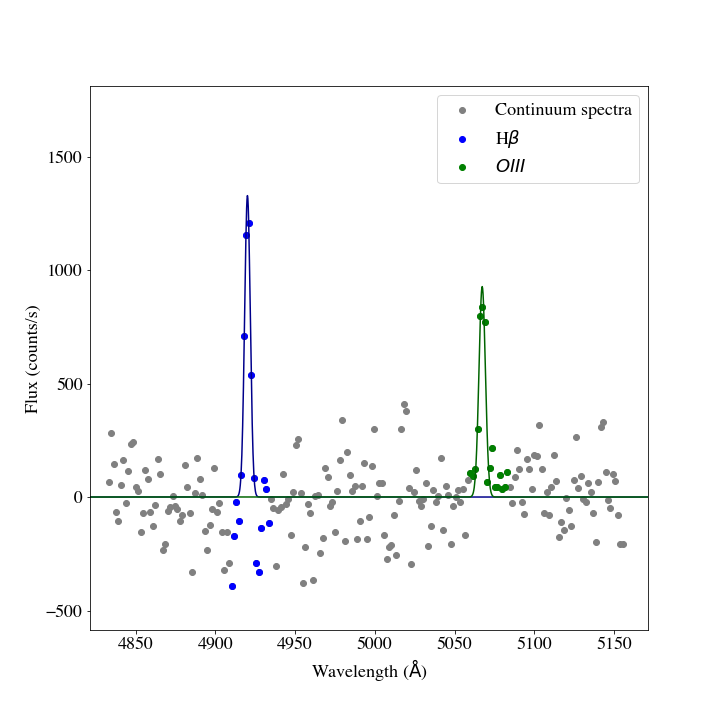}
        \caption{The corrections and fits applied to archival spectra for H$\beta\rm{\lambda 4863}$ and [\rm{OIII}]$\rm{\lambda 5008}$ emission lines. The top panel, shows the original spectra in green, the scaled \citet{BC2003} is shown in black and the subtracted spectra is shown in purple. The corrected spectra are then modelled using Gaussians as shown in the bottom panel, and hence the emission line ratios can be measured.}
        \label{fig:Example_spectra_tilting_OIII_lowerSN}
    \end{figure}

    Using the \citet{BC2003} model with $Z = 0.008$ and an age of $11$ Gyr, we scale it to account for the redshift at which each galaxy spectrum is measured. We then measure the mean continuum flux of the galaxy in a region (6485\AA-6510\AA) that does not have emission lines, but which is close enough to H$\alpha\rm{\lambda 6563}$ and [\rm{NII}]$\rm{\lambda 6583}$ to represent the continuum level around these emission lines. Similarly for H$\beta\rm{\lambda 4863}$ and [\rm{OIII}]$\rm{\lambda 5008}$ we measure the continuum, however as H$\beta\rm{\lambda 4863}$ and [\rm{OIII}]$\rm{\lambda 5008}$ are further apart, we measure the continuum flux in between the emission lines at 4925\AA-4950\AA, so that it is more representative of the continuum at both emission lines respectively. We then scale the \citet{BC2003} model to fit this measured continuum flux, which can be seen in the top panel of Figures~\ref{fig:Example_spectra_tilting} and \ref{fig:Example_spectra_tilting_OIII_lowerSN}. This scaled continuum model is then subtracted from the original galaxy spectra. However, in order to account for any tilt in the galaxy spectra, which may be the result of flux calibration errors, we measure the continuum on either side of the emission lines and then correct for this tilt. We then have spectra that has a continuum level of zero, is not tilted, and for which partial in-fill of absorption lines by nebula emission are accounted for. We do not account for absorption line broadening from velocity dispersion in galaxies when modelling the continuum, as the spectral resolution of our spectra is low (6dFGS $R\sim1000$) and so the broadening from the instrument and pressure broadening will dominate over velocity dispersion broadening \citep{Jones2004}. 
    
    While we are fitting three Gaussians to the H$\alpha\rm{\lambda 6563}$, [\rm{NII}]$\rm{\lambda 6583}$, and [\rm{NII}]$\rm{\lambda 6548}$ lines, 9 free-parameters (amplitude, $\lambda_{rest frame}$, and $\sigma$ of each gaussian) are not required to model these emission lines. We use the lab wavelengths and one redshift for all three lines. We adopt a flux ratio of [$6583$ \rm{\AA}]/[$6548$ \rm{\AA}] $\sim 2.95$ which is consistent with the emission line ratio we would expect from theory \citep{NII1989}, and we assume the two NII lines have the same $\sigma$. We thus have 5 free parameters for the 3 emission lines ([\rm{NII}]$\rm{\lambda 6548}$, H$\alpha\rm{\lambda 6563}$ and [\rm{NII}]$\rm{\lambda 6583}$). From the corrected spectra we then measure the flux, by simultaneously fitting 3 Gaussians to the [\rm{NII}]$\rm{\lambda 6548}$, H$\alpha\rm{\lambda 6563}$ and [\rm{NII}]$\rm{\lambda 6583}$ emission lines, with our initial guesses for amplitude reflecting that we expect the ratio of [\rm{NII}]$\rm{\lambda 6583}$/[\rm{NII}]$\rm{\lambda 6548}$ to be $\sim 2.95$. We have found that fitting the three Gaussians simultaneously provides better individual fits for the emission lines as it prevents the [\rm{NII}]$\rm{\lambda 6550}$ peak from being included in or confused with the H$\alpha\rm{\lambda 6563}$ peak. Similarly, we fit 2 independent Gaussians to the H$\beta$ $\rm{\lambda 4863}$, and [\rm{OIII}]$\rm{\lambda 5008}$ emission lines as they are well separated, Figures~\ref{fig:Example_spectra_tilting} and \ref{fig:Example_spectra_tilting_OIII_lowerSN} respectively. 

 
    To determine which combination of emission lines measurements we can use for the BPT classification criteria, we use a signal-to-noise threshold of 3 for each line. We determine the noise per pixel using the median absolute deviation in the same region for which we measure the continuum flux and set a noise limit for our emission lines such that we only measure emission line ratios if signal$>3\times  \sigma_{MAD}$, where $\sigma_{MAD}$ is the median absolute deviation of the continuum flux. When classifying galaxies using [\rm{NII}]$\rm{\lambda 6583}$/H$\alpha\rm{\lambda 6563}$ we only require that [\rm{NII}]$\rm{\lambda 6583}$ has a signal$>3\times  \sigma_{MAD}$, the reasoning behind not requiring H$\alpha\rm{\lambda 6563}$ to also have this condition is that LINERs must have [\rm{NII}]$\rm{\lambda 6583}$/H$\alpha\rm{\lambda 6563}$ $ \geq 0.6$ and so having a low H$\alpha\rm{\lambda 6563}$ doesn't exclude galaxies from being LINERs. For galaxies that do have H$\alpha\rm{\lambda 6563} < 3\times  \sigma_{MAD}$ we set a limit of H$\alpha\rm{\lambda 6563} = 3\times  \sigma_{MAD}$, which gives us a minimum value of [\rm{NII}]$\rm{\lambda 6583}$/H$\alpha\rm{\lambda 6563}$. Similarly if [\rm{NII}]$\rm{\lambda 6583}$ is in emission and passes our signal to noise criteria, but H$\alpha\rm{\lambda 6563}$ is in absorption, we take this same limit to get a minimum positive emission line ratio. 
    
    For [\rm{OIII}]$\rm{\lambda 5008}$/H$\beta \rm{\lambda 4863}$ we require that both emission lines have signal$>3\times  \sigma_{MAD}$. However, if one emission line meets this condition and the other does not, we set a limit for the weaker emission line of $3\times  \sigma_{MAD}$, which allows us to give emission line ratios for galaxies with significant [\rm{OIII}]$\rm{\lambda 5008}$ or H$\beta\rm{\lambda 4863}$, even if the spectra is noisy.  We have found that this produces a well fitting, scaled model, and seems to reliably measure the noise.  We have also observed that there is far greater noise at bluer wavelengths than around H$\alpha\rm{\lambda 6563}$ and [\rm{NII}]$\rm{\lambda 6583}$, which is expected as scattered light contributes to noise more at bluer wavelengths due to lower overall counts \citep{Jones2004}.

    Some of the galaxies studied in this work do not have spectra with sufficient signal-to-noise at bluer wavelengths and hence reliably measuring [\rm{OIII}]$\rm{\lambda 5008}$/H$\beta\rm{\lambda 4863}$ is difficult. Therefore, we pragmatically use the \citet{Kauffmann2003LINER} criteria to classify galaxies where only [\rm{NII}]$\rm{\lambda 6583}$/H$\alpha\rm{\lambda 6563}$ is measurable. We classify galaxies based on the criteria in Figure ~\ref{fig:Example_BPT} using the \citet{Kauffmann2003LINER} line, corresponding to a [\rm{NII}]$\rm{\lambda 6583}$/H$\alpha\rm{\lambda 6563}$ ratio of $0.6$, which is also consistent with the criteria used by \citet{Kewley2006} and \citet{Kauffmann2003LINER}. 

\section{Results}
    
\begin{table*}
    \centering
    \small
    \setlength\tabcolsep{1pt}
    \begin{tabular}{|p{2.3cm}|p{1.4cm}|p{1.3cm}|p{0.7cm}|p{1.3cm}|p{1.2cm}|p{0.90cm}|p{0.9cm}|p{1cm}|p{1.3cm}|p{1.4cm}|p{1.5cm}|p{1.3cm}|}
         \hline
         Name & RA & DEC & z & SFR$_{12}$ & $\sigma$ & Log & Dist & Morph & Spectra & [NII]/H$\alpha$ & [OIII]/H$\beta$ & Spec. \\
         & (J2000) & (J2000) &  & $(\rm{M_{\odot}\rm{yr}^{-1}})$ & (SFR$_{12}$) & Mass & (Mpc) &  &  source & & & Class \\
         \hline\hline
         ESO018-G002 & 124.8091 & -78.6961 & 0.02 & 4.25 & 1.47 & 11.34  & 85.33 & Sbc & 6dFGS & 0.98 & 0.76 & LINER\\  \hline
         UGCA145 & 131.8208 & -20.0357 & 0.02 & 8.3 & 2.88 & 11.31 & 74.35 & Sbc & 6dFGS & 0.85 & 0.94 & LINER\\ \hline
         NGC2713 & 134.3355 & 2.9214 & 0.01 & 4.25 & 1.47 & 11.48 & 59.99 & SBb & SDSS & 1.35 & 0.58 & LINER\\ \hline
         UGC04869 & 138.6428 & 30.1409 & 0.02 & 4.68 & 1.62 & 11.34 & 98.65 & S0-a & SDSS & 0.76 & -99 & NII LINER\\ \hline
         NGC2876 & 141.3074 & -6.7166 & 0.02 & 1.43 & 0.5 & 11.39  & 93.73 & S0 & 2MRS-Fast & 0.30 & -99.0 & NII SF\\ \hline
         MCG-01-25-005 & 143.9858 & -7.7267 & 0.02 & 1.4 & 0.48 & 11.35 & 91.21 & S0-a & 2MRS-Fast & 1.07 & -99.0 & NII LINER\\ \hline
         MCG-02-25-006 & 144.1169 & -11.3305 & 0.02 & 3.24 & 1.13 & 11.4 & 83.08 & Sa & N/A & -99.0 & -99.0 & N/A\\ \hline
         NGC2945 & 144.4214 & -22.035 & 0.02 & 1.77 & 0.61 & 11.36 & 81.66 & E-S0 & 6dFGS & 1.43 & 1.95 & LINER\\ \hline
         MRK1239 & 148.0796 & -1.612 & 0.02 & 37.17 & 12.87 & 11.36 & 91.92 & E & 6dFGS & 0.39 & 1.54 & Seyfert*\\ \hline
         IC0575 & 148.6372 & -6.8576 & 0.02 & 3.37 & 1.17 & 11.46 & 91.92 & Sa & N/A & -99.0 & -99.0 & N/A\\ \hline
         NGC3275 & 157.7158 & -36.737 & 0.01 & 2.7 & 0.94 & 11.33 & 50.94 & Sab & 6dFGS & 1.83 & -99 & NII LINER \\ 
        \hline
    \end{tabular}
    \captionof{table}{Example of available data table - the data table provides information for all 126 galaxies in the local sample. This data includes, RA, DEC, redshift, star-formation rate and the associated error, mass in stars, distance in Mpc, $W_{1}$ flux, $W_{2}$ flux, $W3_{PaH}$ flux, $W4_{dust}$ flux, $W_{1}$-$W_{2}$, $W_{2}$-$W_{3}$, $M_{W_{1}}$, the LEDA morphologies, the source of the available spectra, the [NII]/H$\alpha$ ratio, the [OIII]/H$\beta$ ratio, and the resulting classification. This example table shows a selection of galaxies and their key data. The error in the galaxy mass (in log scale) predominantly comes from the calibration as these are bright nearby galaxies, the resulting error is $\sim 0.11$ for each galaxy. The 67 galaxies with no available archival spectra are also shown in this table with the spectral class $\rm{N\backslash{A}}$. \rm{*MRK1239} has broad emission lines and so is classified as a Seyfert 1 and excluded from the BPT diagram.}
    \label{results}
\end{table*}    

    Of the 59 galaxies with archival spectra, $79 \pm 6\%$ have [\rm{NII}]$\rm{\lambda 6583}$/H$\alpha\rm{\lambda 6563}$ emission line ratios greater than 0.6 \citep{Kauffmann2003LINER} consistent with them being LINERs, and a further $12 \pm 6\%$ are made up of star-forming galaxies. The remaining $9 \pm 6\%$ of these galaxies have insufficient emission lines to measure relevant emission line ratios, which is likely made up of galaxies that are genuinely passive, or that just have very noisy spectra. The data for all galaxies, both with and without spectra, is available in a data table, an example of which is shown in Table \ref{results}.
       
    For the galaxies where it was possible to measure H$\beta\rm{\lambda 4863}$ and [\rm{OIII}]$\rm{\lambda 5008}$ we have also classified them using the \citet{Schawinski2007} criteria, corresponding to a ratio of [\rm{OIII}]$\rm{\lambda 5008}$/H$\beta\rm{\lambda 4863}<3$ for LINER galaxies and [\rm{OIII}]$\rm{\lambda 5008}$/H$\beta\rm{\lambda 4863} \geq3$ for Seyfert galaxies. The median [\rm{OIII}]$\rm{\lambda 5008}$/H$\beta\rm{\lambda 4863}$ value is 0.85, which reveals that the majority of our galaxies that have available spectra and measurable emission lines are LINERs, as expected from the [\rm{NII}]$\rm{\lambda 6583}$/H$\alpha\rm{\lambda 6563}$ emission line ratios. 

    \begin{figure}
        \centering
        \includegraphics[width=0.47\textwidth,angle=0]{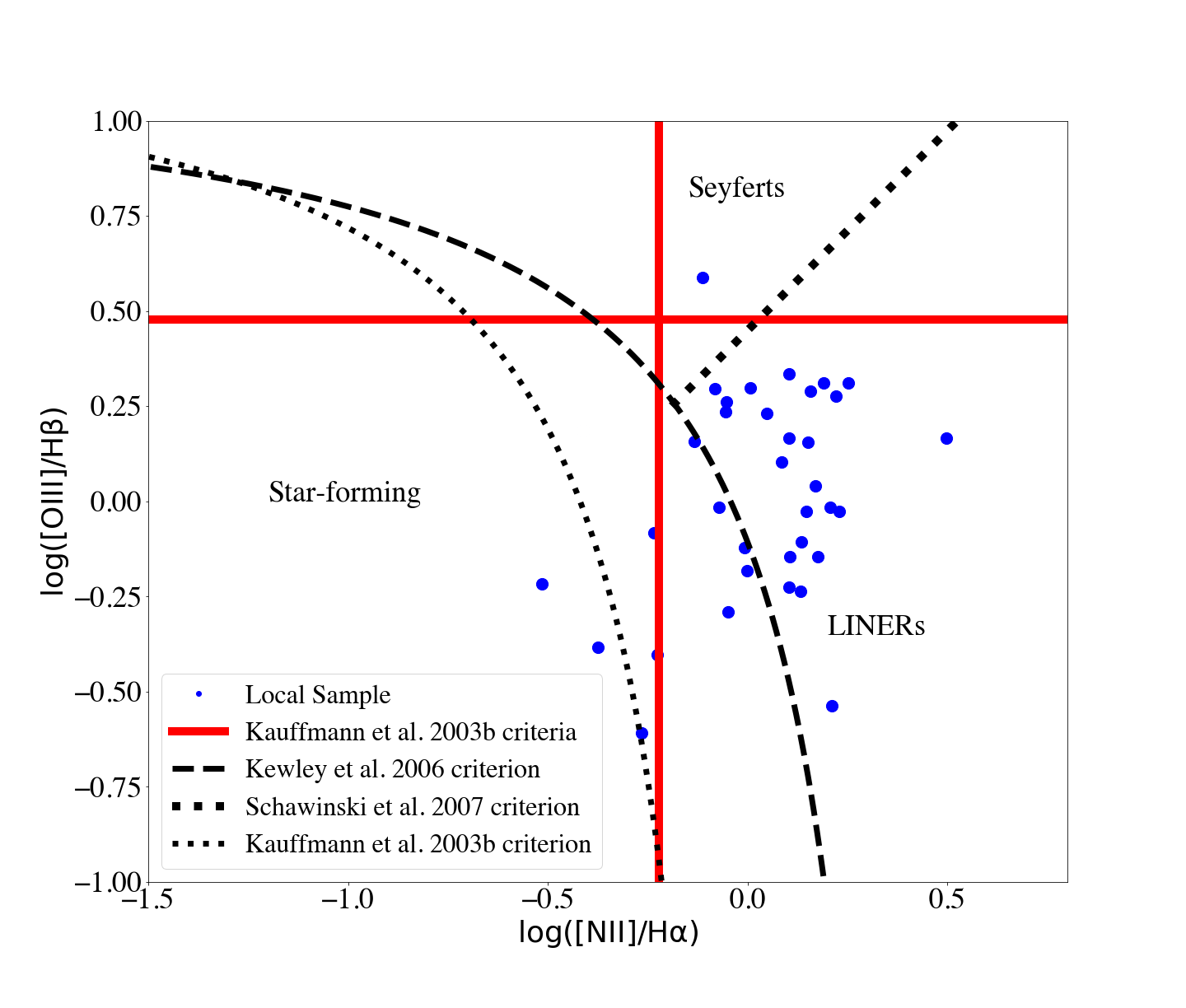}
        \caption{The BPT diagram for the 35 galaxies in our sample that have measurable emission lines from the available spectra. The black dashed line is the \citet{Kewley2006} criteria and the dotted thin black line shows the criteria from \citet{Kauffmann2003LINER}, both of which separate HII galaxies from Seyferts and LINERs. The black dotted line separates Seyferts from LINERs from \citet{Schawinski2007}. As can be seen in the diagram, 29 of the galaxies in our sample are LINERs and there are only 5 star-forming and 1 Seyfert galaxies.}
        \label{fig:BPT_diagram}
    \end{figure}
 
    Additionally, for the galaxies for which [\rm{NII}]$\rm{\lambda 6583}$, H$\alpha\rm{\lambda 6563}$, H$\beta\rm{\lambda 4863}$ and [\rm{OIII}]$\rm{\lambda 5008}$ are all measurable, we plotted a BPT diagram, shown in Figure~\ref{fig:BPT_diagram}, which shows both the \citet{Kewley2006, Kauffmann2003LINER} and \citet{Schawinski2007} criteria. This diagram reveals that while the use of different criteria changes the percentage of LINERs in the sample, the majority of galaxies in the sample are LINERs, regardless of the literature criteria used. 35 galaxies have all four emission lines with sufficient signal to noise and of these galaxies 83\% (29 galaxies) are LINERs, 3\% (1 galaxy) is a Seyfert, and 14\% (5 galaxies) are star-forming. We have used the \citet{Kauffmann2003LINER} criterion for our sample due to limitations of the archival spectra. However, the bulk of the objects selected as LINERs with the \citet{Kauffmann2003LINER} criterion would also be selected as LINERs by the \citet{Kewley2006} and \citet{Schawinski2007} criteria. We use binomial statistics to estimate the error in our sample statistics and find that our sample is comprised of $83 \pm 6\%$ LINERs from the BPT diagram and $79 \pm 6\%$ LINERs classified with [\rm{NII}]$\rm{\lambda 6583}$/H$\alpha\rm{\lambda 6563}$.

    \begin{figure}
        \centering
        \includegraphics[width=0.47\textwidth,angle=0]{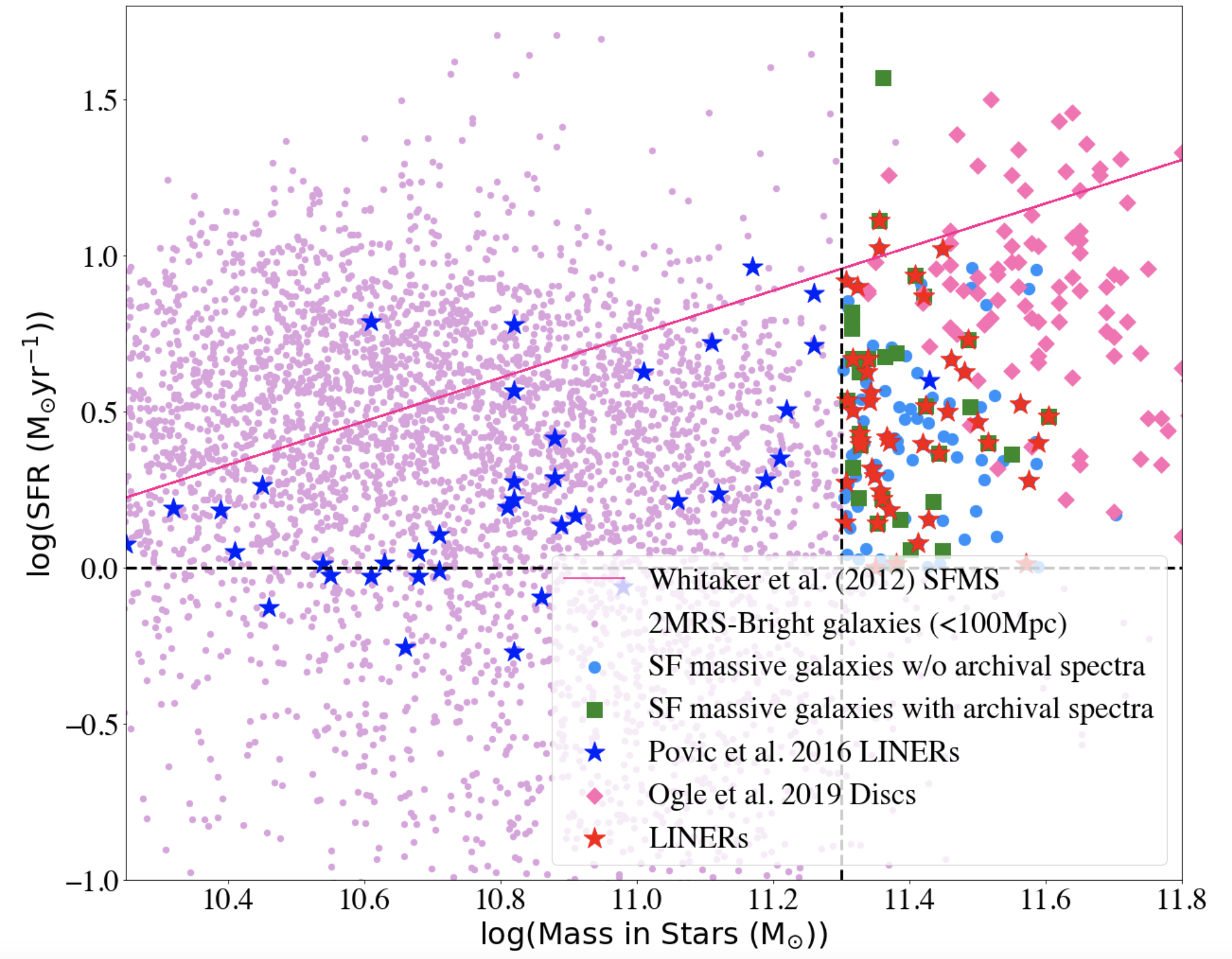}
        \caption{Our sample of galaxies against the wider galaxy population with our local sample divided into those with and without spectra and then further into those with that are LINERs. The LINERs clearly make up the majority of those galaxies with spectra. The SFMS by \citet{Whitaker2012} is shown in pink in the Figure and demonstrates that, as expected, our galaxies largely fall below the SFMS. We also show the \citet{Povic2016} LINERs which have far lower masses than our sample and the \citet{Ogle2019} sample of galaxies which we compare our sample to. Interestingly in all three samples the LINERs largely sit below the star-forming main sequence.}
        \label{fig:Archival_spectra_LINER}
    \end{figure}
  
    In Figure~\ref{fig:Archival_spectra_LINER} we show the location of the emission line classifications in the stellar mass-SFR plane and we observe that the LINERs, comprising $79 \pm 6\%$ of the sample with archival spectra, are well distributed throughout our sample space. The local sample has a much higher percentage ($79 \pm 6\%$) of LINERs than one might naively expect (30\% in massive galaxy populations \citep{Belfiore2016}). This vast difference in LINER prevalence suggests there is a link between our selection criteria - massive, star-forming galaxies, and the prevalence of LINERs or that the star-formation seen in these galaxies is producing the observed LINER emission. 
 
\begin{figure}
        \centering
        \includegraphics[width=0.45\textwidth,angle=0]{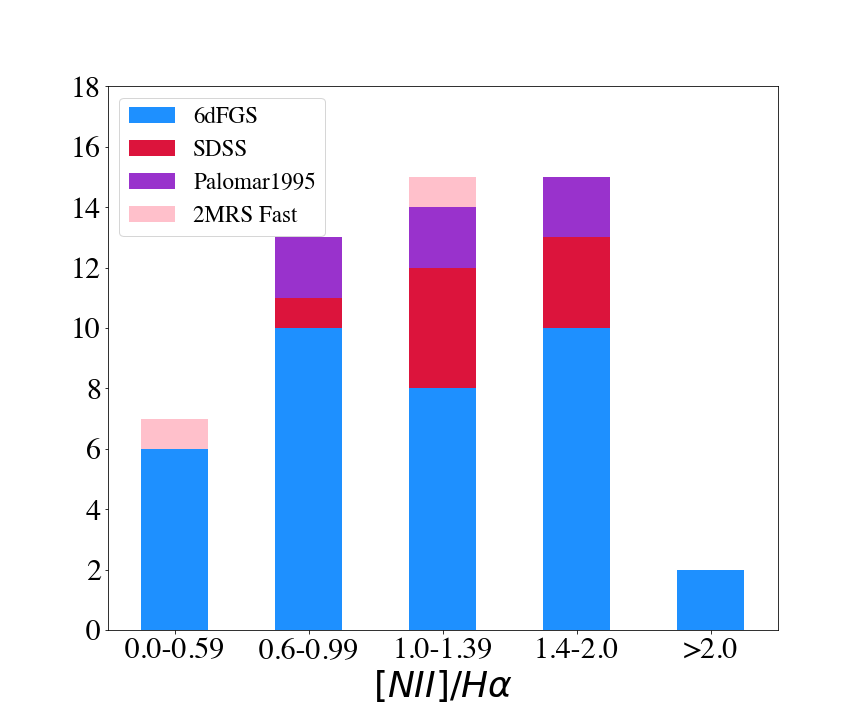}
        \caption{The distribution of [\rm{NII}]$\rm{\lambda 6583}$ to H$\alpha\rm{\lambda 6563}$ ratio that we use to classify LINERs is shown as well as the breakdown of this ratio based on the survey they originated from. As can be seen, the emission line ratio has no strong dependence on the survey from which the spectra is from, as expected, due to the corrections applied to all spectra. We can also see that the majority of galaxies have [\rm{NII}]$\rm{\lambda 6583}$/H$\alpha\rm{\lambda 6563} \geq 0.6$, making them LINERs.}
        \label{fig:LINER_hist}
    \end{figure}

    As one of our criteria for classifying LINERs comes from [\rm{NII}]$\rm{\lambda 6583}$/H$\alpha\rm{\lambda 6563}$, we have investigated the spread of [\rm{NII}]$\rm{\lambda 6583}$/H$\alpha\rm{\lambda 6563}$ values over the sample. Figure~\ref{fig:LINER_hist} shows the histogram of the emission line ratios broken down by the origin of the spectra. This reveals that the classification of LINER or star-forming is not strongly dependent on the source of the spectra. It also shows that the sample as a whole has very high values of [\rm{NII}]$\rm{\lambda 6583}$/H$\alpha\rm{\lambda 6563}$, with a median [\rm{NII}]$\rm{\lambda 6583}$/H$\alpha\rm{\lambda 6563}$ of $1.03$, which is well above the LINER criterion of [\rm{NII}]$\rm{\lambda 6583}$/H$\alpha\rm{\lambda 6563} \geq 0.6$, reflecting that the majority of the sample are LINERs. 
    
    While our initial inspection of the \citet{Ogle2019} super spiral population was qualitative, we have since used the SDSS spectra available for these galaxies and our code to measure the emission line ratios for \citet{Ogle2019} in exactly the same way we did for our local sample. 
    
    We also compared the percentage of the population classified as LINERs, using the SDSS emission line measurements and our own measurements and they agree exactly, that is every galaxy classified as a LINER by the SDSS emission line measurements is also classified as a LINER by our measurements. 
    By measuring [\rm{NII}]$\rm{\lambda 6583}$/H$\alpha\rm{\lambda 6563}$ ratios for spectral classification we found that 64\%, of the \citet{Ogle2019} super spirals are LINERs, with a mean value of [\rm{NII}]$\rm{\lambda 6583}$/H$\alpha\rm{\lambda 6563} = 1.1$, confirming that a large fraction of this population are also LINERs. While this abundance of LINERs is not quite as high as in our local sample, it is considerably greater than the expected $\approx 30\%$ in massive galaxies and strengthens a link between massive, star-forming galaxies and LINERs \citep{Belfiore2016}. 
    As the \citet{Ogle2019} sample is more distant, with a mean redshift of $z=0.22$ compared to our sample with a mean redshift of $z=0.0175$, the extent of the galaxy included in the nuclear spectra will be greater in the \citet{Ogle2019} sample. Specifically, the mean projected diameter of the \citet{Ogle2019} sample is $\sim 10.7$ kpc while for our sample using 6dFGS spectra, the average projected diameter is $\sim3.2$ kpc. This means that the \citet{Ogle2019} study were more likely to see star-formation in their spectra than we are in our sample and we may be more likely to detect faint AGNs, although given the high percentage of LINERs in both samples, this effect does not appear to be overwhelming. 

\section{Discussion}   
\begin{figure*}
        \centering
        \includegraphics[width=0.23\textwidth,angle=0]{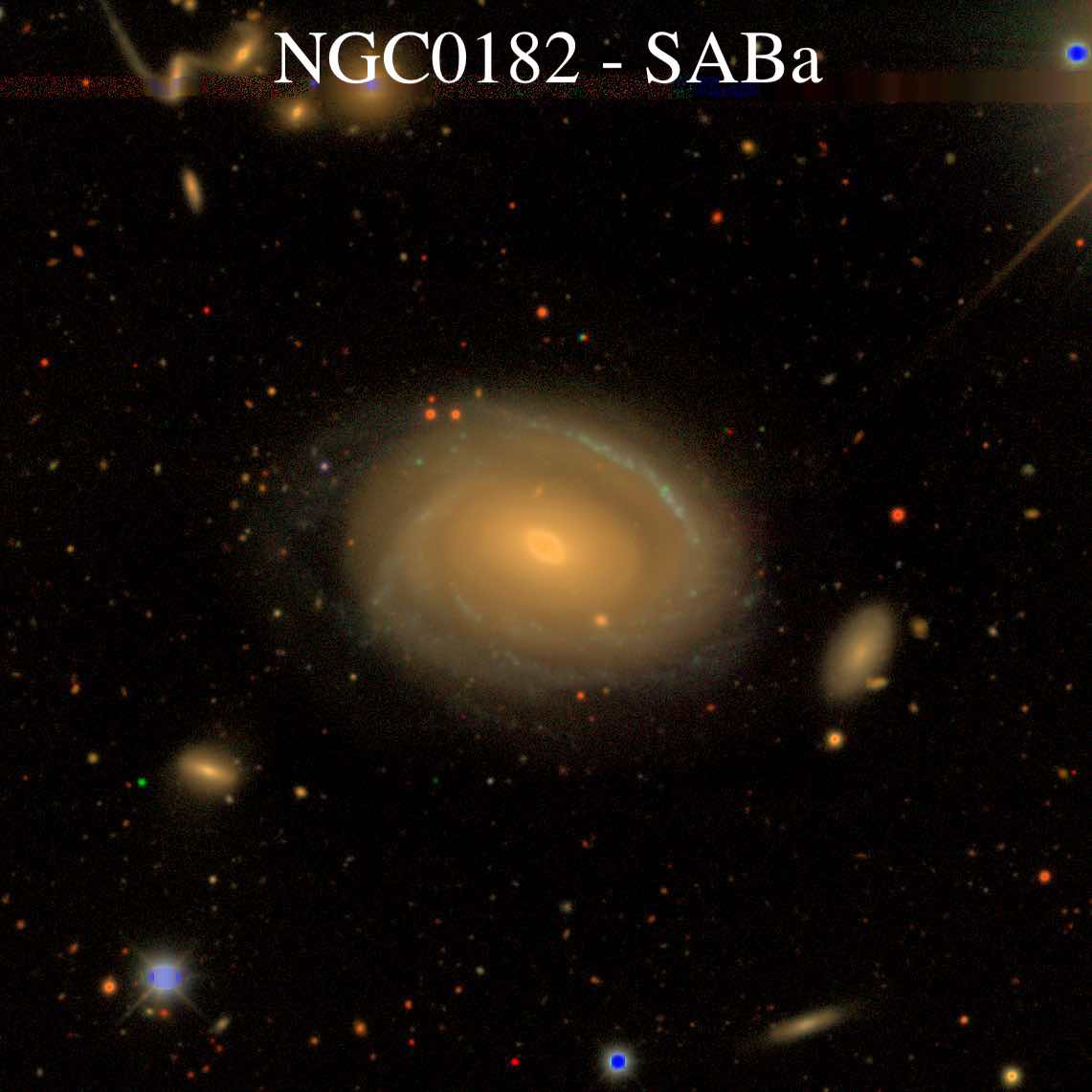}
        \includegraphics[width=0.23\textwidth,angle=0]{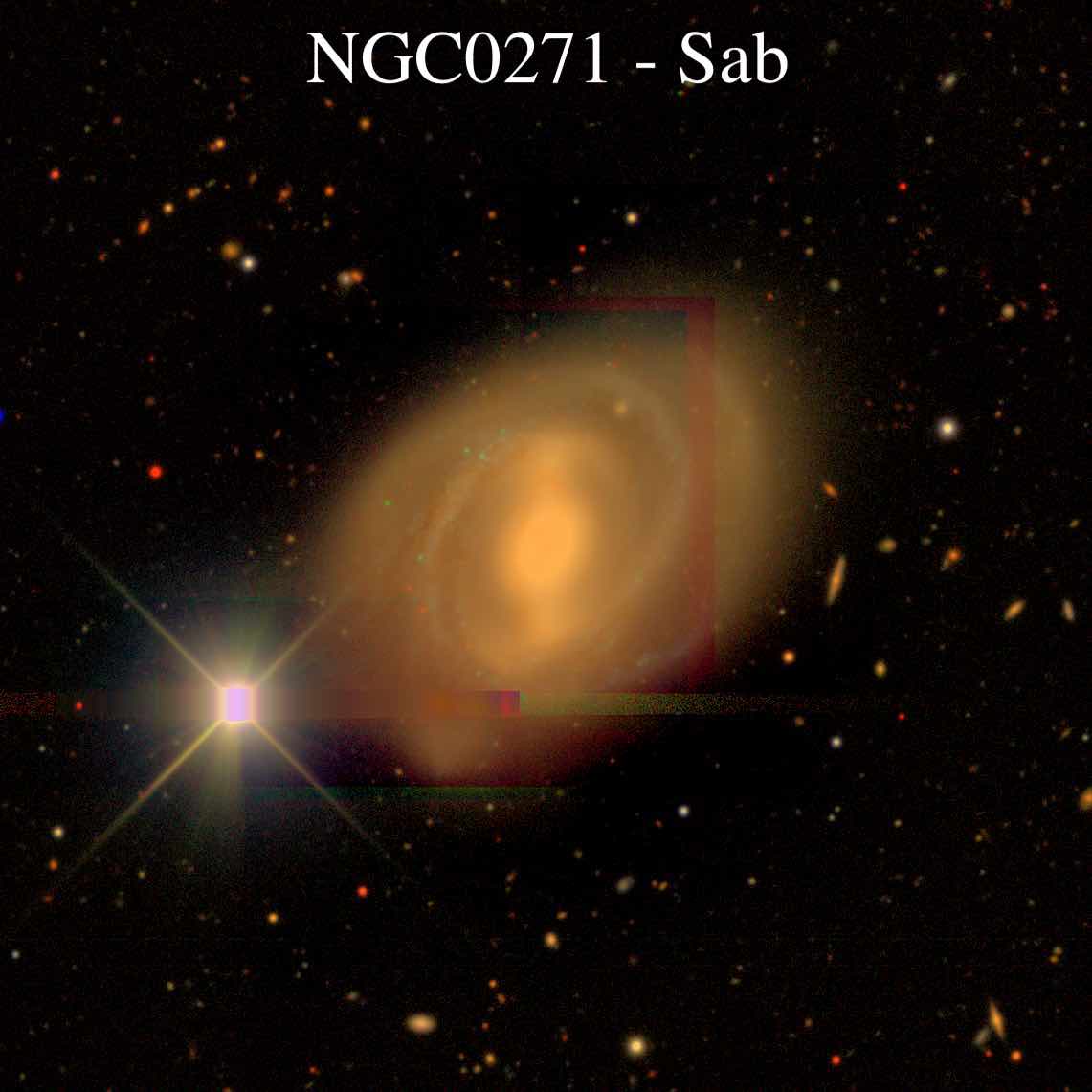}
        \includegraphics[width=0.23\textwidth,angle=0]{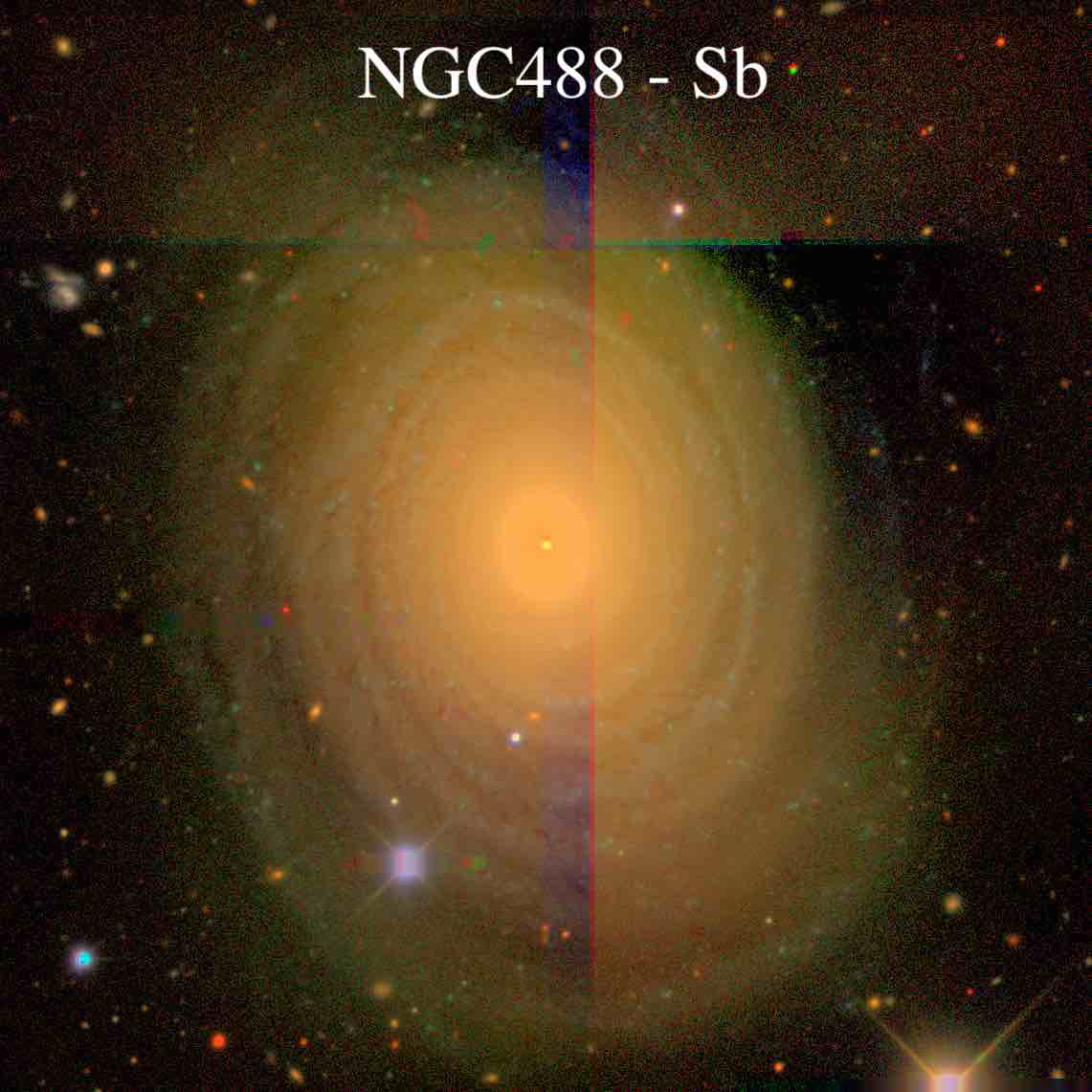}
        \includegraphics[width=0.23\textwidth,angle=0]{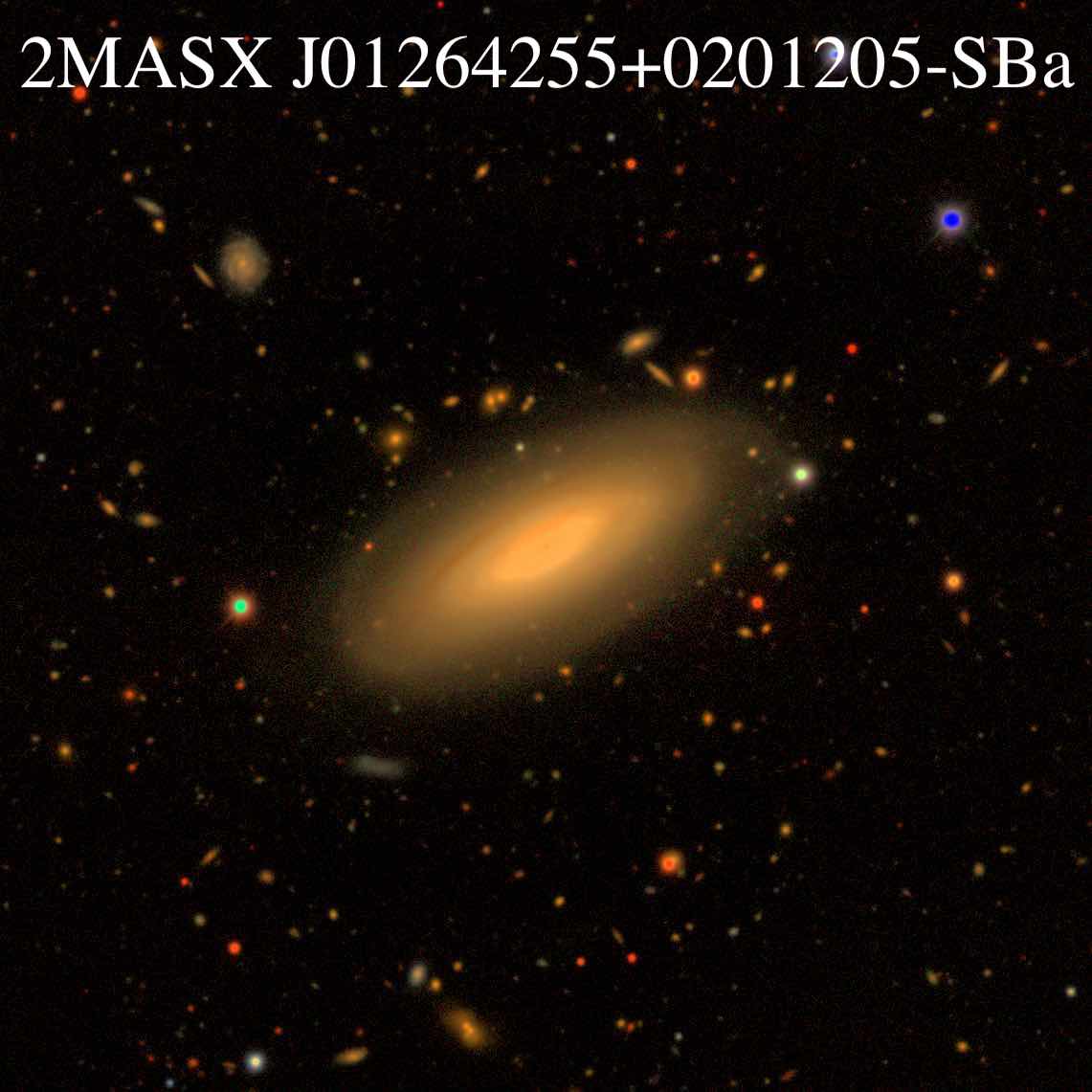}
        \includegraphics[width=0.23\textwidth,angle=0]{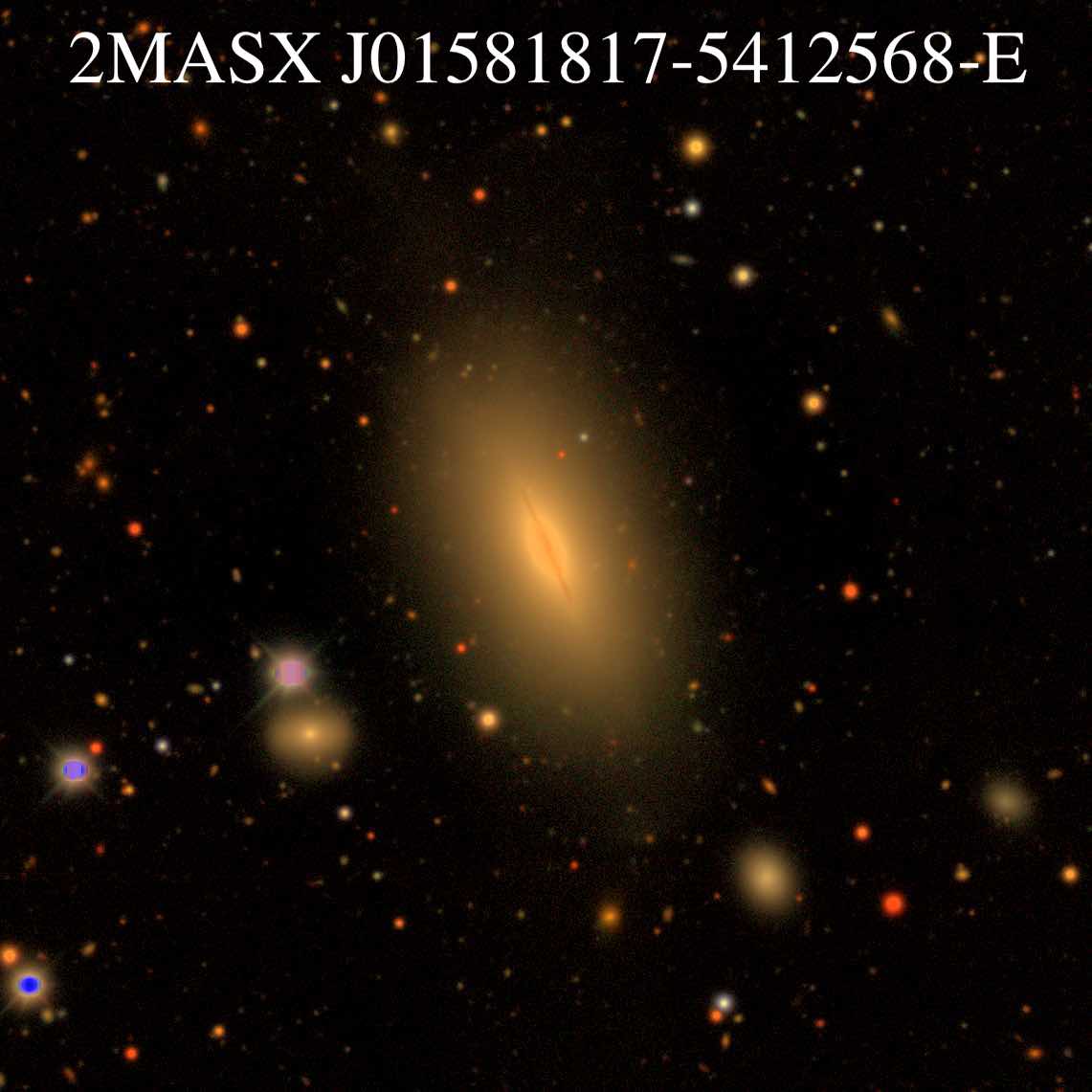}
        \includegraphics[width=0.23\textwidth,angle=0]{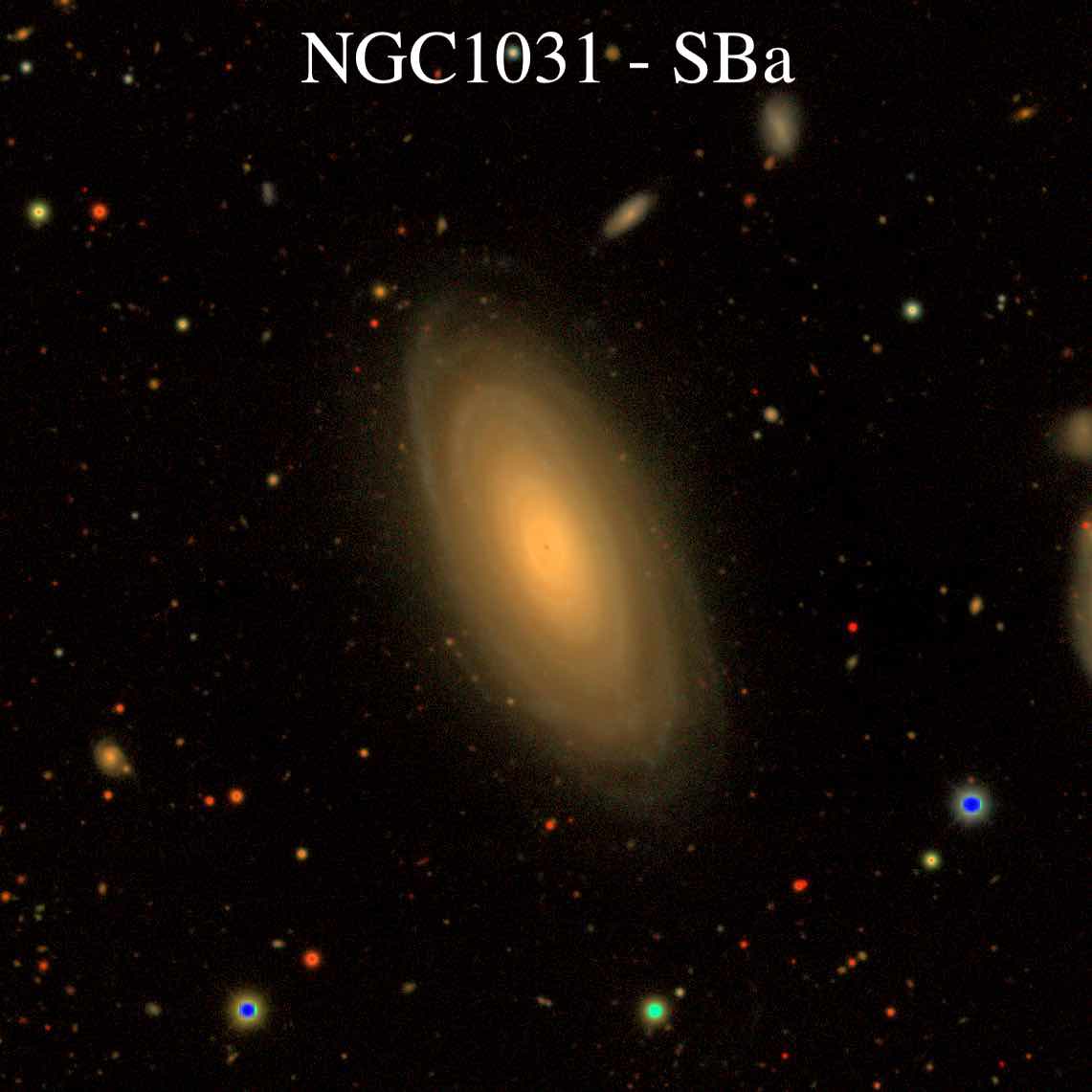}
        \includegraphics[width=0.23\textwidth,angle=0]{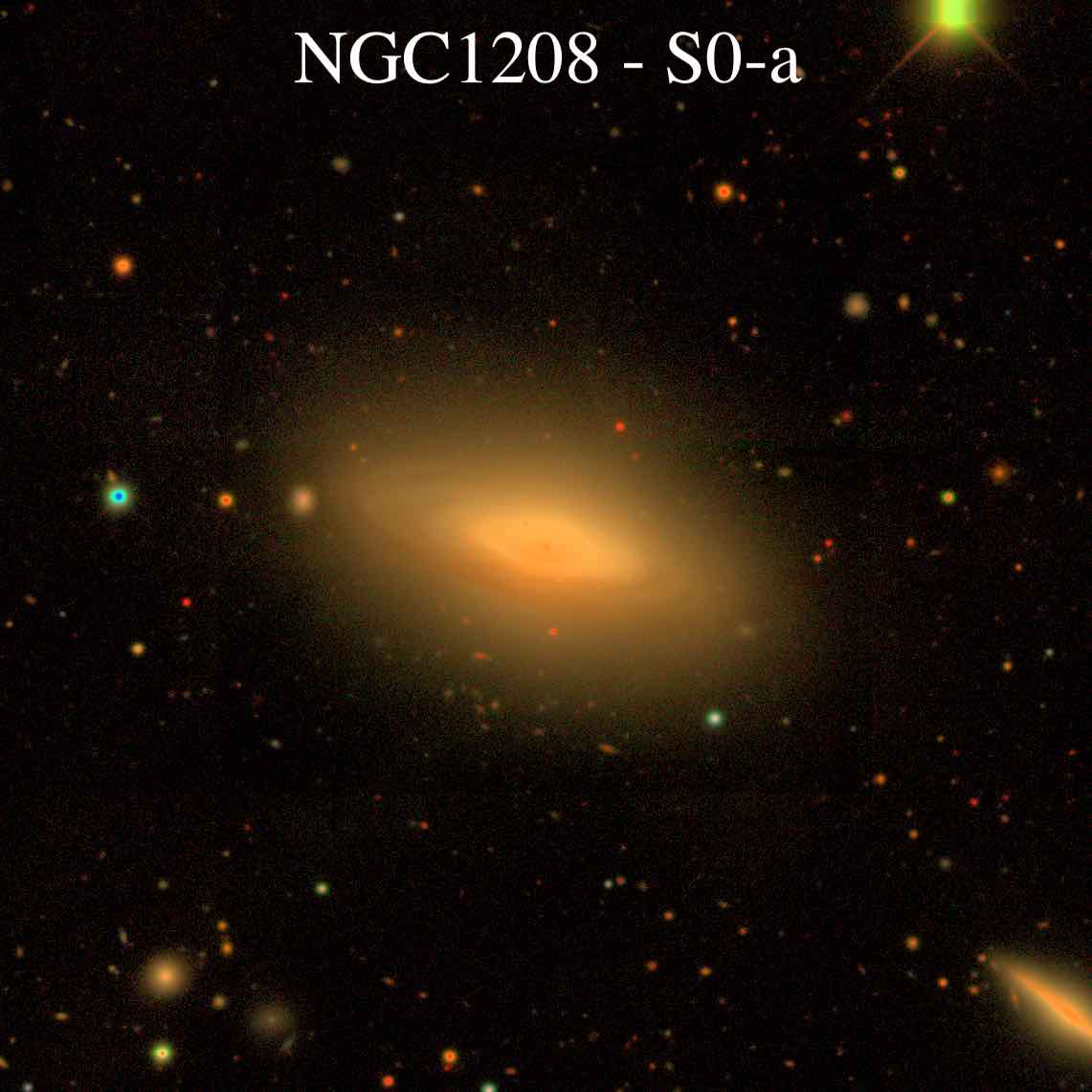}
        \includegraphics[width=0.23\textwidth,angle=0]{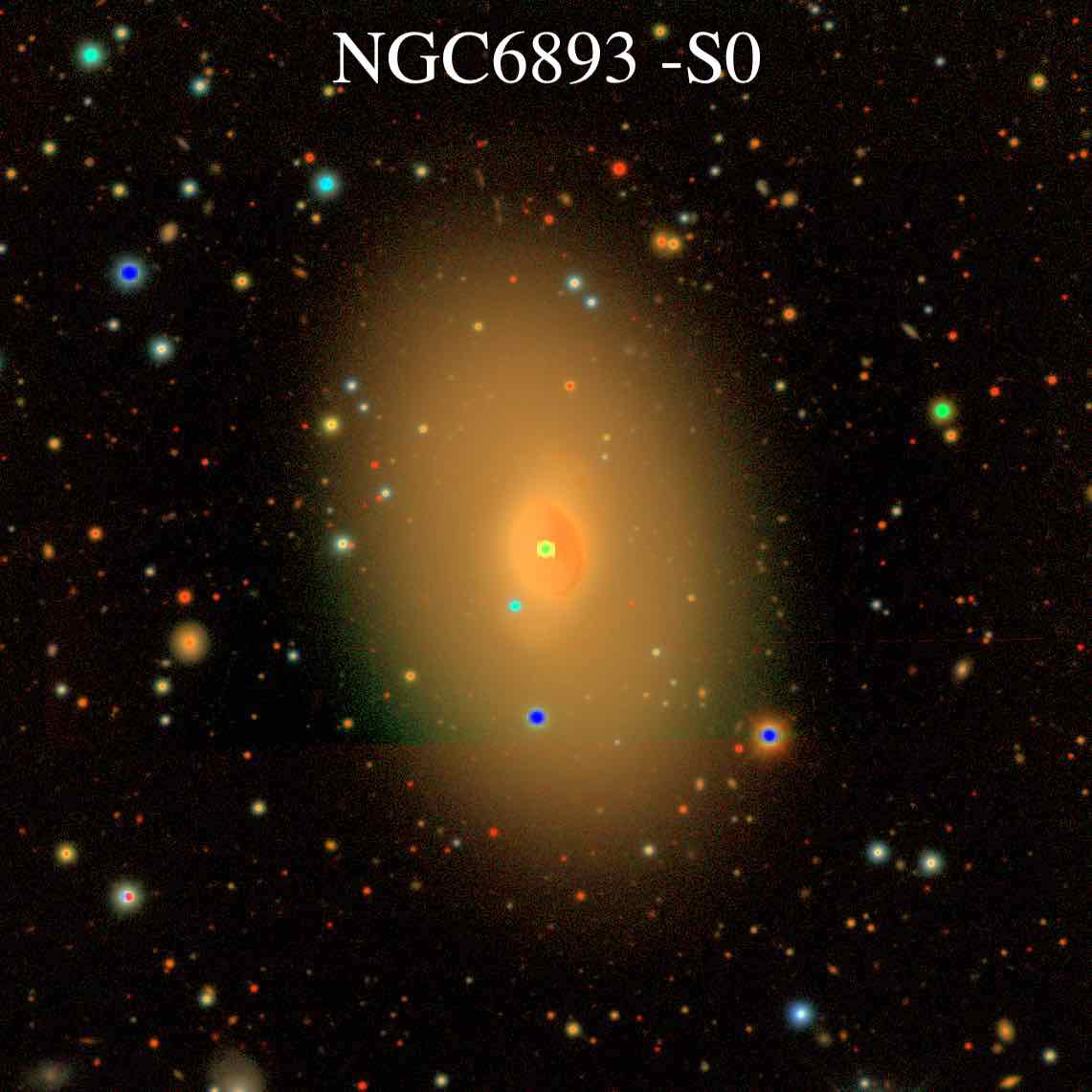}
        \includegraphics[width=0.23\textwidth,angle=0]{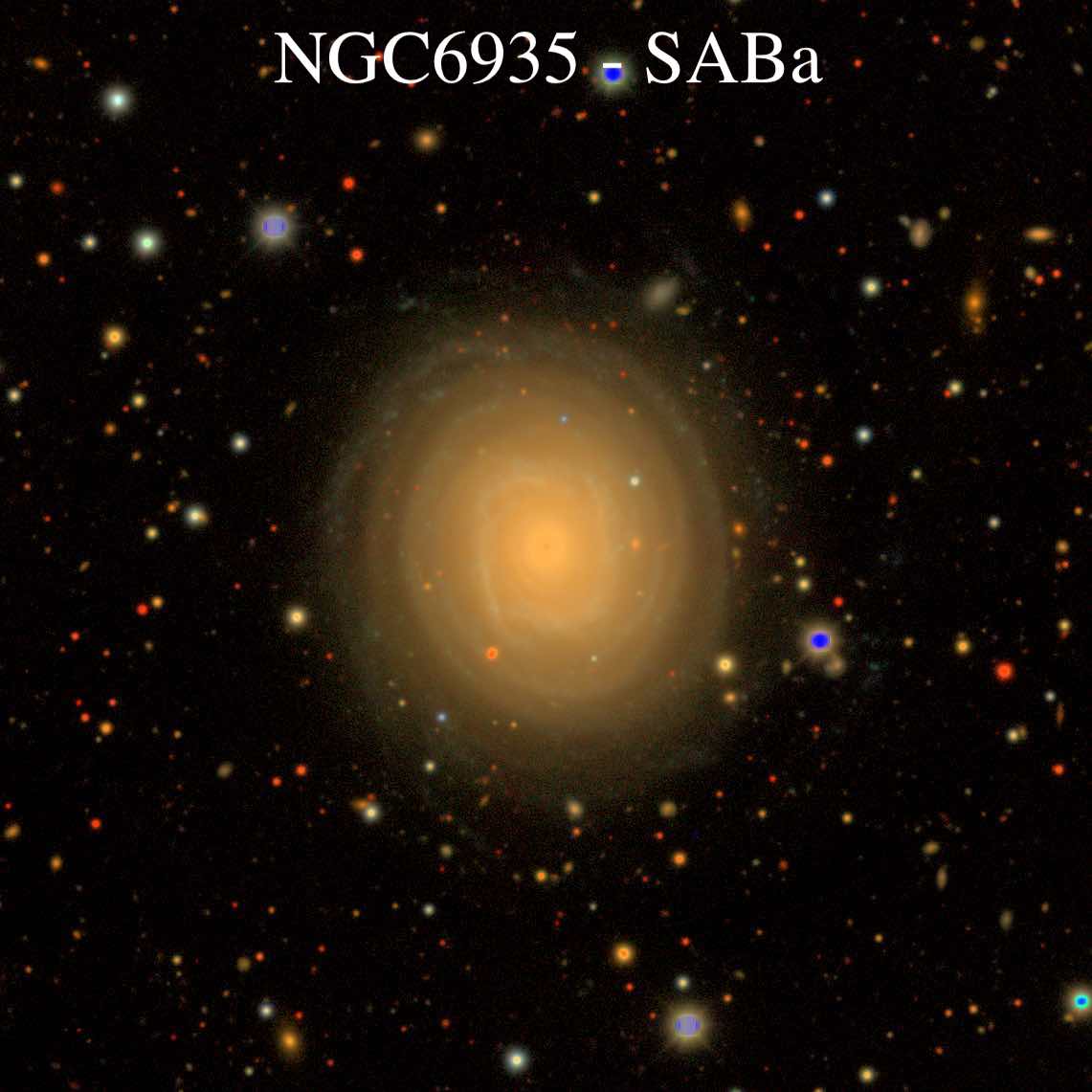}
        \includegraphics[width=0.23\textwidth,angle=0]{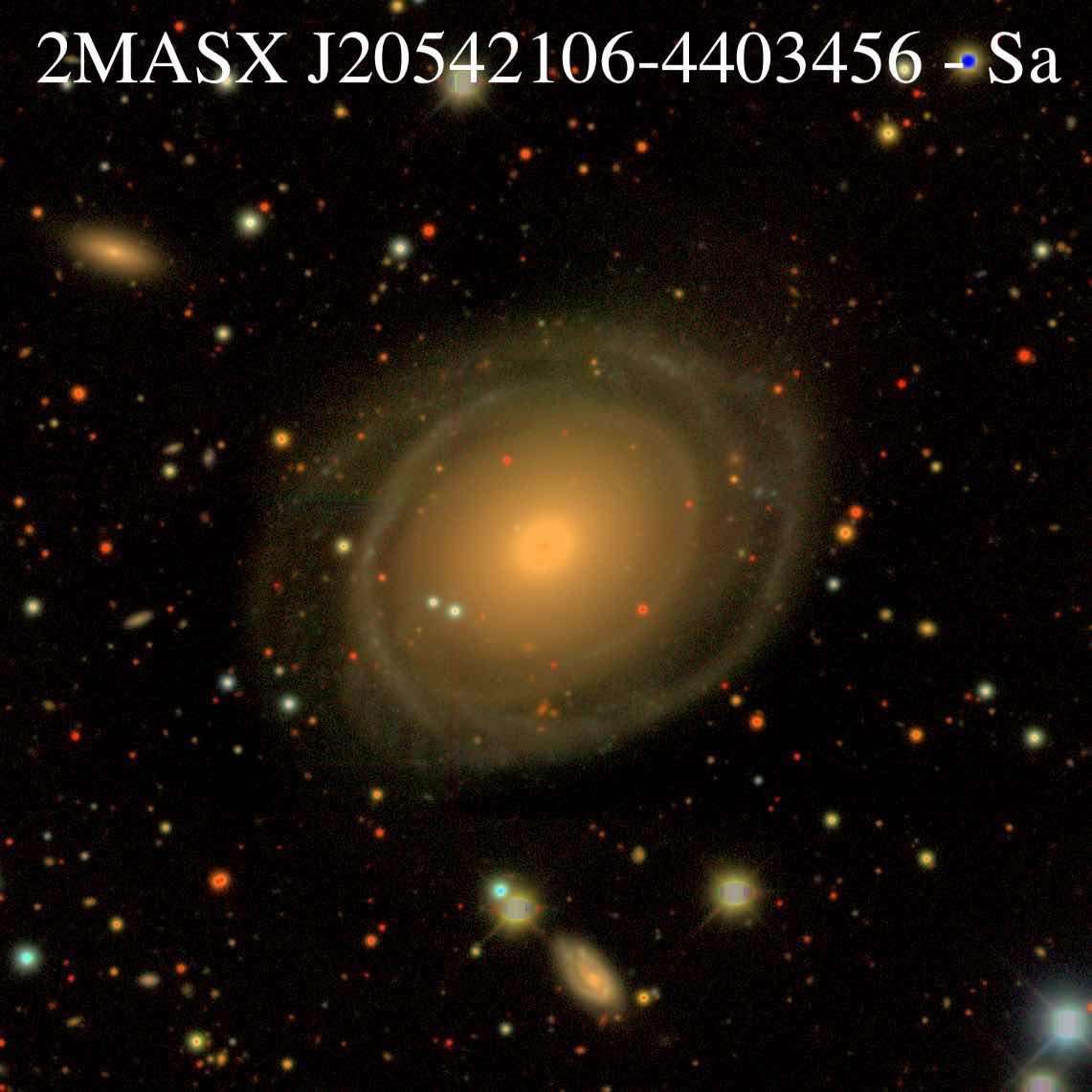}
        \includegraphics[width=0.23\textwidth,angle=0]{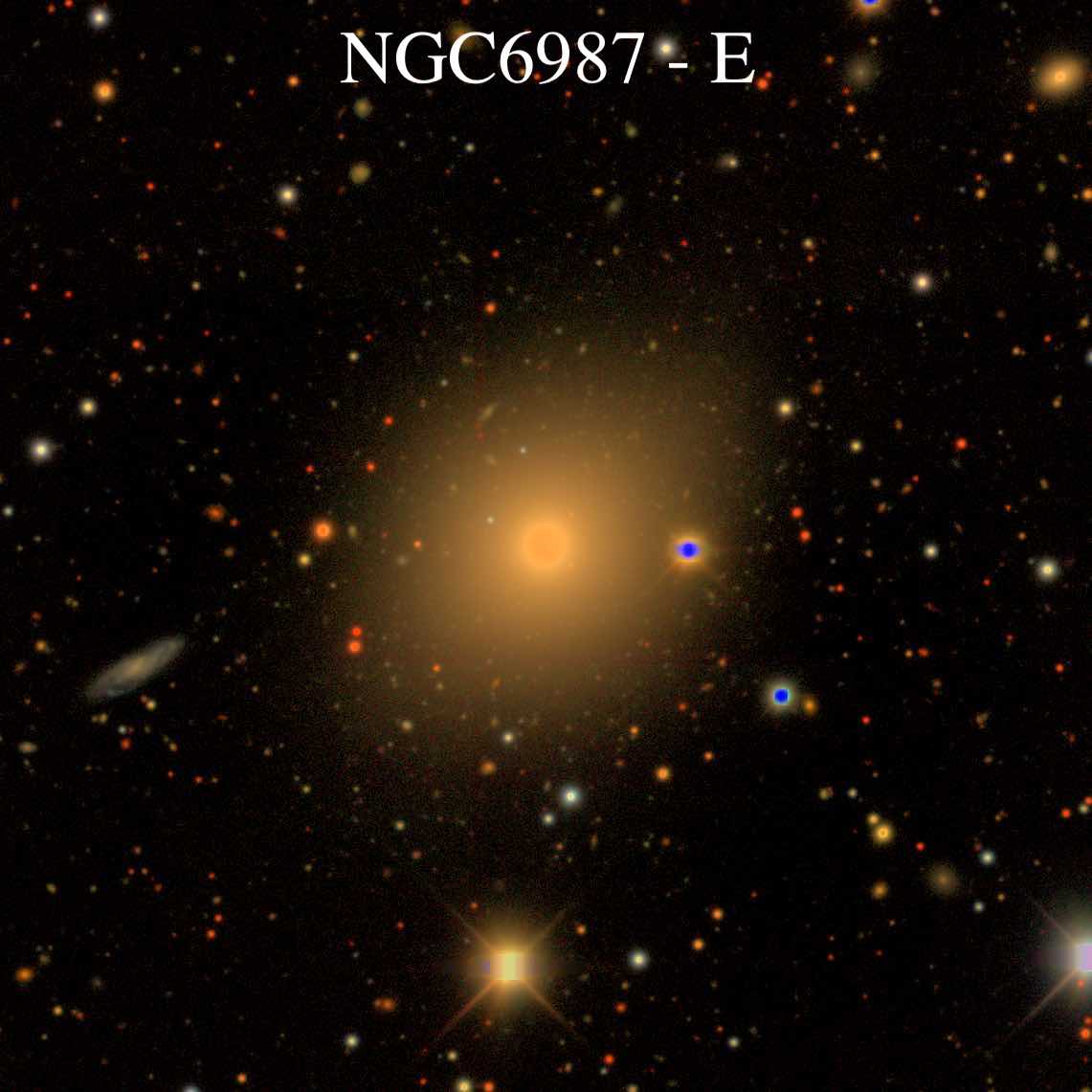}
        \includegraphics[width=0.23\textwidth,angle=0]{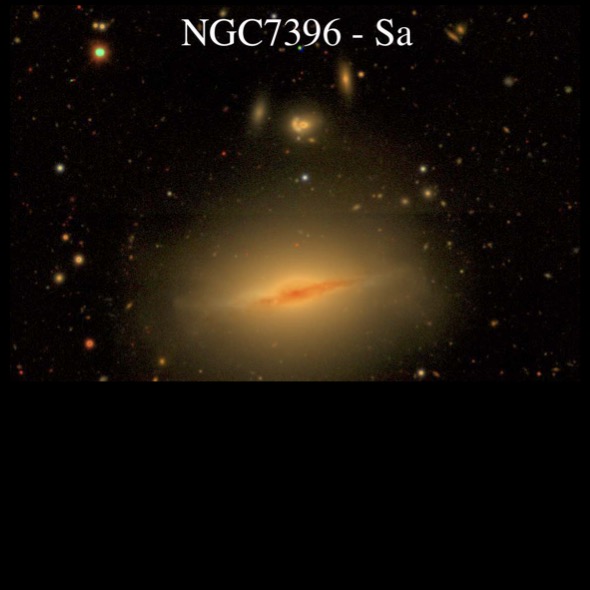}
        \caption{The available Dark Energy Survey (DES) images of galaxies in our sample with LEDA morphologies sourced by Jarrett et al, in prep. As can be seen these galaxies are largely disc galaxies with spiral structure and do not show signs of major mergers. The galaxy in the final panel is cropped as it sits on the edge of the DES visible field.}
        \label{fig:DES}
    \end{figure*}

Most very massive galaxies are elliptical galaxies with negligible star-formation, making up the red portion of the bimodal distribution \citep[e.g.][]{Baldry2004}. We have selected those very massive galaxies that break with this trend and have at least some star-formation. Given the dependence of galaxy quenching on stellar mass \citep[e.g.][]{Nelson2017}  and the correlation between morphology and mass \citep[e.g.][]{Bell2004}, one may naively expect massive star-forming galaxies in this mass range to be LIRGs or mergers (for example, Perseus A). However, our sample has a median $12 {\rm \mu m}$ star formation rate of $2.5 ~{\rm M_\odot ~{yr}^{-1} }$ and are largely undisturbed with only a couple of mergers in the sample. Instead of LIRGs or mergers, we find galaxies similar to the higher redshift super spiral galaxies selected by \citet{Ogle2016, Ogle2019} on the basis of stellar mass, star formation rate and morphology. The DES images and LEDA morphologies of our sample reveal that 63\% have spiral structure rather than major mergers or ellipticals. The remaining sample is comprised of 15\% ellipticals and 22\% lenticular galaxies, however by visually inspecting these galaxies with DES imaging, we note that some of these galaxies have spiral structure and hence the percentage of spiral galaxies is likely a lower limit. Examples of our galaxies are shown in Figure \ref{fig:DES} and we note that when viewing our sample through high quality CCD images, such as DES, we see a higher spiral fraction than photographic images (e.g. LEDA). One example of this is shown in Figure \ref{fig:DES} Panel 5, where 2MASX~J01581817-5412568 is classified as elliptical despite spiral structure. Therefore, LEDA morphologies may overestimate the number of elliptical galaxies in a sample and we expect morphologies derived from photographic data to have more ellipticals than morphologies derived from CCD images due to poorer signal-to-noise and spatial resolution \citep[e.g.][]{Bamford2009}. We also see examples of flocculent spirals in our sample such as NGC~4921 and NGC~4999, which have long been studied as massive galaxies with some ongoing star formation \citep[e.g.][]{Elmegreen1982, Romanishin1985}. 

The majority of galaxies in our sample appear to be disc galaxies, which is surprising given the mass of these galaxies, with \citet{Kauffmann2003, Baldry2004} showing galaxies above $\sim 3 \times 10^{10}~{\rm M_\odot}$ are predominately bulge dominated galaxies with old stellar populations and little star-formation. This raises the question of what has enabled these galaxies to continue to grow in mass without quenching? Examining the distribution of RA and DEC, we see no particular affinity for or absence of clusters and groups. We note that NGC~3313 is in the Hydra I Cluster, NGC~5292 is in the IC 4329 group and NGC~7761 is in a void - and hence there’s no clear trend with environment, although this is not a full environmental study, which is outside the scope of this paper. Studies such as \citet{Kauffmann2003} have found that the threshold for quenching is $\sim 3 \times 10^{10}~{\rm M_\odot}$, well below the masses of our galaxies. The high number of discs in our sample also confirms that this sample is a local analogue to the \citet{Ogle2016, Ogle2019} sample of massive star-forming disc galaxies, which also overwhelmingly show spiral structure, 85\% of their "Super disc" galaxies are spiral galaxies. The similarities between \citet{Ogle2016, Ogle2019} and our own sample are partially the result of comparable selection criteria but it isn’t necessarily expected that spiral structures should be so prevalent in samples separated by $1.3-3.4$~\rm{Gyr} in cosmic time. While we see no major mergers in our sample, \citet{Ogle2016} identify 4 major merger candidates and more potential minor mergers, but this difference between the two samples isn’t unexpected given the major merger rate rapidly increases with redshift \citep[e.g. ][]{Conselice2003}. Given the morphological composition of both samples, we believe we are seeing star formation in disks rather than star formation triggered by a merger with a relatively massive galaxy, examples of which are shown in the DES images in Figure \ref{fig:DES}. This is also supported by recent work examining the WISE properties of nearby galaxies, specifically the S4G sample (Jarrett et al., in prep), which reveals a link between WISE $W3-W4$ colour and the geometry of star formation. Galaxies with centrally-concentrated star-formation are found to have WISE colours where $W3-W4 > 2$, while galaxies with star formation spread across their disks, such as Messier~86 have $W3-W4 < 2$. The mean WISE $W3-W4$ colours of our selected sample is 1.66, which supports a picture of star-formation distributed throughout the disk of our galaxies and not associated with nuclear starbursts. Given this and the DES morphologies, the majority (at least 63\%) of these galaxies appear to be calm discs which are not star-bursts, or in the process of dramatic transformation.

 The galaxies in our sample, are by construction, far more massive than typical spiral galaxies and have star formation, indicating the presence of cold gas. Like the \citet{Ogle2019} sample, our SFRs range from $1-13$ \rm{M$_{\odot}$ $\rm{yr}^{-1}$} (Mrk~1239 has a SFR$_{12\mu m}$ of $37$~\rm{M$_{\odot}$ $\rm{yr}^{-1}$} but we believe this is due to AGN dust contamination, supported by its literature classification as a Seyfert \citep{VCV2006}), and hence most of our sample will fall on or just below the star-forming main sequence, as shown in Figure \ref{fig:Archival_spectra_LINER}. With a mean sSFR $ = 1.06\times 10^{-11} ~\rm{yr^{-1}}$ they have comparable sSFRs to galaxies in the green valley, however, the blue cloud and green valley are not well defined for galaxies at high mass ranges. The galaxies in our sample resemble green valley galaxies in specific star-formation rate and colour but are not necessarily undergoing rapid transformation. Unlike green valley galaxies they do not have a higher star-formation rate counterpart of similar mass from which they might transition to a quiescent galaxy, nor do they appear to be rapidly changing morphologically. Hence, while they may have the optical properties of green valley galaxies, such as trickles of ongoing star-formation in otherwise red galaxies, they may not be involved in the same evolutionary processes. 

 Modelling of the colours of super spiral galaxies by \citet{Ogle2019} indicates that they have colours and sSFRs consistent with a mixture of ongoing star-formation and old stellar populations. Our sample selects for very massive galaxies, which we expect to have older stellar populations, with some ongoing star-formation and hence we have very similar sSFRs to the \citet{Ogle2019} sample. However, the emission line ratios and spectra indicate that the gas present is being ionised by something other than young, hot stars \citep[][and references therein]{Heckman1980, Ho2008, Belfiore2016, Percival2020}, which raises the question of what the ionisation source is?

To provide insight to this question we studied the nuclear spectra of our sample, which suggests that many massive, star-forming galaxies are LINERs. While $\sim30\%$ of very massive galaxies are LINERs, we find that $\sim83\pm6\%$ of very massive star-forming galaxies are LINERs. This suggests there is a connection between LINERs in very massive galaxies and the presence of gas that fuels star formation which is consistent with prior literature \citep[e.g. ][]{Belfiore2016, Coldwell2018, Graves2007}. 

The 46 LINERs are well distributed throughout our sample space in mass and SFR. Although 17 of the 46 LINERs are classified with [\rm{NII}]$\rm{\lambda 6583}$ and H$\alpha\rm{\lambda 6563}$ alone, of the 35 galaxies with sufficient signal-to-noise to measure H$\beta\rm{\lambda 4863}$ and [\rm{OIII}]$\rm{\lambda 5008}$  29 show BPT ratios consistent with LINERs. These results can be seen in the BPT diagram, Figure~\ref{fig:BPT_diagram}, which shows the [\rm{NII}]$\rm{\lambda 6583}$/H$\alpha\rm{\lambda 6563}$ against [\rm{OIII}]$\rm{\lambda 5008}$/H$\beta\rm{\lambda 4863}$ for all galaxies in the sample with available spectra and sufficient signal-to-noise. 

The BPT diagram shows that most galaxies in our sample are LINERs regardless of the literature criteria we use. Of course, passive galaxies do not have sufficient emission lines to be apart of this diagram, but from our [\rm{NII}]$\rm{\lambda 6583}$/H$\alpha\rm{\lambda 6563}$ ratio alone, we know that the upper limit of galaxies with passive nuclear spectra in the sample is 9\%, due to our star-forming selection criteria. Using the \citet{Fernandes2011} WHAN diagram criteria, we find only 2 galaxies in our sample meet the requirements for a passive galaxy, with equivalent widths for both [\rm{NII}]$\rm{\lambda 6583}$ and H$\alpha\rm{\lambda 6563} <0.5$. These massive, star-forming galaxies with passive, nuclear spectra are exceptions in our sample and are likely due to the projected aperture missing star-formation. However, the overall sample is made up of star-forming galaxies of which the majority are LINERs, pointing to a correlation between star-formation and the presence of LINER emission. 

A breakdown of the BPT diagram shows there is one Seyfert classified by BPT emission lines and several galaxies that would be considered composite LINER-HII or Seyfert-HII galaxies by the \citet{Kewley2006} criteria, but which would be considered LINERs by the \citet{Kauffmann2003LINER} criteria. Our primary criteria are the red \citet{Kauffmann2003LINER} lines shown in Figure \ref{fig:BPT_diagram}, which represent [\rm{NII}]$\rm{\lambda 6583}$/H$\alpha\rm{\lambda 6563} > 0.6$ and [\rm{OIII}]$\rm{\lambda 5008}$/H$\beta\rm{\lambda 4863} < 3$ and by this criteria we have 5 star-forming galaxies, 1 Seyfert galaxy and 29 LINERs that are classified by both sets of emission line ratios. This means $83\pm6\%$ of galaxies on the BPT are LINERs, consistent with our proportion of LINERs classified by [\rm{NII}]$\rm{\lambda 6583}$/H$\alpha\rm{\lambda 6563}$ alone.


Our sample of galaxies generally have narrow emission line widths, typically $\lesssim 230 ~\rm{kms^{-1}}$, which is consistent with these galaxies being LINERs and not Seyferts. There are two exceptions to this, one of which is Mrk~1239 which has been excluded from the BPT diagram as it was flagged as having wide emission lines and upon inspection \rm{NII}]$\rm{\lambda 6583}$ and H$\alpha\rm{\lambda 6563}$ are not well separated. It has been previously classified as a Seyfert 1.5 and hence we would expect the broadened emission line region to suppress forbidden emission lines such as \rm{NII}]$\rm{\lambda 6583}$ and so we manually classify this galaxy as a Seyfert due to emission line width \citep{Peterson2006, VCV2006}. The SFR of MRK1239 should also be treated as an upper limit due to AGN dust contamination. The other Seyfert in the sample comes from the \citet{Ho1995} sample, the original purpose of which was to search for nuclear activity in galaxies. Galaxies within this survey often have broad emission line widths and are well known Seyferts in the literature, such as NGC~4565 and our Seyfert UGC~2487. Another galaxy from this survey, NGC~6500, has emission lines broader than 5\AA, likely due to it's the strong, central radio source, however \rm{NII}]$\rm{\lambda 6583}$ and H$\alpha\rm{\lambda 6563}$ are able to be separately measured and our emission line ratios agree with the literature that this is a LINER galaxy \citep{Filho2002, Healey2007}. While we know many galaxies from this survey would appear on our BPT diagram as Seyferts, only a small fraction of \citet{Ho1995} galaxies meet our mass and SFR selection criteria and so we only see one Seyfert from this source. The other galaxies on the BPT diagram have narrow emission line widths and [\rm{OIII}]$\rm{\lambda 5008}$/H$\beta\rm{\lambda 4863}$ emission line ratios inconsistent with Seyfert galaxies. Therefore, with two exceptions that have been outlined, we do not believe that many of the galaxies in our sample are Seyferts. If there are central AGNs in these galaxies, they are very weak, as the H-alpha is often (88\% of galaxies with spectra) not evident until after continuum subtraction. There are also no point sources in Figure 10, which shows all available DES images of our galaxies, further indicating that if there are central AGNs in our sample they are weak and not significantly contaminating the masses or SFRs. We have not calculated the SFRs from H-alpha as the spectroscopic apertures are too small and the emission line ratios indicate H-alpha isn’t the result of star formation. Figure 10 also shows any star formation is largely beyond the galaxy cores and so we would not get an accurate galaxy SFR.

As $79\pm6\%$ of our galaxies have [\rm{NII}]$\rm{\lambda 6583}$ and H$\alpha\rm{\lambda 6563}$ which indicate they are LINERs, have narrow emission line widths, and that $83\pm6\%$ of the galaxies classified with H$\beta\rm{\lambda 4863}$ and [\rm{OIII}]$\rm{\lambda 5008}$ as well as [\rm{NII}]$\rm{\lambda 6583}$ and H$\alpha\rm{\lambda 6563}$ have emission line ratios indicating they are LINERs, we are confident that the 17 galaxies that are classified with [\rm{NII}]$\rm{\lambda 6583}$ and H$\alpha\rm{\lambda 6563}$ alone are also genuinely LINERs. Additionally of these 17 galaxies UGC508 and NGC5078 both have literature classifications as LINERs (\citep{VCV2006, Healey2007}. Therefore, the abundance of LINERs in this sample is remarkable and indicates that some component of our sample selection criteria is selecting LINERs.

There are three components to our sample selection - mass, SFR and redshift, which are directly comparable to the mass and SFR criteria of the \citet{Ogle2019} sample. Previous work shows that the percentage of LINERs increases with mass, with \citet{Belfiore2016} finding that $\sim$ 30\% of galaxies in their highest mass range, $10.5<$ log(\rm{M$^{*}$/M$_{\odot}$}) $<11.5$ are LINERs. However, this mass range is lower than that of our sample, beginning at log(\rm{M$^{*}$/M$_{\odot}$}) $={11.3}$, and our sample has a significantly higher fraction of LINERs.  The galaxy mass vs percentage LINERs presented in \citet{Belfiore2016} is therefore not sufficient to explain our sample's proportion of LINERs. Hence, while mass is likely a contributing factor for the high percentage of LINERs in this sample, it doesn't entirely account for the $83\pm6\%$ LINERs we see in our sample. Redshift is also not likely to explain this abundance, as the \citet{Ogle2016, Ogle2019} galaxies also have a high abundance of LINERs and are at a higher redshift. 


As mass and redshift alone are insufficient to explain the high percentage of LINERs in our sample, we look at the other component of our sample selection, star-formation rate. Previous studies have detected LINER emission in massive, old, red galaxies \citep{Kewley2006, Coldwell2018}, with studies by \citet{Graves2007} finding that red galaxies with LINER emission are younger, by $2-3.5$~\rm{Gyr}, than typical red elliptical galaxies. This indicates that red galaxies in which LINERs were detected had more star-formation, producing younger stars or having more cold gas, than an entirely passive galaxy, supporting a connection between star-formation rate and LINERs. \citet{Graves2007} findings align with our work finding LINERs preferentially reside in very massive, star-forming galaxies, but we note their colour criterion may exclude some star-forming galaxies that would be included in our sample. Our results indicate a link between star formation and LINER activity and may help to explain why having a SFR criteria effects the proportion of LINERs in the sample. Together, these studies suggest that LINERs are younger than typical passive galaxies, but older and redder than bright star-forming galaxies, containing evolved stellar populations. This picture of LINERs matches well with galaxies in our sample, which can be characterised as very massive galaxies with some star-formation.

 



We have found that very massive, star-forming galaxies preferentially host LINERs which is qualitatively consistent with work by \citet{Kauffmann2003LINER}. \citet{Kauffmann2003LINER} finds that AGNs (including LINERs) are preferentially found in very massive galaxies that have younger stellar populations than typical passive galaxies (D4000<1.45 vs D4000>1.7). We do caution that quantitatively our results differ from those of \citet{Kauffmann2003LINER} due to sample selection (including redshift and aperture bias), with \citet{Kauffmann2003LINER} finding $\sim 30$\% of very massive galaxies with young stellar populations host AGNs (including LINERs) while we find that $83\pm6\%$ of very massive galaxies with star formation rates higher than 1 \rm{M$_{\odot}$} host LINERs. One contributor to this quantitative difference is that the \citet{Kauffmann2003LINER} sample have masses $\rm{M}_{\rm{stellar}} \sim 10^{11}$ $\rm{M_\odot}$ which are lower than the masses of our sample at $\rm{M}_{\rm{stellar}}>10^{11.3}$ $\rm{M_\odot}$. \citet{Kauffmann2003LINER} also find that more that $80$\% of emission line galaxies with $\rm{M}_{\rm{stellar}}>10^{11}$ $\rm{M_\odot}$ host AGNs including LINERs. While these studies differ in purpose and sample selection, both our study and work by \citet{Kauffmann2003LINER} suggest that in spite of these differences, very massive galaxies with emission lines often host LINERs (or AGNs in \citet{Kauffmann2003LINER}). 

The morphology of our massive star-forming sample is similar to some samples of LINERs and in contrast to others. \citet{Ricci2023} found that $40\pm9\%$ of elliptical galaxies in their DIVING3D sample were LINERs, while $57\pm10\%$ of lenticulars were LINERs, supporting a higher proportion of LINERs in disc galaxies. \citet{Tommasin2012} studied distant $z\approx0.3$ LINERs and found that, like our sample, many of their galaxies were spiral galaxies with star-formation rates $\approx 10$ \rm{M$_{\odot}$ $\rm{yr}^{-1}$}, similar to the galaxies in this paper, although with a lower mass range than that of our sample $M_* \geq 10^{9.1} \rm{M_\odot}$. LINER emission in spirals has also been studied by \citet{Percival2020} who found LINER emission in the central, passive region swept by the bar of star-forming, barred, spiral galaxies, which may demonstrate that it is possible to have LINER emission in the nuclear region of star-forming, massive, disc galaxies, like our own. 

While we find that there are many LINERs in our sample of massive, star-forming galaxies, it doesn't follow that all LINERs are necessarily found in similar samples. While very massive galaxies, with some star-formation seem to be conducive to producing LINERs, LINERs are not unique to this population. Our LINERs are distinguished from other local LINERs such as those in \citet{Povic2016} as our sample have much higher masses, \citet{Povic2016}'s LINERs range from $6-7 \times 10^{10} \rm{M_\odot}$, and have much lower SFRs, SFR $\sim 10 \rm{M_{\odot} \rm{yr}^{-1}}$. \citet{Yan2012} found that 19 of their 59 quiescent, red galaxies were LINERs. LINERs have also been found in some blue early type galaxies \citep{Deshmukh2022}, although it should be noted that this sample selected for star-formation and AGN activity and only found 14 of their 89 (16\%) galaxies with mass $\geq 10^{9.75} \rm{M_\odot}$ to be LINERs. While these studies have LINERs in their samples, they do not make up the large majority of their samples like our galaxies do. The high frequency of LINERs in our sample may provide clues as to the environment/s that can produce LINER emission. Our results may indicate that LINERs are likely to be produced in massive galaxies, with enough gas to have some ongoing star-formation and older stellar populations but this will need to be investigated further with IFU spectroscopy and as such it is outside the scope of this paper to speculate on the ionisation source for LINERs \citep{Ricci2023}. 

 We will continue to study these galaxies by looking at WiFeS IFU spectroscopy \citep{WiFeS2007} of a subset of these galaxies, which we have recently collected. This data will allow us to map emission line gradients in the centre and extended regions of these galaxies, which will allow us to trace LINER emission throughout these galaxies. The distribution of LINER emission in these galaxies will help us determine if the origin of the ionisation follows star-formation, indicating an older stellar population, or is centrally distributed, implying that low-luminosity AGNs (LLAGNs) may be the source of ionisation, responsible for LINER emission.

\section{Conclusion}

    To understand why some very massive galaxies break the expected bimodal distribution of galaxies and are forming stars, we studied a local sample of massive, M$_{\rm{stellar}}$ $\geq10^{11.3}$ \rm{M$_{\odot}$}, star-forming, SFR $>1$ \rm{M$_{\odot}$ $\rm{yr}^{-1}$} galaxies. The DES images of our sample reveal that they are at least 63\% spiral in morphology, with star-forming, undisturbed disks and hence do no appear to be in the process of rapid transformation. These galaxies generally sit just below the star-forming main sequence in sSFR and have colours and sSFRs consistent with a mixture of gas that fuels star-formation, trickles of ongoing star-formation and old stellar populations. By measuring emission lines from the nuclear spectra of these galaxies, we unexpectedly found that $83\pm6\%$ of our sample are classified as LINERs on a BPT diagram. The BPT diagram shows that the majority of our sample are LINERs regardless of the literature criterion used to classify them. The LINER emission observed in our sample is seen throughout our mass and star-formation rate range and is also seen in the analogous, although higher redshift, \citet{Ogle2019} super spiral sample. We further confirmed these results by looking at the [\rm{NII}]$\rm{\lambda 6583}$/H$\alpha\rm{\lambda 6563}$ of our local sample and found $79\pm6\%$ have emission line ratios consistent with LINER emission. Our results indicate that the vast majority of massive star-forming galaxies are also LINERs, and that LINER emission in massive galaxies may be linked to the presence of gas that fuels star formation.

    

\section{Acknowledgements}
Jaimie Sheil acknowledges support by an Australian Government Research Training (RTP) Scholarship. We would like to thank Patrick Ogle for his interesting work which inspired this study. We would also like to thank Vaishali Parkash and Madhooshi Senarath for their advice and guidance. 

We would also like to thank the reviewer for their insightful comments and feedback which have helped improve the paper. 

This publication makes use of data products from the Wide-field Infrared Survey Explorer, which is a joint project of the University of California, Los Angeles, and the Jet
Propulsion Laboratory/California Institute of Technology, funded by the National Aeronautics and Space Administration. 

Funding for SDSS-III has been provided by the Alfred P. Sloan Foundation, the Participating Institutions, the National Science Foundation, and the U.S. Department of Energy Office of Science. The SDSS-III web site is http://www.sdss3.org/.
SDSS-III is managed by the Astrophysical Research Consortium for the Participating Institutions of the SDSS-III Collaboration including the University of Arizona, the Brazilian Participation Group, Brookhaven National Laboratory, Carnegie Mellon University, University of Florida, the French Participation Group, the German Participation Group, Harvard University, the Instituto de Astrofisica de Canarias, the Michigan State/Notre Dame/JINA Participation Group, Johns Hopkins University, Lawrence Berkeley National Laboratory, Max Planck Institute for Astrophysics, Max Planck Institute for Extraterrestrial Physics, New Mexico State University, New York University, Ohio State University, Pennsylvania State University, University of Portsmouth, Princeton University, the Spanish Participation Group, University of Tokyo, University of Utah, Vanderbilt University, University of Virginia, University of Washington, and Yale University. 

This publication makes use of data products of the Two Micron All Sky Survey, which is a joint project of the University of Massachusetts and the Infrared Processing and Analysis Center, funded by the NASA and the National Science Foundation. 


This project used public archival data from the Dark Energy Survey (DES). Funding for the DES Projects has been provided by the U.S. Department of Energy, the U.S. National Science Foundation, the Ministry of Science and Education of Spain, the Science and Technology Facilities Council of the United Kingdom, the Higher Education Funding Council for England, the National Center for Supercomputing Applications at the University of Illinois at Urbana-Champaign, the Kavli Institute of Cosmological Physics at the University of Chicago, the Center for Cosmology and Astro-Particle Physics at the Ohio State University, the Mitchell Institute for Fundamental Physics and Astronomy at Texas A\&M University, Financiadora de Estudos e Projetos, Fundação Carlos Chagas Filho de Amparo à Pesquisa do Estado do Rio de Janeiro, Conselho Nacional de Desenvolvimento Científico e Tecnológico and the Ministério da Ciência, Tecnologia e Inovação, the Deutsche Forschungsgemeinschaft, and the Collaborating Institutions in the Dark Energy Survey. The Collaborating Institutions are Argonne National Laboratory, the University of California at Santa Cruz, the University of Cambridge, Centro de Investigaciones Energéticas, Medioambientales y Tecnológicas-Madrid, the University of Chicago, University College London, the DES-Brazil Consortium, the University of Edinburgh, the Eidgenössische Technische Hochschule (ETH) Zürich, Fermi National Accelerator Laboratory, the University of Illinois at Urbana-Champaign, the Institut de Ciències de l’Espai (IEEC/CSIC), the Institut de Física d’Altes Energies, Lawrence Berkeley National Laboratory, the Ludwig-Maximilians Universität München and the associated Excellence Cluster Universe, the University of Michigan, the National Optical Astronomy Observatory, the University of Nottingham, The Ohio State University, the OzDES Membership Consortium, the University of Pennsylvania, the University of Portsmouth, SLAC National Accelerator Laboratory, Stanford University, the University of Sussex, and Texas A\&M University. Based in part on observations at Cerro Tololo Inter-American Observatory, National Optical Astronomy Observatory, which is operated by the Association of Universities for Research in Astronomy (AURA) under a cooperative agreement with the National Science Foundation.

This research has made use of the NASA/IPAC Extragalactic Database (NED) which is operated by the Jet Propulsion Laboratory, California Institute of Technology,
under contract with the National Aeronautics and Space Administration and the HyperLeda database (http://leda.univlyon1.fr). 
This research made use of Astropy, a community developed core Python package for Astronomy (Astropy Collaboration, 2018).

\section{Data Availability}
The data underlying this article are available in its online supplementary material.



\bibliographystyle{pasa-mnras}
\bibliography{biblio} 

\begin{thebibliography}{}
\makeatletter
\relax
\def\mn@urlcharsother{\let\do\@makeother \do\$\do\&\do\#\do\^\do\_\do\%\do\~}
\definecolor{darkblue}{rgb}{0,0,0.597656}
\def\mndoi{\begingroup\mn@urlcharsother \@ifnextchar [ {\mndoi@} {\mndoi@[]}}
\def\mndoi@[#1]#2{\def\@tempa{#1}\ifx\@tempa\@empty \href {http://dx.doi.org/#2} {\textcolor{darkblue}{doi:#2}}\else \href {http://dx.doi.org/#2} {\textcolor{darkblue}{#1}}\fi \endgroup}
\def\mn@eprint#1#2{\mn@eprint@#1:#2::\@nil}
\def\mn@eprint@arXiv#1{\href {http://arxiv.org/abs/#1} {{\tt arXiv:#1}}}
\def\mn@eprint@dblp#1{\href {http://dblp.uni-trier.de/rec/bibtex/#1.xml} {dblp:#1}}
\def\mn@eprint@#1:#2:#3:#4\@nil{\def\@tempa {#1}\def\@tempb {#2}\def\@tempc {#3}\ifx \@tempc \@empty \let \@tempc \@tempb \let \@tempb \@tempa \fi \ifx \@tempb \@empty \def\@tempb {arXiv}\fi \@ifundefined {mn@eprint@\@tempb}{\@tempb:\@tempc}{\expandafter \expandafter \csname mn@eprint@\@tempb\endcsname \expandafter{\@tempc}}}

\bibitem[\protect\citeauthoryear{{Acker}, {K{\"o}ppen}, {Samland}  \& {Stenholm}}{{Acker} et~al.}{1989}]{NII1989}
{Acker} A.,  {K{\"o}ppen} J.,  {Samland} M.,   {Stenholm} B.,  1989, The Messenger, \href {https://ui.adsabs.harvard.edu/abs/1989Msngr..58...44A} {58, 44}

\bibitem[\protect\citeauthoryear{{Alam} et~al.,}{{Alam} et~al.}{2015}]{SDSSDR12}
{Alam} S.,  et~al., 2015, \mndoi [\apjs] {10.1088/0067-0049/219/1/12}, \href {https://ui.adsabs.harvard.edu/abs/2015ApJS..219...12A} {219, 12}

\bibitem[\protect\citeauthoryear{{Baldry}, {Glazebrook}, {Brinkmann}, {Ivezi{\'c}}, {Lupton}, {Nichol}  \& {Szalay}}{{Baldry} et~al.}{2004}]{Baldry2004}
{Baldry} I.~K.,  {Glazebrook} K.,  {Brinkmann} J.,  {Ivezi{\'c}} {\v{Z}}.,  {Lupton} R.~H.,  {Nichol} R.~C.,   {Szalay} A.~S.,  2004, \mndoi [\apj] {10.1086/380092}, \href {https://ui.adsabs.harvard.edu/abs/2004ApJ...600..681B} {600, 681}

\bibitem[\protect\citeauthoryear{{Baldwin}, {Phillips}  \& {Terlevich}}{{Baldwin} et~al.}{1981}]{BPT1981}
{Baldwin} J.~A.,  {Phillips} M.~M.,   {Terlevich} R.,  1981, \mndoi [\pasp] {10.1086/130766}, \href {https://ui.adsabs.harvard.edu/abs/1981PASP...93....5B} {93, 5}

\bibitem[\protect\citeauthoryear{Bamford et~al.,}{Bamford et~al.}{2009}]{Bamford2009}
Bamford S.~P.,  et~al., 2009, \mndoi [Monthly Notices of the Royal Astronomical Society] {10.1111/j.1365-2966.2008.14252.x}, 393, 1324

\bibitem[\protect\citeauthoryear{{Belfiore} et~al.,}{{Belfiore} et~al.}{2016}]{Belfiore2016}
{Belfiore} F.,  et~al., 2016, \mndoi [\mnras] {10.1093/mnras/stw1234}, \href {https://ui.adsabs.harvard.edu/abs/2016MNRAS.461.3111B} {461, 3111}

\bibitem[\protect\citeauthoryear{{Bell}, {McIntosh}, {Katz}  \& {Weinberg}}{{Bell} et~al.}{2003}]{Bell2003}
{Bell} E.~F.,  {McIntosh} D.~H.,  {Katz} N.,   {Weinberg} M.~D.,  2003, \mndoi [\apjs] {10.1086/378847}, \href {https://ui.adsabs.harvard.edu/abs/2003ApJS..149..289B} {149, 289}

\bibitem[\protect\citeauthoryear{{Bell} et~al.,}{{Bell} et~al.}{2004}]{Bell2004}
{Bell} E.~F.,  et~al., 2004, \mndoi [\apj] {10.1086/420778}, \href {https://ui.adsabs.harvard.edu/abs/2004ApJ...608..752B} {608, 752}

\bibitem[\protect\citeauthoryear{{Binette}, {Magris}, {Stasi{\'n}ska}  \& {Bruzual}}{{Binette} et~al.}{1994}]{Binette1994}
{Binette} L.,  {Magris} C.~G.,  {Stasi{\'n}ska} G.,   {Bruzual} A.~G.,  1994, \aap, \href {https://ui.adsabs.harvard.edu/abs/1994A&A...292...13B} {292, 13}

\bibitem[\protect\citeauthoryear{{Birnboim} \& {Dekel}}{{Birnboim} \& {Dekel}}{2003}]{Birnboim2003}
{Birnboim} Y.,  {Dekel} A.,  2003, \mndoi [\mnras] {10.1046/j.1365-8711.2003.06955.x}, \href {https://ui.adsabs.harvard.edu/abs/2003MNRAS.345..349B} {345, 349}

\bibitem[\protect\citeauthoryear{{Blanton} et~al.,}{{Blanton} et~al.}{2003}]{Blanton2003}
{Blanton} M.~R.,  et~al., 2003, \mndoi [\apj] {10.1086/375528}, \href {https://ui.adsabs.harvard.edu/abs/2003ApJ...594..186B} {594, 186}

\bibitem[\protect\citeauthoryear{{Bluck} et~al.,}{{Bluck} et~al.}{2020}]{bluck2020}
{Bluck} A. F.~L.,  et~al., 2020, \mndoi [\mnras] {10.1093/mnras/staa2806}, \href {https://ui.adsabs.harvard.edu/abs/2020MNRAS.499..230B} {499, 230}

\bibitem[\protect\citeauthoryear{{Bruzual} \& {Charlot}}{{Bruzual} \& {Charlot}}{2003}]{BC2003}
{Bruzual} G.,  {Charlot} S.,  2003, \mndoi [\mnras] {10.1046/j.1365-8711.2003.06897.x}, \href {https://ui.adsabs.harvard.edu/abs/2003MNRAS.344.1000B} {344, 1000}

\bibitem[\protect\citeauthoryear{{Buta}}{{Buta}}{2013}]{Buta2011}
{Buta} R.~J.,  2013, in {Falc{\'o}n-Barroso} J.,  {Knapen} J.~H.,  eds, , Secular Evolution of Galaxies.
Cambridge University Press, p.~155

\bibitem[\protect\citeauthoryear{Cattaneo, Dekel, Devriendt, Guiderdoni  \& Blaizot}{Cattaneo et~al.}{2006}]{Cattaneo2006}
Cattaneo A.,  Dekel A.,  Devriendt J.,  Guiderdoni B.,   Blaizot J.,  2006, \mndoi [Monthly Notices of the Royal Astronomical Society] {10.1111/j.1365-2966.2006.10608.x}, 370, 1651

\bibitem[\protect\citeauthoryear{{Chang}, {van der Wel}, {da Cunha}  \& {Rix}}{{Chang} et~al.}{2015}]{Chang2015}
{Chang} Y.-Y.,  {van der Wel} A.,  {da Cunha} E.,   {Rix} H.-W.,  2015, \mndoi [\apjs] {10.1088/0067-0049/219/1/8}, \href {https://ui.adsabs.harvard.edu/abs/2015ApJS..219....8C} {219, 8}

\bibitem[\protect\citeauthoryear{Cid~Fernandes, Stasińska, Mateus  \& Vale~Asari}{Cid~Fernandes et~al.}{2011}]{Fernandes2011}
Cid~Fernandes R.,  Stasińska G.,  Mateus A.,   Vale~Asari N.,  2011, \mndoi [Monthly Notices of the Royal Astronomical Society] {10.1111/j.1365-2966.2011.18244.x}, 413, 1687

\bibitem[\protect\citeauthoryear{{Cluver} et~al.,}{{Cluver} et~al.}{2014}]{Cluver2014}
{Cluver} M.~E.,  et~al., 2014, \mndoi [\apj] {10.1088/0004-637X/782/2/90}, \href {https://ui.adsabs.harvard.edu/abs/2014ApJ...782...90C} {782, 90}

\bibitem[\protect\citeauthoryear{{Cluver}, {Jarrett}, {Dale}, {Smith}, {August}  \& {Brown}}{{Cluver} et~al.}{2017}]{Cluver2017}
{Cluver} M.~E.,  {Jarrett} T.~H.,  {Dale} D.~A.,  {Smith} J. D.~T.,  {August} T.,   {Brown} M.~J.~I.,  2017, \mndoi [\apj] {10.3847/1538-4357/aa92c7}, \href {https://ui.adsabs.harvard.edu/abs/2017ApJ...850...68C} {850, 68}

\bibitem[\protect\citeauthoryear{{Coldwell}, {Pereyra}, {Alonso}, {Donoso}  \& {Duplancic}}{{Coldwell} et~al.}{2017}]{Coldwell2017}
{Coldwell} G.~V.,  {Pereyra} L.,  {Alonso} S.,  {Donoso} E.,   {Duplancic} F.,  2017, \mndoi [\mnras] {10.1093/mnras/stx294}, \href {https://ui.adsabs.harvard.edu/abs/2017MNRAS.467.3338C} {467, 3338}

\bibitem[\protect\citeauthoryear{{Coldwell}, {Alonso}, {Duplancic}  \& {Mesa}}{{Coldwell} et~al.}{2018}]{Coldwell2018}
{Coldwell} G.~V.,  {Alonso} S.,  {Duplancic} F.,   {Mesa} V.,  2018, \mndoi [\mnras] {10.1093/mnras/sty395}, \href {https://ui.adsabs.harvard.edu/abs/2018MNRAS.476.2457C} {476, 2457}

\bibitem[\protect\citeauthoryear{Conselice, Bershady, Dickinson  \& Papovich}{Conselice et~al.}{2003}]{Conselice2003}
Conselice C.~J.,  Bershady M.~A.,  Dickinson M.~E.,   Papovich C.,  2003, The Astronomical Journal, 126, 1183

\bibitem[\protect\citeauthoryear{{Deshmukh}, {Vagshette}  \& {Patil}}{{Deshmukh} et~al.}{2022}]{Deshmukh2022}
{Deshmukh} S.~P.,  {Vagshette} N.~D.,   {Patil} M.~K.,  2022, \mndoi [Serbian Astronomical Journal] {10.2298/SAJ2205023D}, \href {https://ui.adsabs.harvard.edu/abs/2022SerAJ.205...23D} {205, 23}

\bibitem[\protect\citeauthoryear{{Dopita}, {Hart}, {McGregor}, {Oates}, {Bloxham}  \& {Jones}}{{Dopita} et~al.}{2007}]{WiFeS2007}
{Dopita} M.,  {Hart} J.,  {McGregor} P.,  {Oates} P.,  {Bloxham} G.,   {Jones} D.,  2007, \mndoi [\apss] {10.1007/s10509-007-9510-z}, \href {https://ui.adsabs.harvard.edu/abs/2007Ap&SS.310..255D} {310, 255}

\bibitem[\protect\citeauthoryear{{Elmegreen}}{{Elmegreen}}{1981}]{Elmegreen1981}
{Elmegreen} D.~M.,  1981, \mndoi [\apjs] {10.1086/190757}, \href {https://ui.adsabs.harvard.edu/abs/1981ApJS...47..229E} {47, 229}

\bibitem[\protect\citeauthoryear{{Elmegreen} \& {Elmegreen}}{{Elmegreen} \& {Elmegreen}}{1982}]{Elmegreen1982}
{Elmegreen} D.~M.,  {Elmegreen} B.~G.,  1982, \mndoi [\mnras] {10.1093/mnras/201.4.1021}, \href {https://ui.adsabs.harvard.edu/abs/1982MNRAS.201.1021E} {201, 1021}

\bibitem[\protect\citeauthoryear{Faber, Worthey  \& Gonzalez}{Faber et~al.}{1992}]{Faber1992}
Faber S.~M.,  Worthey G.,   Gonzalez J.~J.,  1992, in Barbuy B.,  Renzini A.,  eds, The Stellar Populations of Galaxies. Springer Netherlands, Dordrecht, pp 255--265

\bibitem[\protect\citeauthoryear{{Filho}, {Barthel}  \& {Ho}}{{Filho} et~al.}{2002}]{Filho2002}
{Filho} M.~E.,  {Barthel} P.~D.,   {Ho} L.~C.,  2002, \mndoi [\aap] {10.1051/0004-6361:20020138}, \href {https://ui.adsabs.harvard.edu/abs/2002A&A...385..425F} {385, 425}

\bibitem[\protect\citeauthoryear{{Forman}, {Jones}  \& {Tucker}}{{Forman} et~al.}{1985}]{Forman1985}
{Forman} W.,  {Jones} C.,   {Tucker} W.,  1985, \mndoi [\apj] {10.1086/163218}, \href {https://ui.adsabs.harvard.edu/abs/1985ApJ...293..102F} {293, 102}

\bibitem[\protect\citeauthoryear{Fraser-McKelvie, Brown, Pimbblet, Dolley, Crossett  \& Bonne}{Fraser-McKelvie et~al.}{2016}]{McKelvie2016}
Fraser-McKelvie A.,  Brown M. J.~I.,  Pimbblet K.~A.,  Dolley T.,  Crossett J.~P.,   Bonne N.~J.,  2016, \mndoi [Monthly Notices of the Royal Astronomical Society: Letters] {10.1093/mnrasl/slw117}, 462, L11

\bibitem[\protect\citeauthoryear{Fraser-McKelvie, Brown, Pimbblet, Dolley  \& Bonne}{Fraser-McKelvie et~al.}{2017}]{McKelvie2017}
Fraser-McKelvie A.,  Brown M. J.~I.,  Pimbblet K.,  Dolley T.,   Bonne N.~J.,  2017, \mndoi [Monthly Notices of the Royal Astronomical Society] {10.1093/mnras/stx2823}, 474, 1909

\bibitem[\protect\citeauthoryear{{Gallimore}, {Baum}, {O'Dea}, {Pedlar}  \& {Brinks}}{{Gallimore} et~al.}{1999}]{Gallimore1999}
{Gallimore} J.~F.,  {Baum} S.~A.,  {O'Dea} C.~P.,  {Pedlar} A.,   {Brinks} E.,  1999, \mndoi [\apj] {10.1086/307853}, \href {https://ui.adsabs.harvard.edu/abs/1999ApJ...524..684G} {524, 684}

\bibitem[\protect\citeauthoryear{Gonçalves \& Martin}{Gonçalves \& Martin}{2009}]{Goncalves2009}
Gonçalves T.~S.,  Martin D.~C.,  2009, \mndoi [Proceedings of the International Astronomical Union] {10.1017/S1743921310002887}, 5, 261–264

\bibitem[\protect\citeauthoryear{{Graves}, {Faber}, {Schiavon}  \& {Yan}}{{Graves} et~al.}{2007}]{Graves2007}
{Graves} G.~J.,  {Faber} S.~M.,  {Schiavon} R.~P.,   {Yan} R.,  2007, \mndoi [\apj] {10.1086/522325}, \href {https://ui.adsabs.harvard.edu/abs/2007ApJ...671..243G} {671, 243}

\bibitem[\protect\citeauthoryear{Greene, Murphy, Graves, Gunn, Raskutti, Comerford  \& Gebhardt}{Greene et~al.}{2013}]{Greene_2013}
Greene J.~E.,  Murphy J.~D.,  Graves G.~J.,  Gunn J.~E.,  Raskutti S.,  Comerford J.~M.,   Gebhardt K.,  2013, \mndoi [The Astrophysical Journal] {10.1088/0004-637x/776/2/64}, 776, 64

\bibitem[\protect\citeauthoryear{{Groves} \& {Kewley}}{{Groves} \& {Kewley}}{2008}]{Groves2008}
{Groves} B.,  {Kewley} L.,  2008, in {Knapen} J.~H. e.~a.,  ed.,  Astronomical Society of the Pacific Conference Series Vol. 390, Pathways Through an Eclectic Universe. p.~283 (\mn@eprint {arXiv} {0707.0158})

\bibitem[\protect\citeauthoryear{{Healey}, {Romani}, {Taylor}, {Sadler}, {Ricci}, {Murphy}, {Ulvestad}  \& {Winn}}{{Healey} et~al.}{2007}]{Healey2007}
{Healey} S.~E.,  {Romani} R.~W.,  {Taylor} G.~B.,  {Sadler} E.~M.,  {Ricci} R.,  {Murphy} T.,  {Ulvestad} J.~S.,   {Winn} J.~N.,  2007, \mndoi [\apjs] {10.1086/513742}, \href {https://ui.adsabs.harvard.edu/abs/2007ApJS..171...61H} {171, 61}

\bibitem[\protect\citeauthoryear{{Heckman}}{{Heckman}}{1980}]{Heckman1980}
{Heckman} T.~M.,  1980, \aap, \href {https://ui.adsabs.harvard.edu/abs/1980A&A....88..365H} {88, 365}

\bibitem[\protect\citeauthoryear{{Ho}}{{Ho}}{2008}]{Ho2008}
{Ho} L.~C.,  2008, \mndoi [\araa] {10.1146/annurev.astro.45.051806.110546}, \href {https://ui.adsabs.harvard.edu/abs/2008ARA&A..46..475H} {46, 475}

\bibitem[\protect\citeauthoryear{{Ho}, {Filippenko}  \& {Sargent}}{{Ho} et~al.}{1995}]{Ho1995}
{Ho} L.~C.,  {Filippenko} A.~V.,   {Sargent} W.~L.,  1995, \mndoi [\apjs] {10.1086/192170}, \href {https://ui.adsabs.harvard.edu/abs/1995ApJS...98..477H} {98, 477}

\bibitem[\protect\citeauthoryear{{Hopkins}, {Somerville}, {Hernquist}, {Cox}, {Robertson}  \& {Li}}{{Hopkins} et~al.}{2006}]{Hopkins2006}
{Hopkins} P.~F.,  {Somerville} R.~S.,  {Hernquist} L.,  {Cox} T.~J.,  {Robertson} B.,   {Li} Y.,  2006, \mndoi [\apj] {10.1086/508503}, \href {https://ui.adsabs.harvard.edu/abs/2006ApJ...652..864H} {652, 864}

\bibitem[\protect\citeauthoryear{{Huchra} et~al.,}{{Huchra} et~al.}{2012}]{Huchra2012}
{Huchra} J.~P.,  et~al., 2012, \mndoi [\apjs] {10.1088/0067-0049/199/2/26}, \href {https://ui.adsabs.harvard.edu/abs/2012ApJS..199...26H} {199, 26}

\bibitem[\protect\citeauthoryear{{Jarrett} et~al.,}{{Jarrett} et~al.}{2011}]{Jarrett2011}
{Jarrett} T.~H.,  et~al., 2011, \mndoi [\apj] {10.1088/0004-637X/735/2/112}, \href {https://ui.adsabs.harvard.edu/abs/2011ApJ...735..112J} {735, 112}

\bibitem[\protect\citeauthoryear{{Jarrett} et~al.,}{{Jarrett} et~al.}{2013}]{Jarrett2013}
{Jarrett} T.~H.,  et~al., 2013, \mndoi [\aj] {10.1088/0004-6256/145/1/6}, \href {https://ui.adsabs.harvard.edu/abs/2013AJ....145....6J} {145, 6}

\bibitem[\protect\citeauthoryear{{Jarrett}, {Cluver}, {Brown}, {Dale}, {Tsai}  \& {Masci}}{{Jarrett} et~al.}{2019}]{Jarrett2019}
{Jarrett} T.~H.,  {Cluver} M.~E.,  {Brown} M.~J.~I.,  {Dale} D.~A.,  {Tsai} C.~W.,   {Masci} F.,  2019, \mndoi [\apjs] {10.3847/1538-4365/ab521a}, \href {https://ui.adsabs.harvard.edu/abs/2019ApJS..245...25J} {245, 25}

\bibitem[\protect\citeauthoryear{{Jones} et~al.,}{{Jones} et~al.}{2004}]{Jones2004}
{Jones} D.~H.,  et~al., 2004, \mndoi [\mnras] {10.1111/j.1365-2966.2004.08353.x}, \href {https://ui.adsabs.harvard.edu/abs/2004MNRAS.355..747J} {355, 747}

\bibitem[\protect\citeauthoryear{{Jones}, {Peterson}, {Colless}  \& {Saunders}}{{Jones} et~al.}{2006}]{6df2006}
{Jones} D.~H.,  {Peterson} B.~A.,  {Colless} M.,   {Saunders} W.,  2006, \mndoi [\mnras] {10.1111/j.1365-2966.2006.10291.x}, \href {https://ui.adsabs.harvard.edu/abs/2006MNRAS.369...25J} {369, 25}

\bibitem[\protect\citeauthoryear{Jones et~al.,}{Jones et~al.}{2009}]{Jones2009}
Jones D.~H.,  et~al., 2009, \mndoi [Monthly Notices of the Royal Astronomical Society] {10.1111/j.1365-2966.2009.15338.x}, 399, 683

\bibitem[\protect\citeauthoryear{{Kauffmann} et~al.,}{{Kauffmann} et~al.}{2003a}]{Kauffmann2003}
{Kauffmann} G.,  et~al., 2003a, \mndoi [\mnras] {10.1046/j.1365-8711.2003.06292.x}, \href {https://ui.adsabs.harvard.edu/abs/2003MNRAS.341...54K} {341, 54}

\bibitem[\protect\citeauthoryear{{Kauffmann} et~al.,}{{Kauffmann} et~al.}{2003b}]{Kauffmann2003LINER}
{Kauffmann} G.,  et~al., 2003b, \mndoi [\mnras] {10.1111/j.1365-2966.2003.07154.x}, \href {https://ui.adsabs.harvard.edu/abs/2003MNRAS.346.1055K} {346, 1055}

\bibitem[\protect\citeauthoryear{{Kennicutt}}{{Kennicutt}}{1992}]{Kennicutt1992}
{Kennicutt} Robert~C. J.,  1992, \mndoi [\apjs] {10.1086/191653}, \href {https://ui.adsabs.harvard.edu/abs/1992ApJS...79..255K} {79, 255}

\bibitem[\protect\citeauthoryear{{Kennicutt}}{{Kennicutt}}{1998}]{kennicutt1998}
{Kennicutt} Robert~C. J.,  1998, \mndoi [\araa] {10.1146/annurev.astro.36.1.189}, \href {https://ui.adsabs.harvard.edu/abs/1998ARA&A..36..189K} {36, 189}

\bibitem[\protect\citeauthoryear{{Kewley}, {Groves}, {Kauffmann}  \& {Heckman}}{{Kewley} et~al.}{2006}]{Kewley2006}
{Kewley} L.~J.,  {Groves} B.,  {Kauffmann} G.,   {Heckman} T.,  2006, \mndoi [\mnras] {10.1111/j.1365-2966.2006.10859.x}, \href {https://ui.adsabs.harvard.edu/abs/2006MNRAS.372..961K} {372, 961}

\bibitem[\protect\citeauthoryear{{Kewley}, {Maier}, {Yabe}, {Ohta}, {Akiyama}, {Dopita}  \& {Yuan}}{{Kewley} et~al.}{2013}]{Kewley2013}
{Kewley} L.~J.,  {Maier} C.,  {Yabe} K.,  {Ohta} K.,  {Akiyama} M.,  {Dopita} M.~A.,   {Yuan} T.,  2013, \mndoi [\apjl] {10.1088/2041-8205/774/1/L10}, \href {https://ui.adsabs.harvard.edu/abs/2013ApJ...774L..10K} {774, L10}

\bibitem[\protect\citeauthoryear{{Kewley}, {Nicholls}  \& {Sutherland}}{{Kewley} et~al.}{2019}]{Kewley2019}
{Kewley} L.~J.,  {Nicholls} D.~C.,   {Sutherland} R.~S.,  2019, \mndoi [\araa] {10.1146/annurev-astro-081817-051832}, \href {https://ui.adsabs.harvard.edu/abs/2019ARA&A..57..511K} {57, 511}

\bibitem[\protect\citeauthoryear{{Kroupa}}{{Kroupa}}{2001}]{Kroupa}
{Kroupa} P.,  2001, \mndoi [\mnras] {10.1046/j.1365-8711.2001.04022.x}, \href {https://ui.adsabs.harvard.edu/abs/2001MNRAS.322..231K} {322, 231}

\bibitem[\protect\citeauthoryear{{Lisenfeld}, {Ogle}, {Appleton}, {Jarrett}  \& {Moncada-Cuadri}}{{Lisenfeld} et~al.}{2023}]{Lisenfeld2023}
{Lisenfeld} U.,  {Ogle} P.~M.,  {Appleton} P.~N.,  {Jarrett} T.~H.,   {Moncada-Cuadri} B.~M.,  2023, \mndoi [\aap] {10.1051/0004-6361/202245675}, \href {https://ui.adsabs.harvard.edu/abs/2023A&A...673A..87L} {673, A87}

\bibitem[\protect\citeauthoryear{{Martin} et~al.,}{{Martin} et~al.}{2005}]{Martin2005}
{Martin} D.~C.,  et~al., 2005, \mndoi [\apjl] {10.1086/426387}, \href {https://ui.adsabs.harvard.edu/abs/2005ApJ...619L...1M} {619, L1}

\bibitem[\protect\citeauthoryear{{Martin} et~al.,}{{Martin} et~al.}{2007}]{Martin2007}
{Martin} D.~C.,  et~al., 2007, \mndoi [\apjs] {10.1086/516639}, \href {https://ui.adsabs.harvard.edu/abs/2007ApJS..173..342M} {173, 342}

\bibitem[\protect\citeauthoryear{{Masters} et~al.,}{{Masters} et~al.}{2010}]{Masters2010}
{Masters} K.~L.,  et~al., 2010, \mndoi [\mnras] {10.1111/j.1365-2966.2010.16503.x}, \href {https://ui.adsabs.harvard.edu/abs/2010MNRAS.405..783M} {405, 783}

\bibitem[\protect\citeauthoryear{{Mehlert}, {Thomas}, {Saglia}, {Bender}  \& {Wegner}}{{Mehlert} et~al.}{2003}]{Mehlert2003}
{Mehlert} D.,  {Thomas} D.,  {Saglia} R.~P.,  {Bender} R.,   {Wegner} G.,  2003, \mndoi [\aap] {10.1051/0004-6361:20030886}, \href {https://ui.adsabs.harvard.edu/abs/2003A&A...407..423M} {407, 423}

\bibitem[\protect\citeauthoryear{{Mihos} \& {Hernquist}}{{Mihos} \& {Hernquist}}{1996}]{Mihos1996}
{Mihos} J.~C.,  {Hernquist} L.,  1996, \mndoi [\apj] {10.1086/177353}, \href {https://ui.adsabs.harvard.edu/abs/1996ApJ...464..641M} {464, 641}

\bibitem[\protect\citeauthoryear{Márquez, Masegosa, González-Martin, Hernández-Garcia, Pović, Netzer, Cazzoli  \& del Olmo}{Márquez et~al.}{2017}]{Marques2017}
Márquez I.,  Masegosa J.,  González-Martin O.,  Hernández-Garcia L.,  Pović M.,  Netzer H.,  Cazzoli S.,   del Olmo A.,  2017, \mndoi [Frontiers in Astronomy and Space Sciences] {10.3389/fspas.2017.00034}, 4

\bibitem[\protect\citeauthoryear{Nelson et~al.,}{Nelson et~al.}{2017}]{Nelson2017}
Nelson D.,  et~al., 2017, \mndoi [Monthly Notices of the Royal Astronomical Society] {10.1093/mnras/stx3040}, 475, 624

\bibitem[\protect\citeauthoryear{{Nogueira-Cavalcante}, {Gon{\c{c}}alves}, {Men{\'e}ndez-Delmestre}  \& {Sheth}}{{Nogueira-Cavalcante} et~al.}{2018}]{Nogueira2018}
{Nogueira-Cavalcante} J.~P.,  {Gon{\c{c}}alves} T.~S.,  {Men{\'e}ndez-Delmestre} K.,   {Sheth} K.,  2018, \mndoi [\mnras] {10.1093/mnras/stx2399}, \href {https://ui.adsabs.harvard.edu/abs/2018MNRAS.473.1346N} {473, 1346}

\bibitem[\protect\citeauthoryear{{Ogle}, {Lanz}, {Nader}  \& {Helou}}{{Ogle} et~al.}{2016}]{Ogle2016}
{Ogle} P.~M.,  {Lanz} L.,  {Nader} C.,   {Helou} G.,  2016, \mndoi [\apj] {10.3847/0004-637X/817/2/109}, \href {https://ui.adsabs.harvard.edu/abs/2016ApJ...817..109O} {817, 109}

\bibitem[\protect\citeauthoryear{{Ogle}, {Lanz}, {Appleton}, {Helou}  \& {Mazzarella}}{{Ogle} et~al.}{2019}]{Ogle2019}
{Ogle} P.~M.,  {Lanz} L.,  {Appleton} P.~N.,  {Helou} G.,   {Mazzarella} J.,  2019, VizieR Online Data Catalog, \href {https://ui.adsabs.harvard.edu/abs/2019yCat..22430014O} {p. J/ApJS/243/14}

\bibitem[\protect\citeauthoryear{{Olsson}, {Aalto}, {Thomasson}  \& {Beswick}}{{Olsson} et~al.}{2010}]{Olsson2010}
{Olsson} E.,  {Aalto} S.,  {Thomasson} M.,   {Beswick} R.,  2010, \mndoi [\aap] {10.1051/0004-6361/200811538}, \href {https://ui.adsabs.harvard.edu/abs/2010A&A...513A..11O} {513, A11}

\bibitem[\protect\citeauthoryear{{Parkash}, {Brown}, {Jarrett}  \& {Bonne}}{{Parkash} et~al.}{2019a}]{Parkash2019}
{Parkash} V.,  {Brown} M.~J.~I.,  {Jarrett} T.~H.,   {Bonne} N.~J.,  2019a, VizieR Online Data Catalog, \href {https://ui.adsabs.harvard.edu/abs/2019yCat..18640040P} {p. J/ApJ/864/40}

\bibitem[\protect\citeauthoryear{{Parkash}, {Brown}, {Jarrett}, {Fraser-McKelvie}  \& {Cluver}}{{Parkash} et~al.}{2019b}]{Parkash2018}
{Parkash} V.,  {Brown} M. J.~I.,  {Jarrett} T.~H.,  {Fraser-McKelvie} A.,   {Cluver} M.~E.,  2019b, \mndoi [\mnras] {10.1093/mnras/stz593}, \href {https://ui.adsabs.harvard.edu/abs/2019MNRAS.485.3169P} {485, 3169}

\bibitem[\protect\citeauthoryear{{Percival} \& {James}}{{Percival} \& {James}}{2020}]{Percival2020}
{Percival} S.~M.,  {James} P.~A.,  2020, \mndoi [\mnras] {10.1093/mnras/staa1369}, \href {https://ui.adsabs.harvard.edu/abs/2020MNRAS.496...36P} {496, 36}

\bibitem[\protect\citeauthoryear{{Peterson}}{{Peterson}}{2006}]{Peterson2006}
{Peterson} B.~M.,  2006, in {Alloin} D.,  ed., , Vol.~693, Physics of Active Galactic Nuclei at all Scales.
Springer Berlin, Heidelberg, p.~77, \mndoi{10.1007/3-540-34621-X_3}

\bibitem[\protect\citeauthoryear{{Povi{\'c}}, {M{\'a}rquez}, {Netzer}, {Masegosa}, {Nordon}, {P{\'e}rez}  \& {Schoenell}}{{Povi{\'c}} et~al.}{2016}]{Povic2016}
{Povi{\'c}} M.,  {M{\'a}rquez} I.,  {Netzer} H.,  {Masegosa} J.,  {Nordon} R.,  {P{\'e}rez} E.,   {Schoenell} W.,  2016, \mndoi [\mnras] {10.1093/mnras/stw1842}, \href {https://ui.adsabs.harvard.edu/abs/2016MNRAS.462.2878P} {462, 2878}

\bibitem[\protect\citeauthoryear{Ricci, Steiner, Menezes, Slodkowski~Clerici  \& da Silva}{Ricci et~al.}{2023}]{Ricci2023}
Ricci T.~V.,  Steiner J.~E.,  Menezes R.~B.,  Slodkowski~Clerici K.,   da Silva M.~D.,  2023, \mndoi [Monthly Notices of the Royal Astronomical Society] {10.1093/mnras/stad1130}, 522, 2207

\bibitem[\protect\citeauthoryear{{Romanishin}}{{Romanishin}}{1985}]{Romanishin1985}
{Romanishin} W.,  1985, \mndoi [\apj] {10.1086/162917}, \href {https://ui.adsabs.harvard.edu/abs/1985ApJ...289..570R} {289, 570}

\bibitem[\protect\citeauthoryear{{Savchenko}, {Marchuk}, {Mosenkov}  \& {Grishunin}}{{Savchenko} et~al.}{2020}]{Sergey2020}
{Savchenko} S.,  {Marchuk} A.,  {Mosenkov} A.,   {Grishunin} K.,  2020, \mndoi [\mnras] {10.1093/mnras/staa258}, \href {https://ui.adsabs.harvard.edu/abs/2020MNRAS.493..390S} {493, 390}

\bibitem[\protect\citeauthoryear{{Schawinski}, {Thomas}, {Sarzi}, {Maraston}, {Kaviraj}, {Joo}, {Yi}  \& {Silk}}{{Schawinski} et~al.}{2007}]{Schawinski2007}
{Schawinski} K.,  {Thomas} D.,  {Sarzi} M.,  {Maraston} C.,  {Kaviraj} S.,  {Joo} S.-J.,  {Yi} S.~K.,   {Silk} J.,  2007, \mndoi [\mnras] {10.1111/j.1365-2966.2007.12487.x}, \href {https://ui.adsabs.harvard.edu/abs/2007MNRAS.382.1415S} {382, 1415}

\bibitem[\protect\citeauthoryear{{Schawinski} et~al.,}{{Schawinski} et~al.}{2014}]{Schawinski2014}
{Schawinski} K.,  et~al., 2014, \mndoi [\mnras] {10.1093/mnras/stu327}, \href {https://ui.adsabs.harvard.edu/abs/2014MNRAS.440..889S} {440, 889}

\bibitem[\protect\citeauthoryear{{Sheth} et~al.,}{{Sheth} et~al.}{2010}]{Sheth2010}
{Sheth} K.,  et~al., 2010, \mndoi [\pasp] {10.1086/657638}, \href {https://ui.adsabs.harvard.edu/abs/2010PASP..122.1397S} {122, 1397}

\bibitem[\protect\citeauthoryear{{Shimakawa}, {Tanaka}, {Bottrell}, {Wu}, {Chang}, {Toba}  \& {Ali}}{{Shimakawa} et~al.}{2022}]{Shimakawa2022}
{Shimakawa} R.,  {Tanaka} M.,  {Bottrell} C.,  {Wu} P.-F.,  {Chang} Y.-Y.,  {Toba} Y.,   {Ali} S.,  2022, \mndoi [\pasj] {10.1093/pasj/psac023}, \href {https://ui.adsabs.harvard.edu/abs/2022PASJ...74..612S} {74, 612}

\bibitem[\protect\citeauthoryear{{Stern} et~al.,}{{Stern} et~al.}{2012}]{Stern2012}
{Stern} D.,  et~al., 2012, \mndoi [\apj] {10.1088/0004-637X/753/1/30}, \href {https://ui.adsabs.harvard.edu/abs/2012ApJ...753...30S} {753, 30}

\bibitem[\protect\citeauthoryear{{Strateva} et~al.,}{{Strateva} et~al.}{2001}]{Strateva2001}
{Strateva} I.,  et~al., 2001, \mndoi [\aj] {10.1086/323301}, \href {https://ui.adsabs.harvard.edu/abs/2001AJ....122.1861S} {122, 1861}

\bibitem[\protect\citeauthoryear{{Tinsley}}{{Tinsley}}{1968}]{Tinsley1968}
{Tinsley} B.~M.,  1968, \mndoi [\apj] {10.1086/149455}, \href {https://ui.adsabs.harvard.edu/abs/1968ApJ...151..547T} {151, 547}

\bibitem[\protect\citeauthoryear{{Tommasin} et~al.,}{{Tommasin} et~al.}{2012}]{Tommasin2012}
{Tommasin} S.,  et~al., 2012, \mndoi [\apj] {10.1088/0004-637X/753/2/155}, \href {https://ui.adsabs.harvard.edu/abs/2012ApJ...753..155T} {753, 155}

\bibitem[\protect\citeauthoryear{{Toomre} \& {Toomre}}{{Toomre} \& {Toomre}}{1972}]{Toomre1972}
{Toomre} A.,  {Toomre} J.,  1972, \mndoi [\apj] {10.1086/151823}, \href {https://ui.adsabs.harvard.edu/abs/1972ApJ...178..623T} {178, 623}

\bibitem[\protect\citeauthoryear{{V{\'e}ron-Cetty} \& {V{\'e}ron}}{{V{\'e}ron-Cetty} \& {V{\'e}ron}}{2006}]{VCV2006}
{V{\'e}ron-Cetty} M.~P.,  {V{\'e}ron} P.,  2006, \mndoi [\aap] {10.1051/0004-6361:20065177}, \href {https://ui.adsabs.harvard.edu/abs/2006A&A...455..773V} {455, 773}

\bibitem[\protect\citeauthoryear{{Whitaker}, {van Dokkum}, {Brammer}  \& {Franx}}{{Whitaker} et~al.}{2012}]{Whitaker2012}
{Whitaker} K.~E.,  {van Dokkum} P.~G.,  {Brammer} G.,   {Franx} M.,  2012, \mndoi [\apjl] {10.1088/2041-8205/754/2/L29}, \href {https://ui.adsabs.harvard.edu/abs/2012ApJ...754L..29W} {754, L29}

\bibitem[\protect\citeauthoryear{{Worthey}, {Faber}  \& {Gonzalez}}{{Worthey} et~al.}{1992}]{Worthey1992}
{Worthey} G.,  {Faber} S.~M.,   {Gonzalez} J.~J.,  1992, \mndoi [\apj] {10.1086/171836}, \href {https://ui.adsabs.harvard.edu/abs/1992ApJ...398...69W} {398, 69}

\bibitem[\protect\citeauthoryear{Wyder et~al.,}{Wyder et~al.}{2007}]{Wyder_2007}
Wyder T.~K.,  et~al., 2007, \mndoi [The Astrophysical Journal Supplement Series] {10.1086/521402}, 173, 293

\bibitem[\protect\citeauthoryear{{Yan} \& {Blanton}}{{Yan} \& {Blanton}}{2012}]{Yan2012}
{Yan} R.,  {Blanton} M.~R.,  2012, \mndoi [\apj] {10.1088/0004-637X/747/1/61}, \href {https://ui.adsabs.harvard.edu/abs/2012ApJ...747...61Y} {747, 61}

\bibitem[\protect\citeauthoryear{{Yao} et~al.,}{{Yao} et~al.}{2020}]{Yao2020}
{Yao} H.~F.~M.,  et~al., 2020, \mndoi [\apj] {10.3847/1538-4357/abba1a}, \href {https://ui.adsabs.harvard.edu/abs/2020ApJ...903...91Y} {903, 91}

\bibitem[\protect\citeauthoryear{{Zucker}, {Walker}, {Johnson}, {Gallagher}, {Alatalo}  \& {Tzanavaris}}{{Zucker} et~al.}{2016}]{Zucker2016}
{Zucker} C.,  {Walker} L.~M.,  {Johnson} K.,  {Gallagher} S.,  {Alatalo} K.,   {Tzanavaris} P.,  2016, \mndoi [\apj] {10.3847/0004-637X/821/2/113}, \href {https://ui.adsabs.harvard.edu/abs/2016ApJ...821..113Z} {821, 113}

\makeatother
\end{thebibliography}



\clearpage 

\appendix

\begin{figure*}
        \centering
        \includegraphics[width=0.3\textwidth,angle=0]{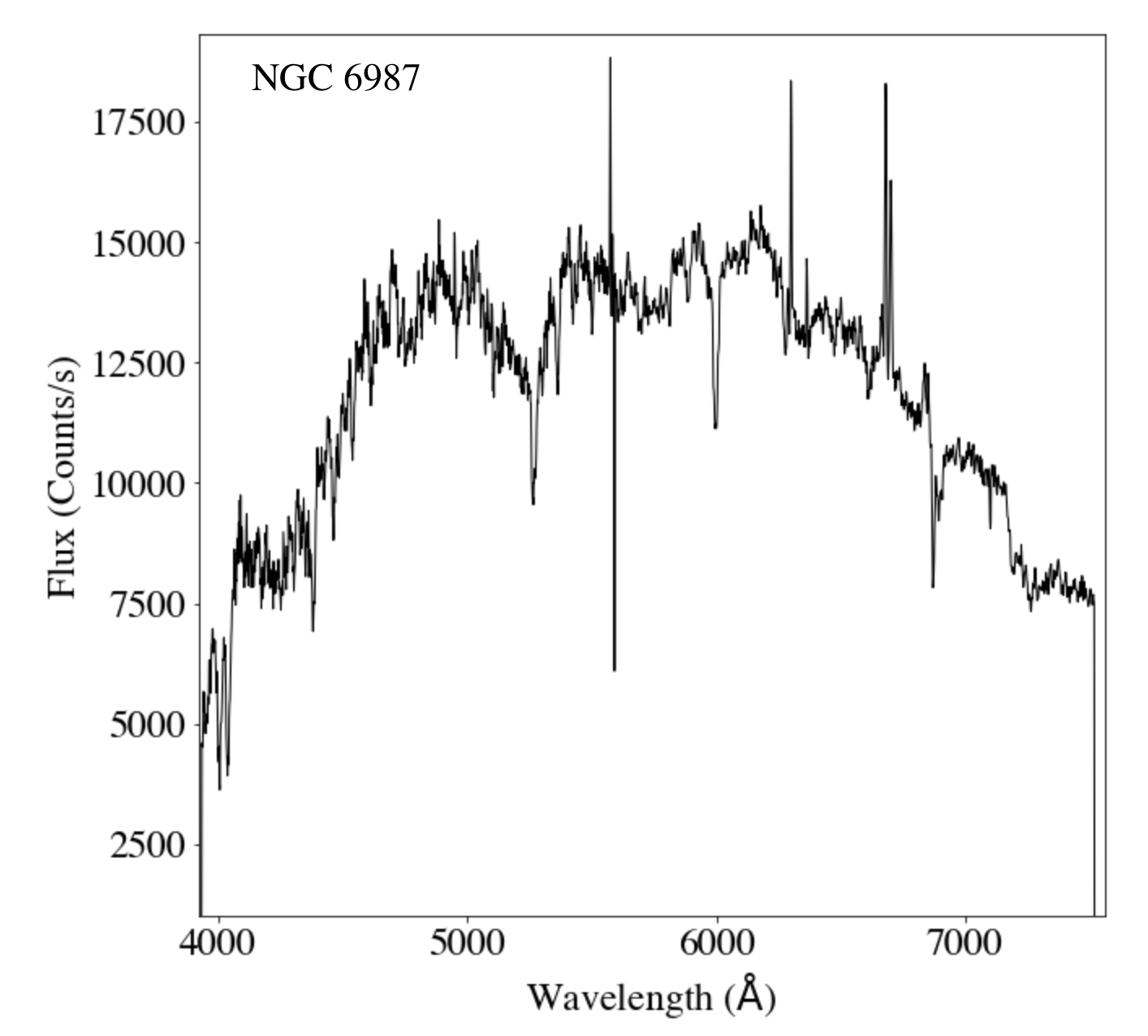}
        \includegraphics[width=0.3\textwidth,angle=0]{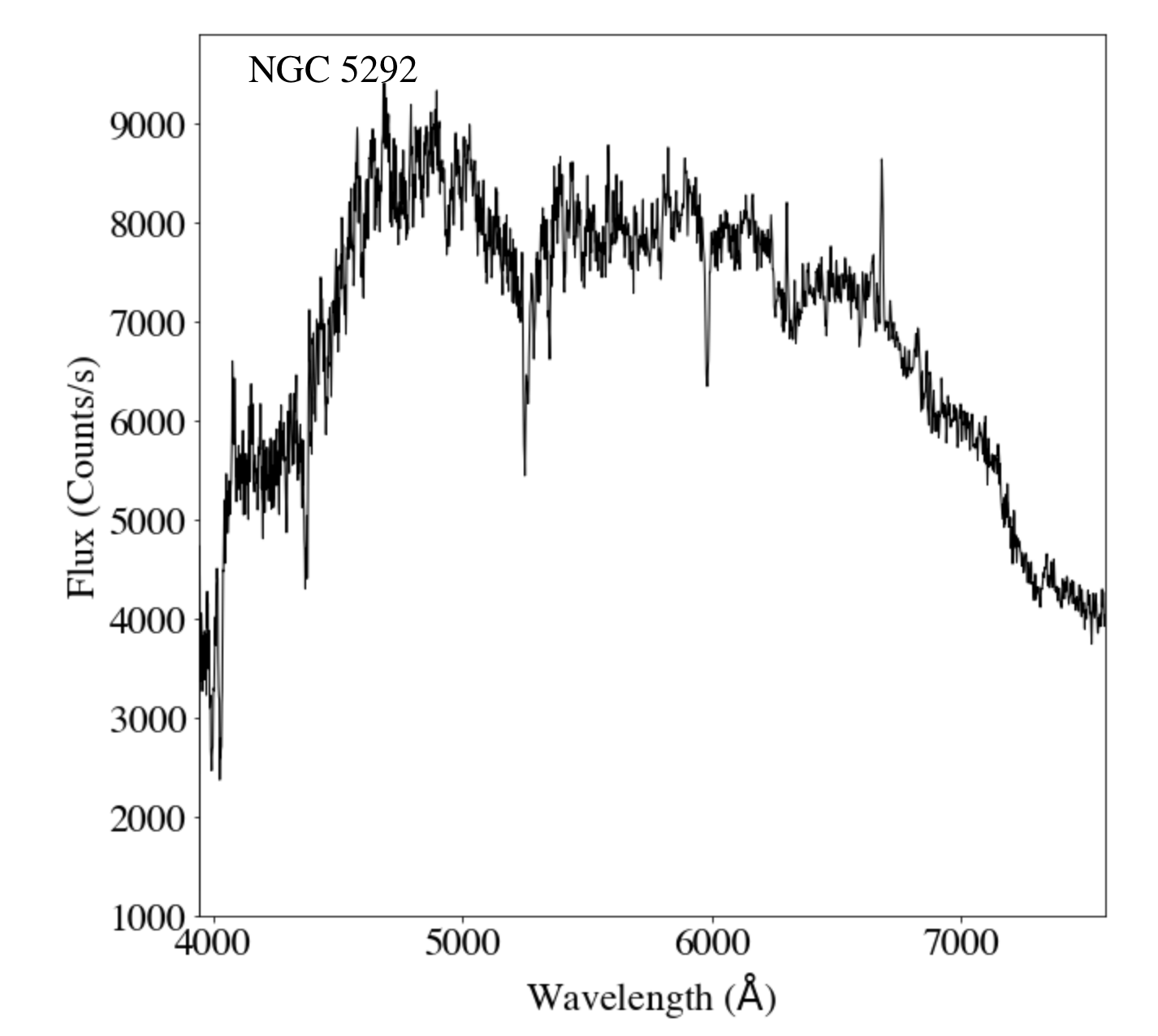}
        \includegraphics[width=0.3\textwidth,angle=0]{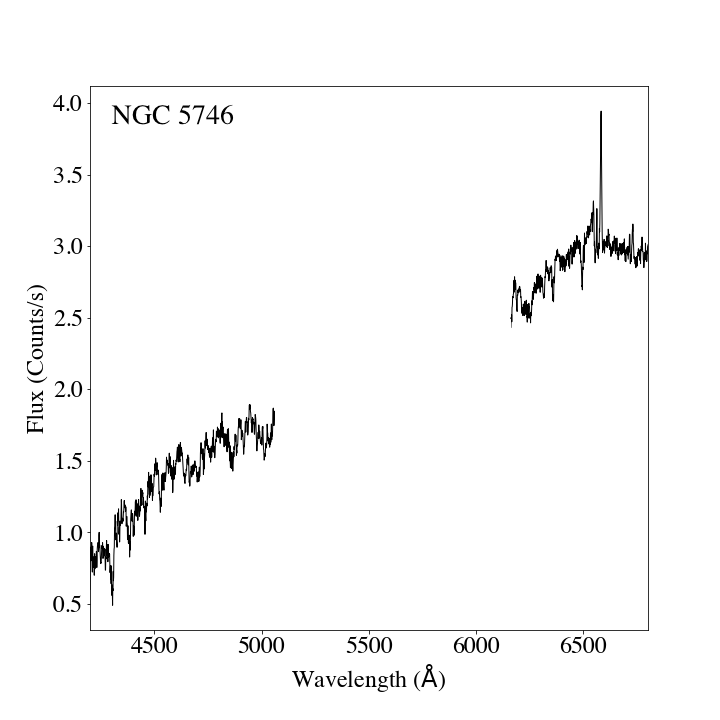}
        \includegraphics[width=0.3\textwidth,angle=0]{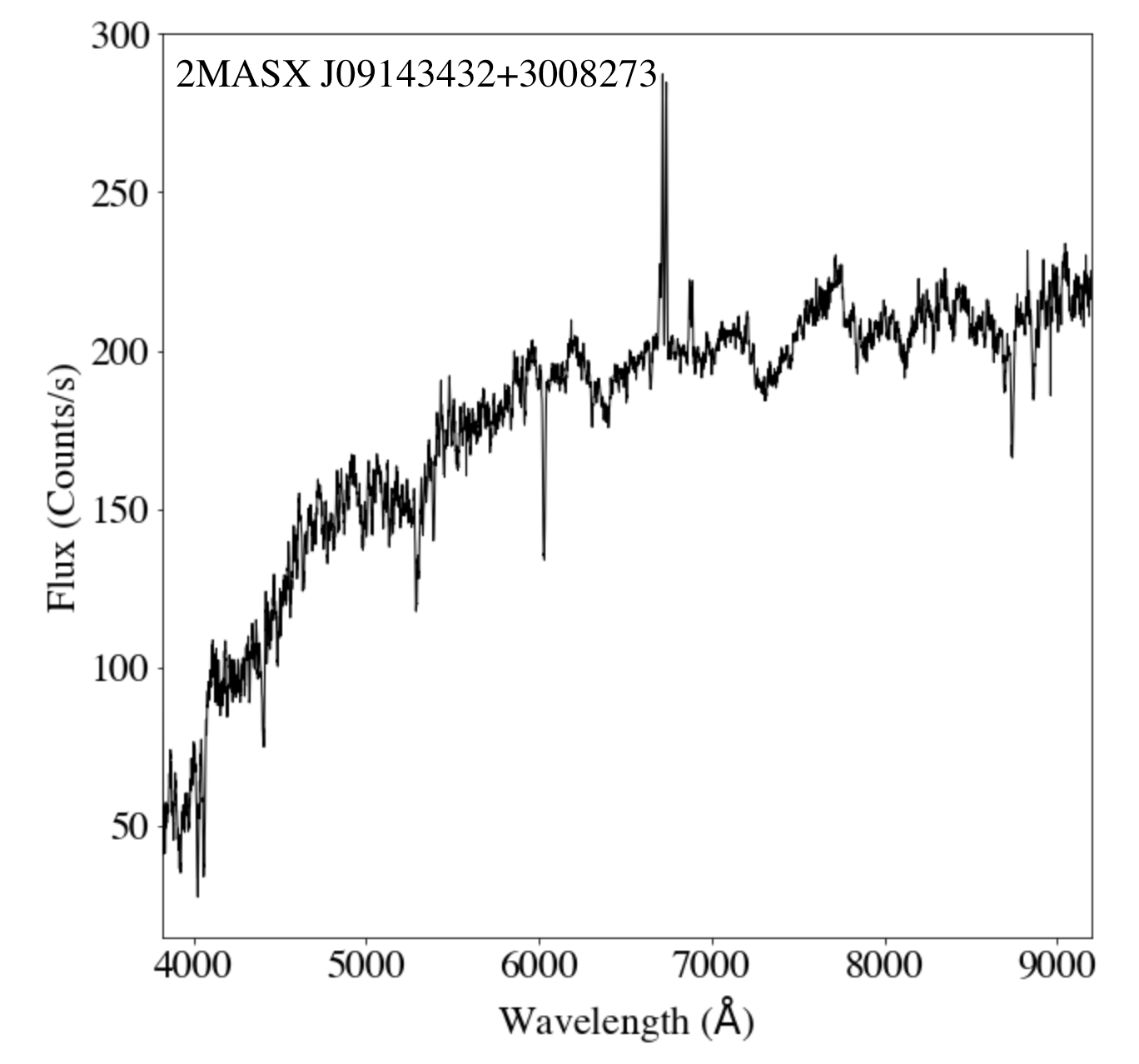}
        \includegraphics[width=0.3\textwidth,angle=0]{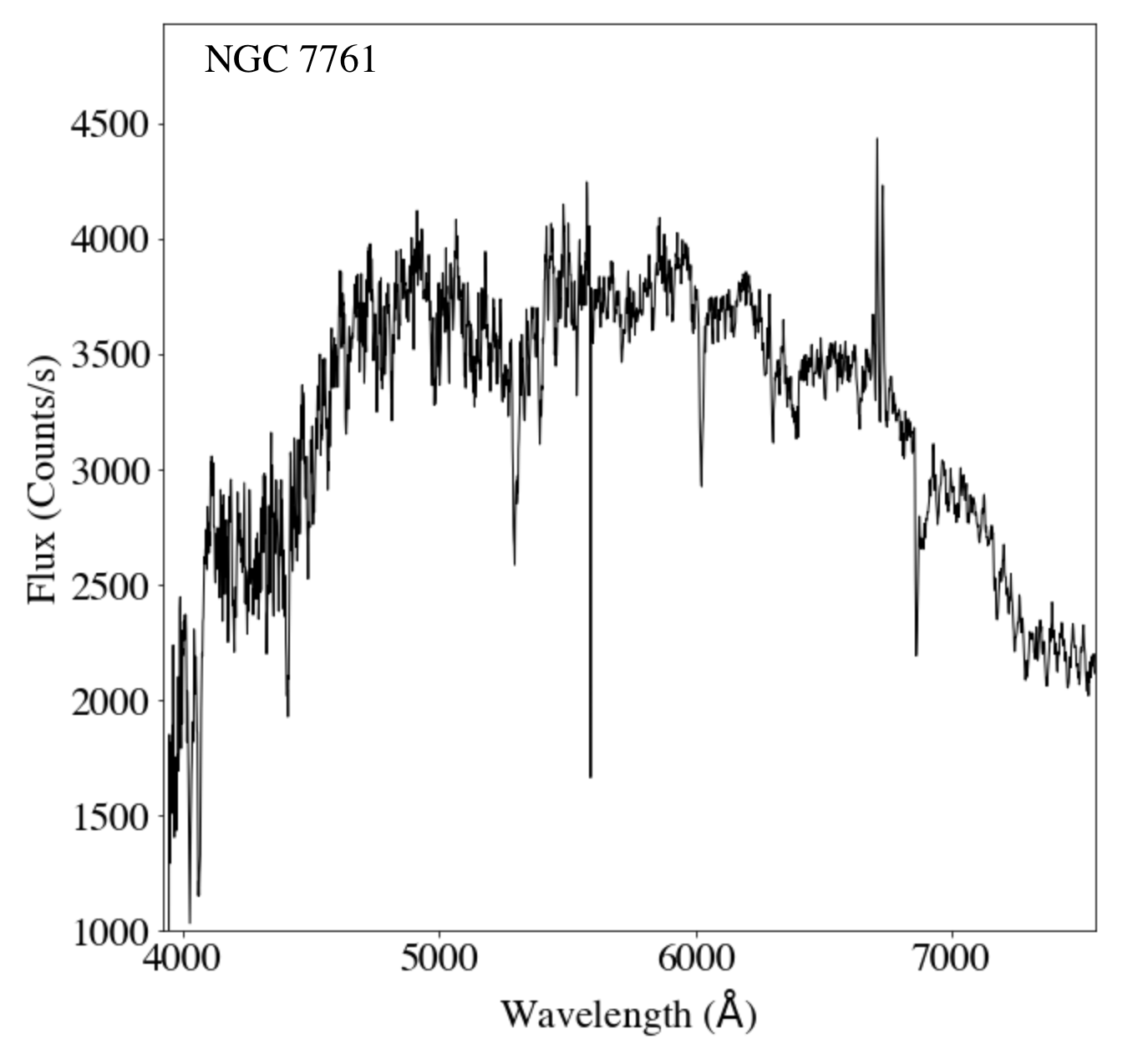}
        \includegraphics[width=0.3\textwidth,angle=0]{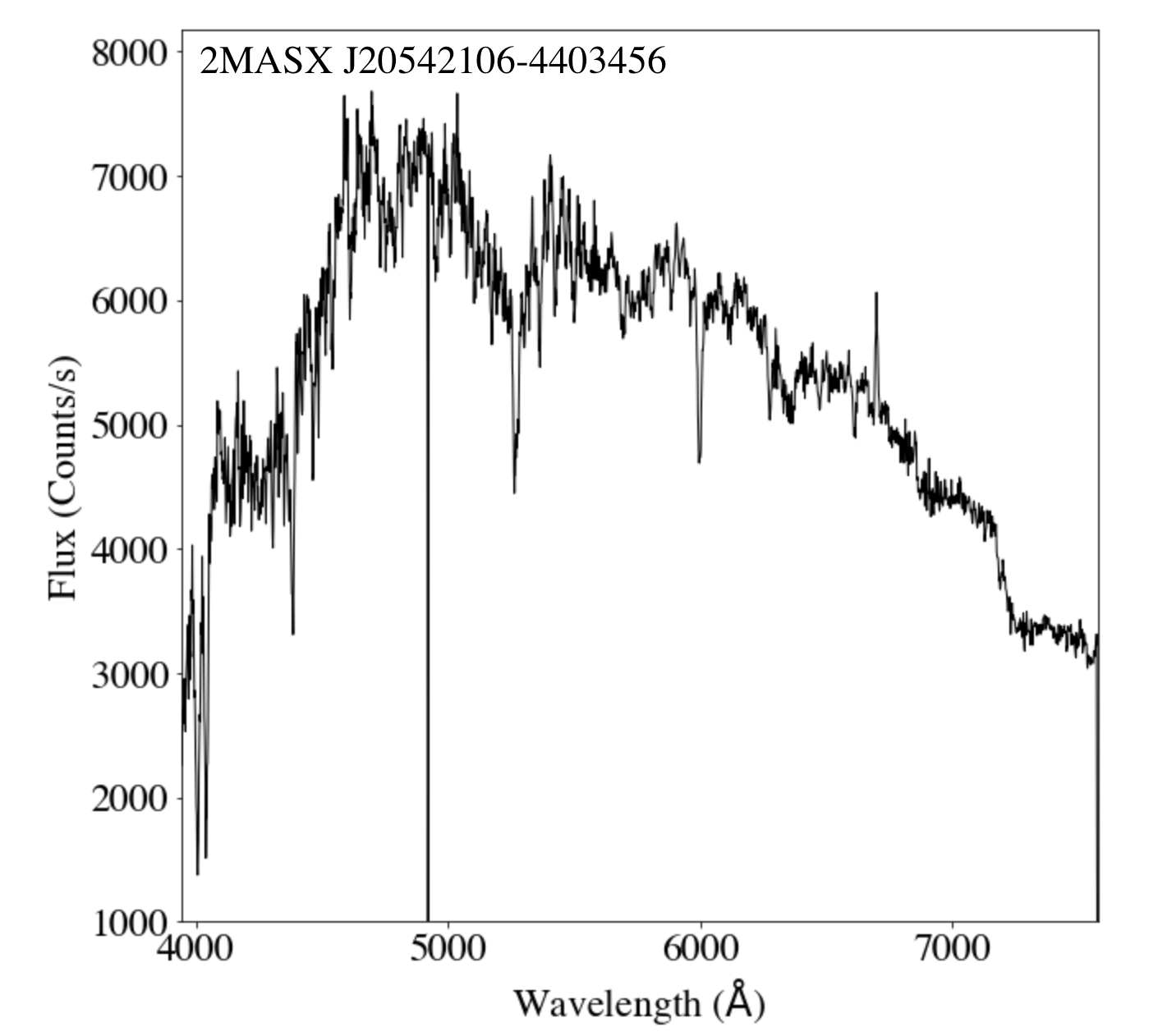}
        \includegraphics[width=0.3\textwidth,angle=0]{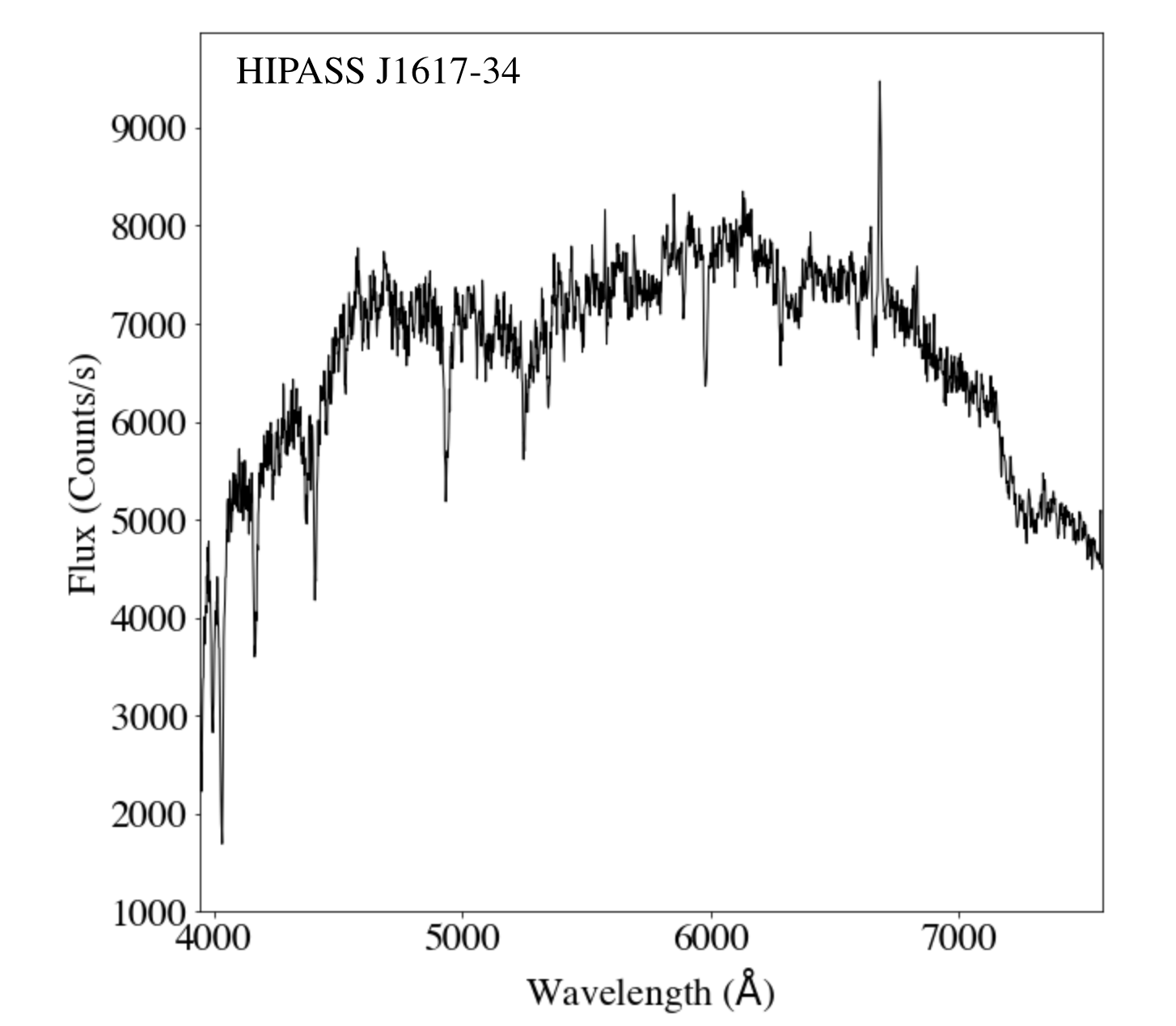}
        \includegraphics[width=0.3\textwidth,angle=0]{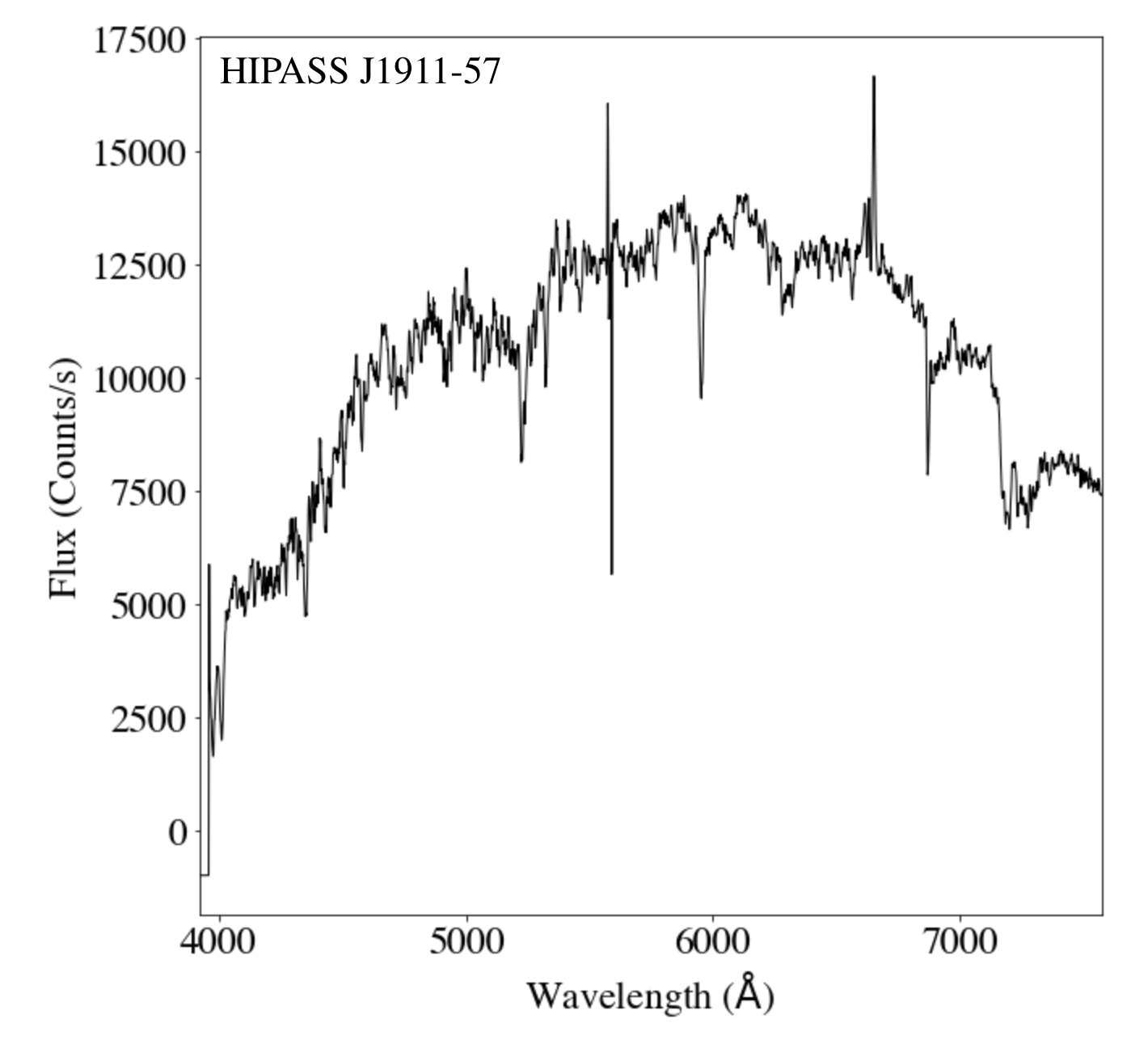}
        \includegraphics[width=0.3\textwidth,angle=0]{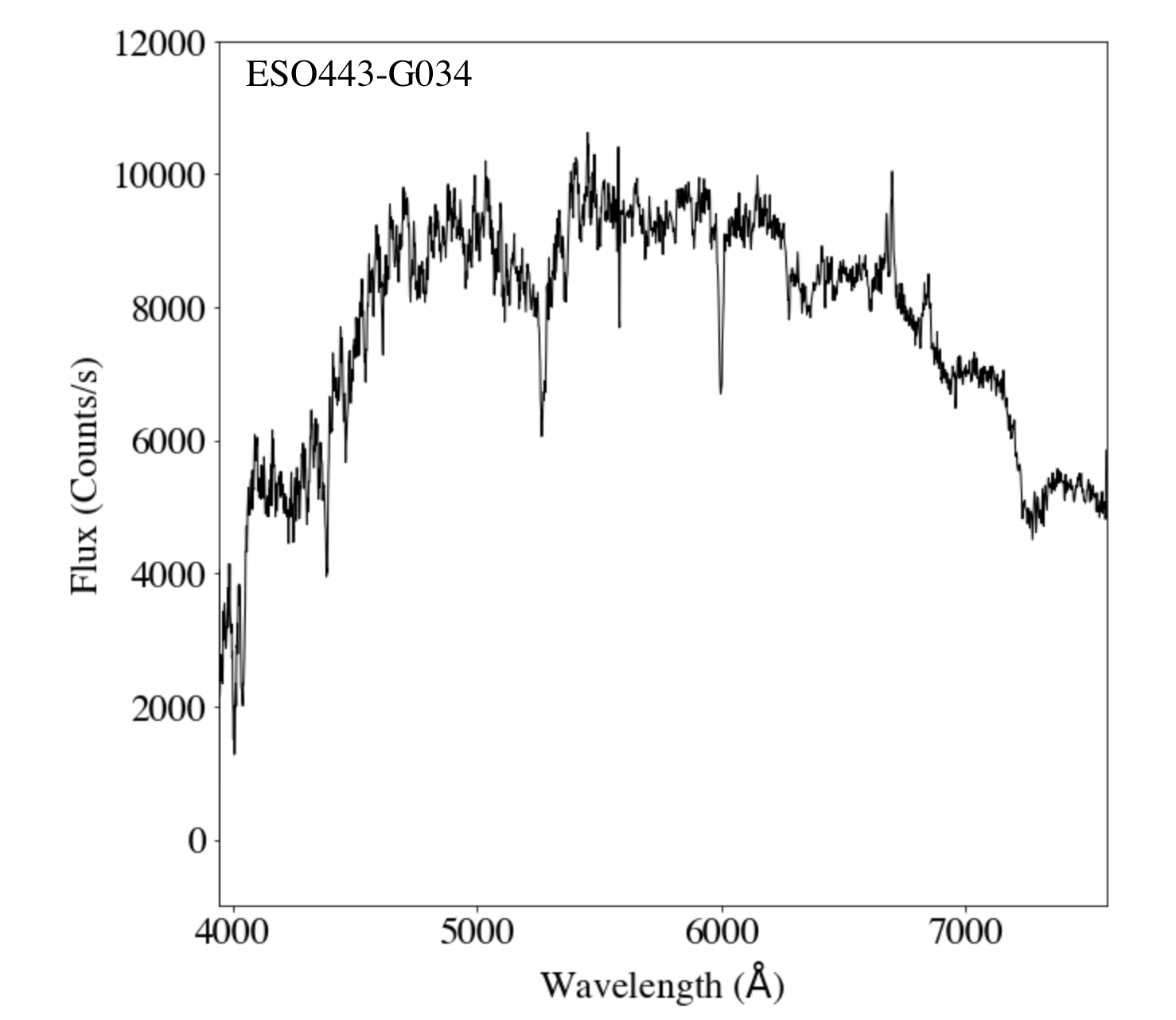}
        \includegraphics[width=0.3\textwidth,angle=0]{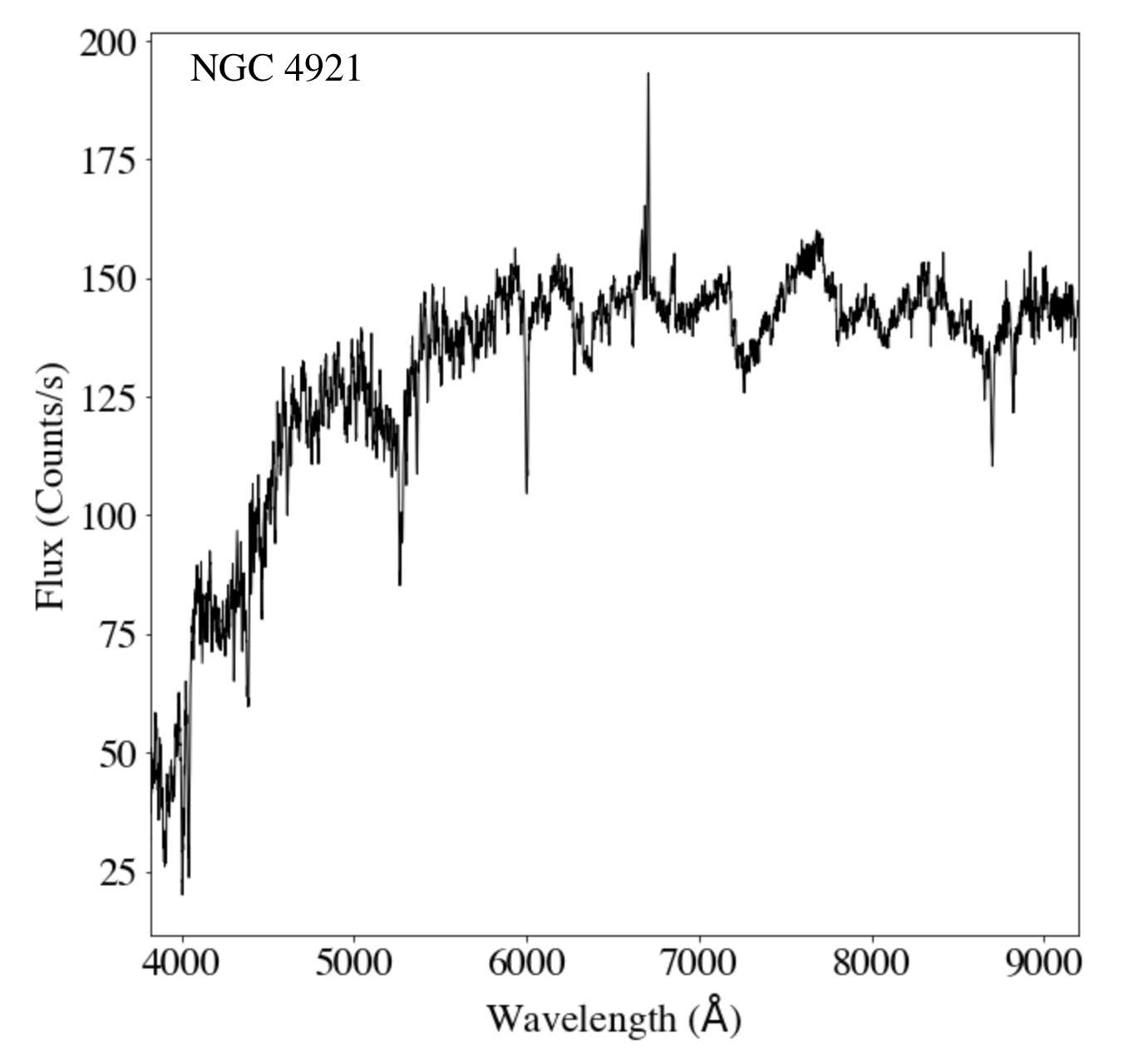}
        \includegraphics[width=0.3\textwidth,angle=0]{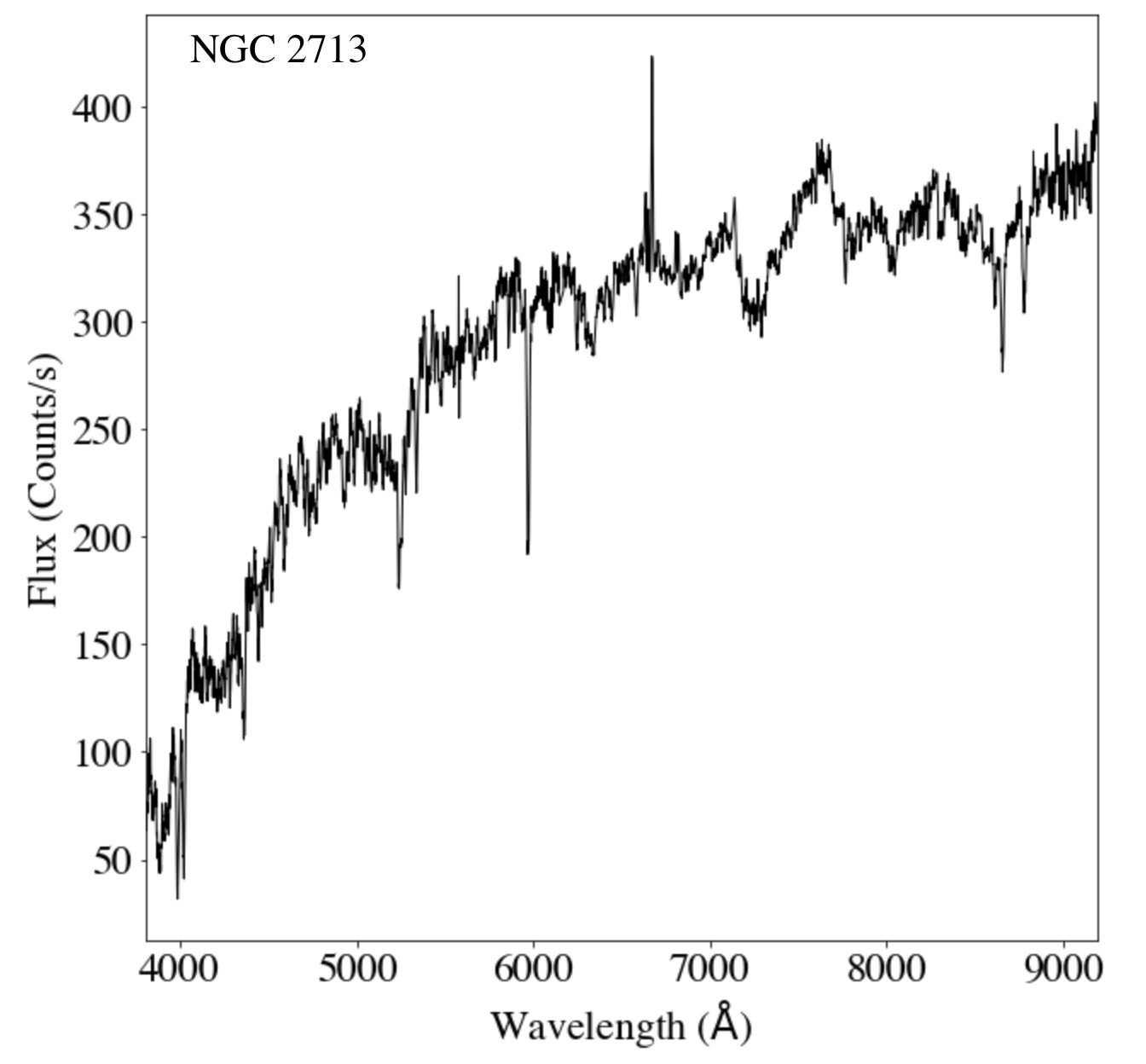}
        \includegraphics[width=0.3\textwidth,angle=0]{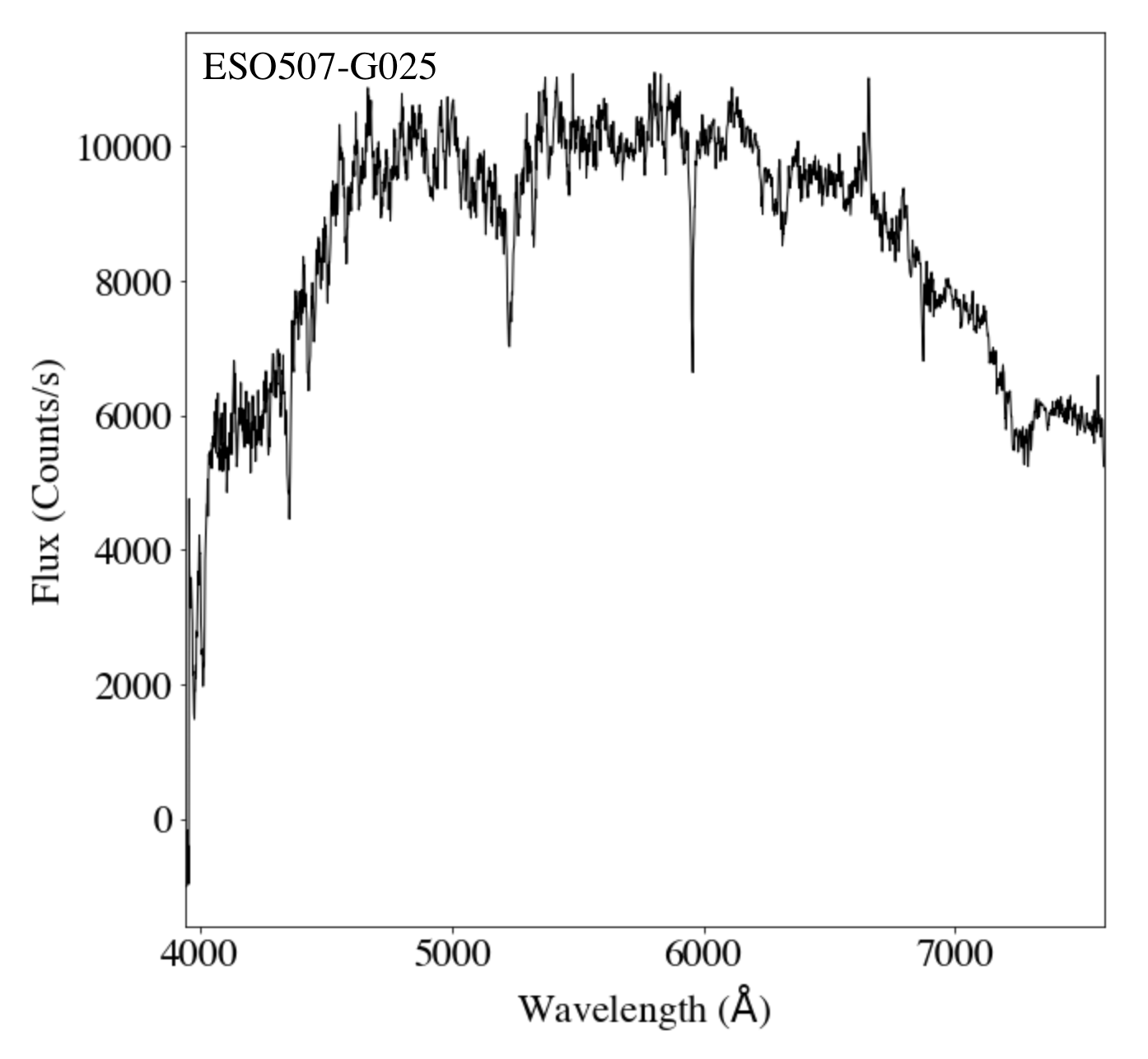}
        \caption{A representative selection of the heterogeneous spectra from our sample of LINERs, comprised primarily of spectra from 6dFGS \citep{6df2006, Jones2009}, as well as additional spectra from SDSS \citep{SDSSDR12}, 2MRS Fast Survey \citep{Huchra2012} and the \citet{Ho1995} sample of nearby active galaxies.}
        \label{fig:spectra}
    \end{figure*}

\begin{figure*}
        \centering
        \includegraphics[width=0.3\textwidth,angle=0]{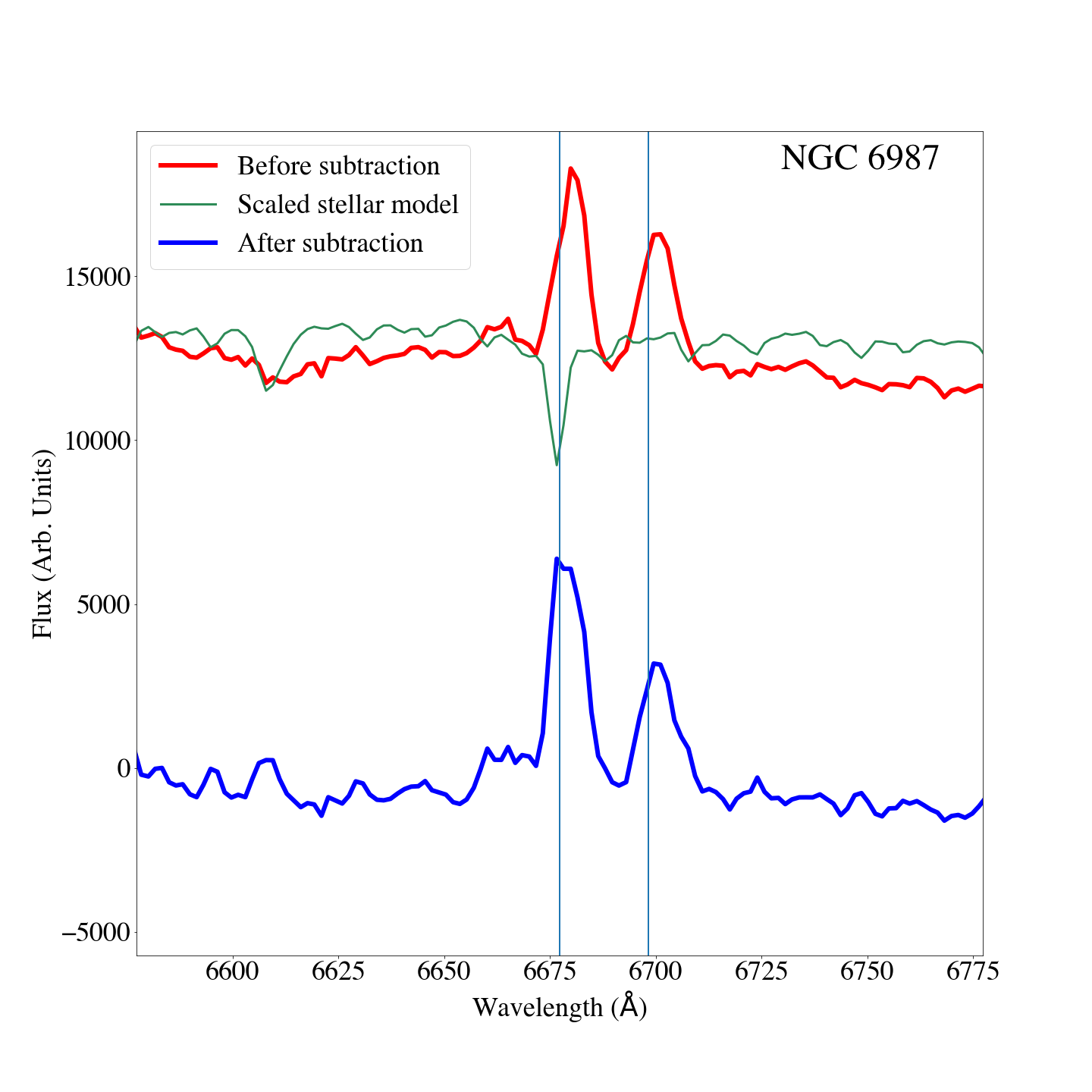}
        \includegraphics[width=0.3\textwidth,angle=0]{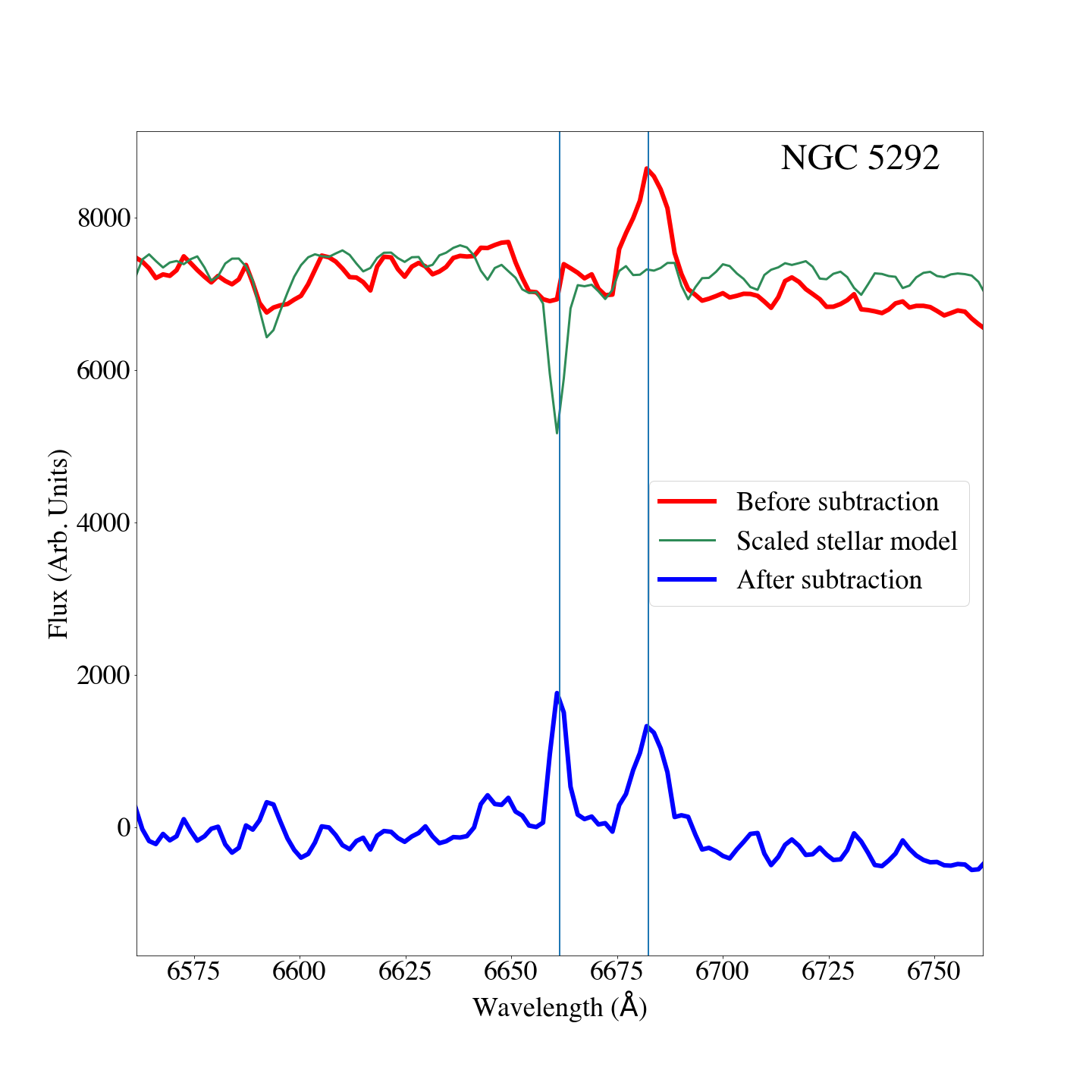}
        \includegraphics[width=0.3\textwidth,angle=0]{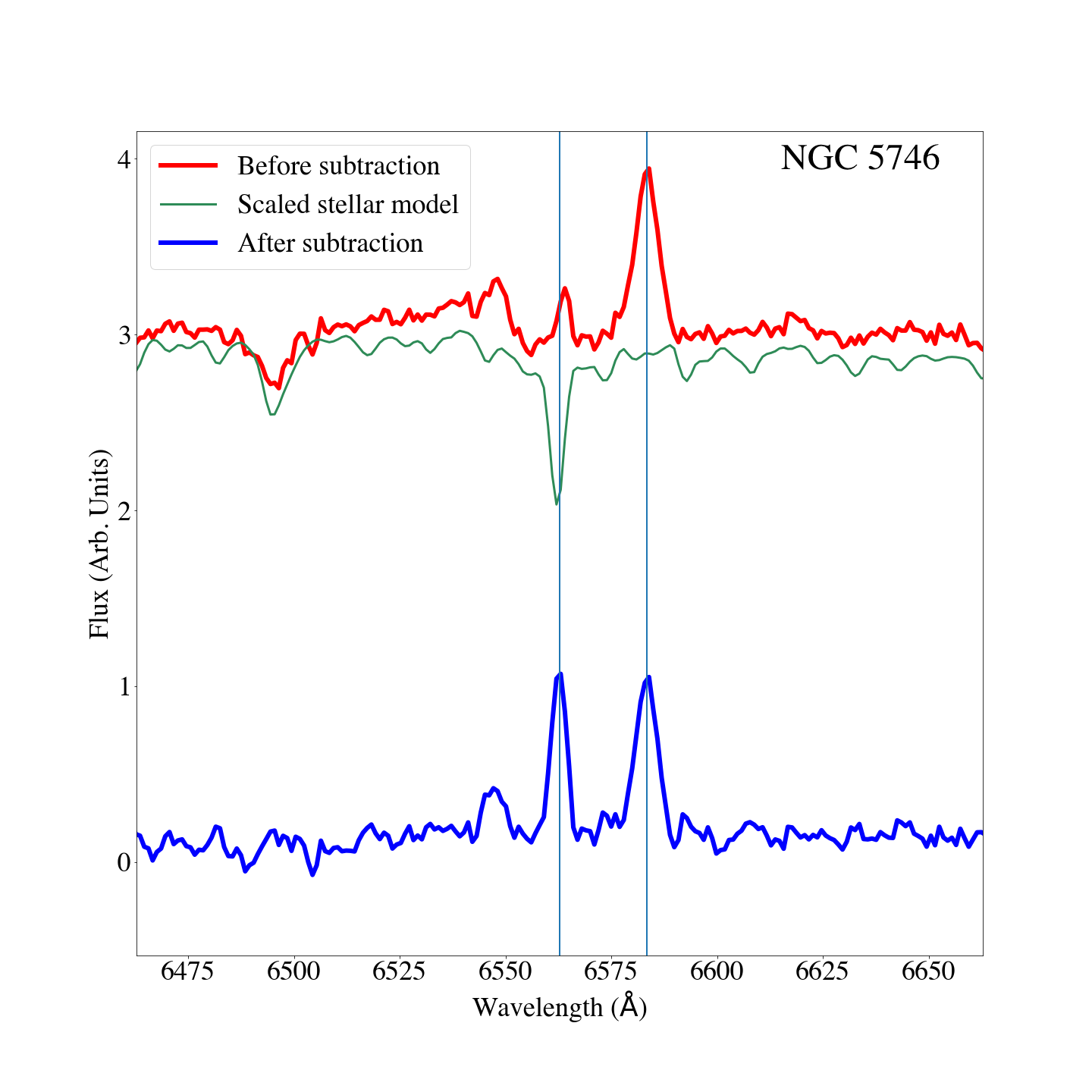}
        \includegraphics[width=0.3\textwidth,angle=0]{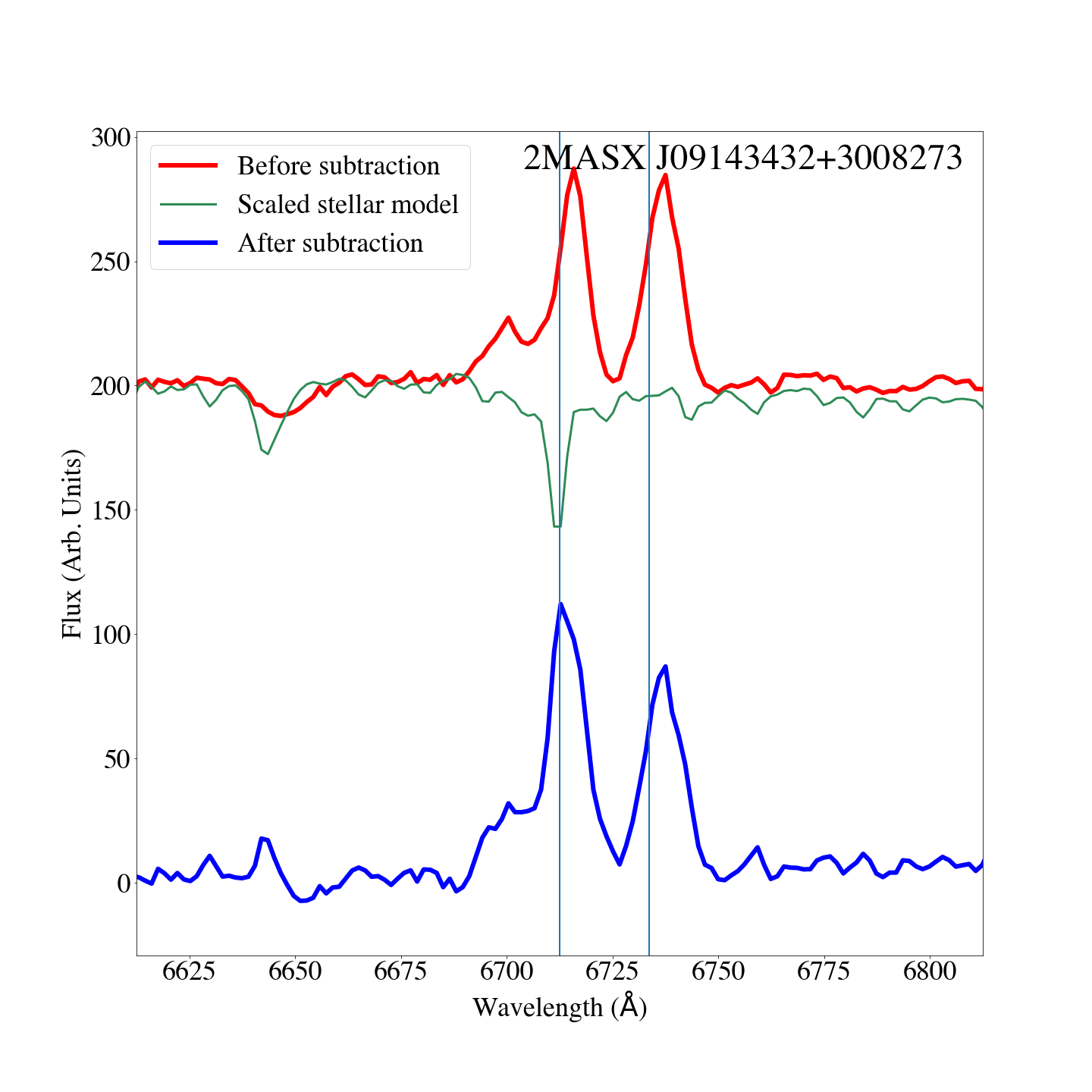}
        \includegraphics[width=0.3\textwidth,angle=0]{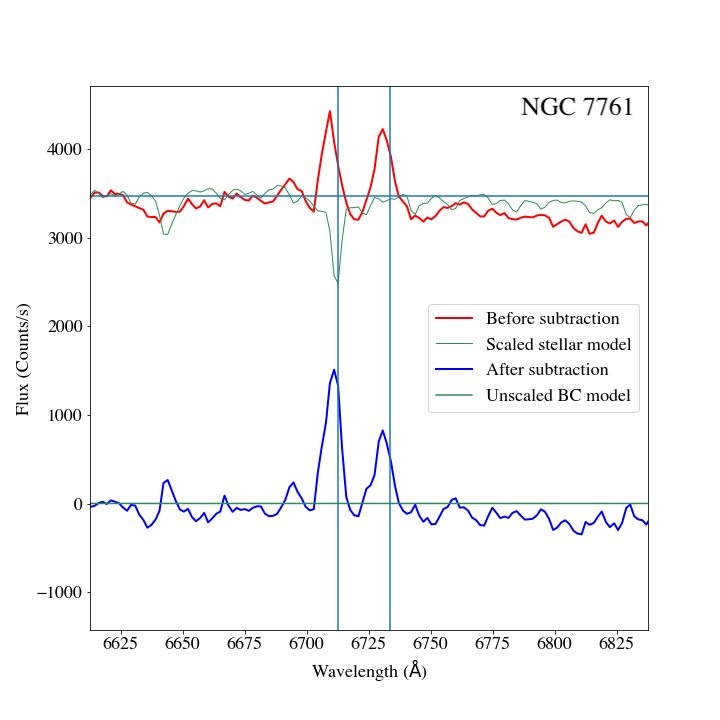}
        \includegraphics[width=0.3\textwidth,angle=0]{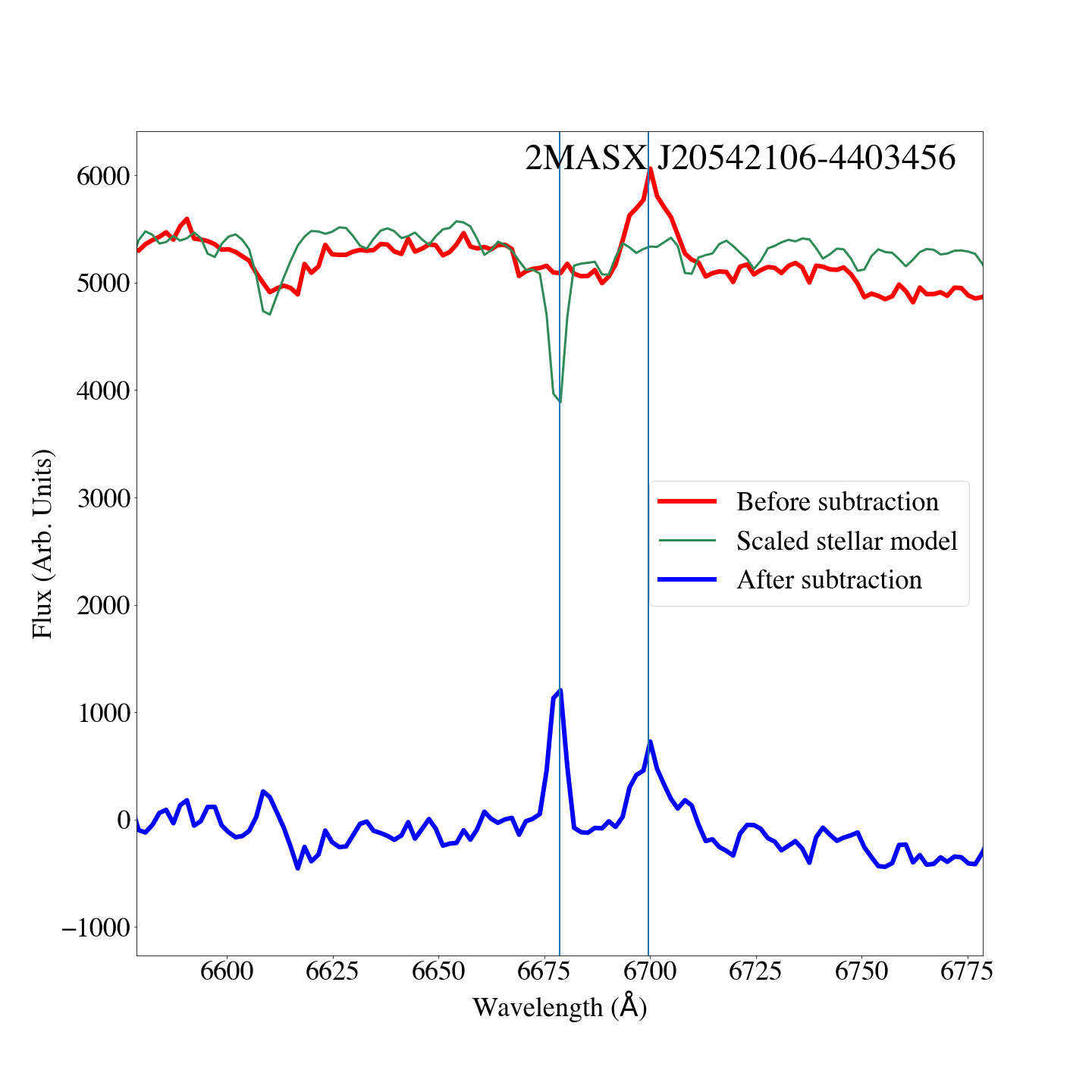}
        \includegraphics[width=0.3\textwidth,angle=0]{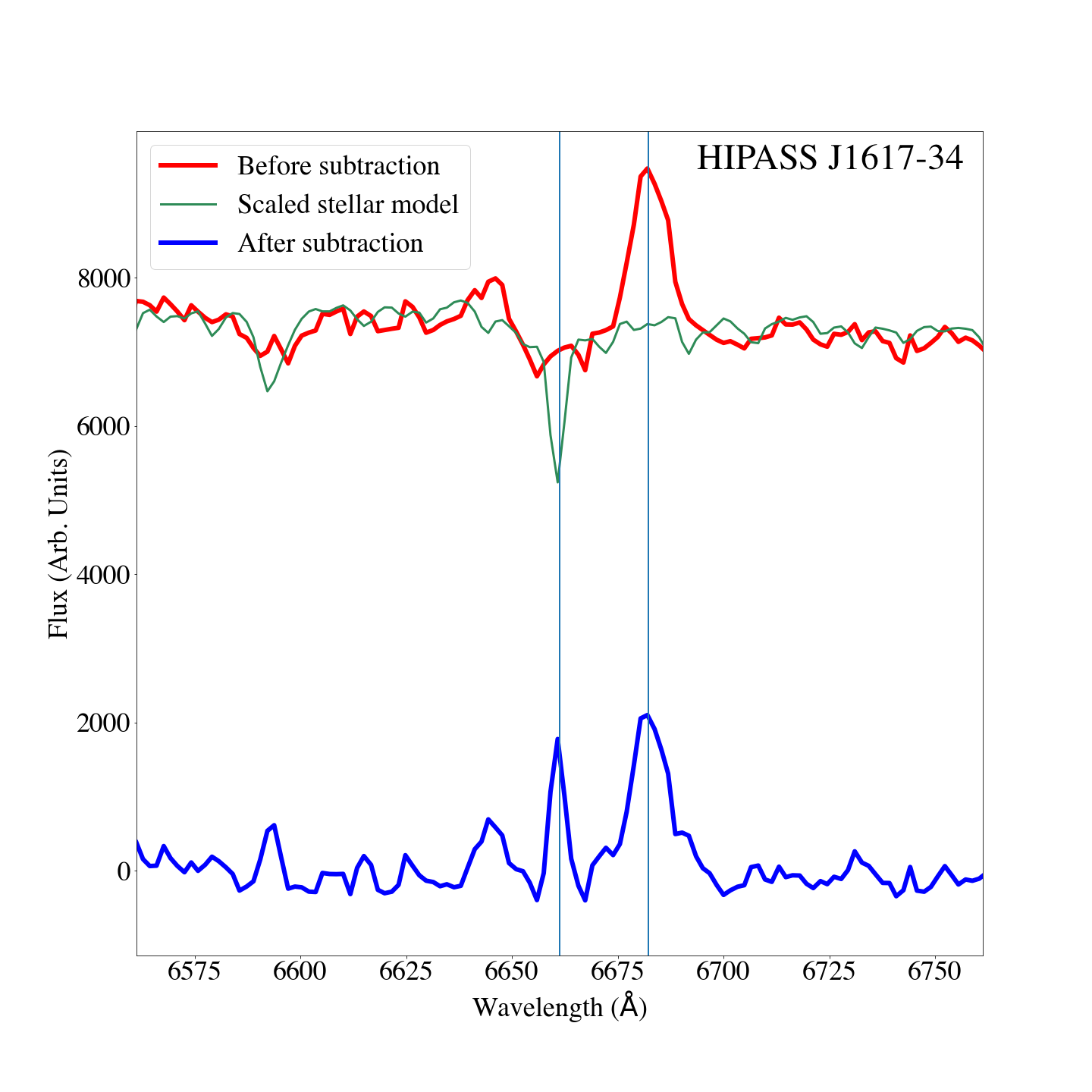}
        \includegraphics[width=0.3\textwidth,angle=0]{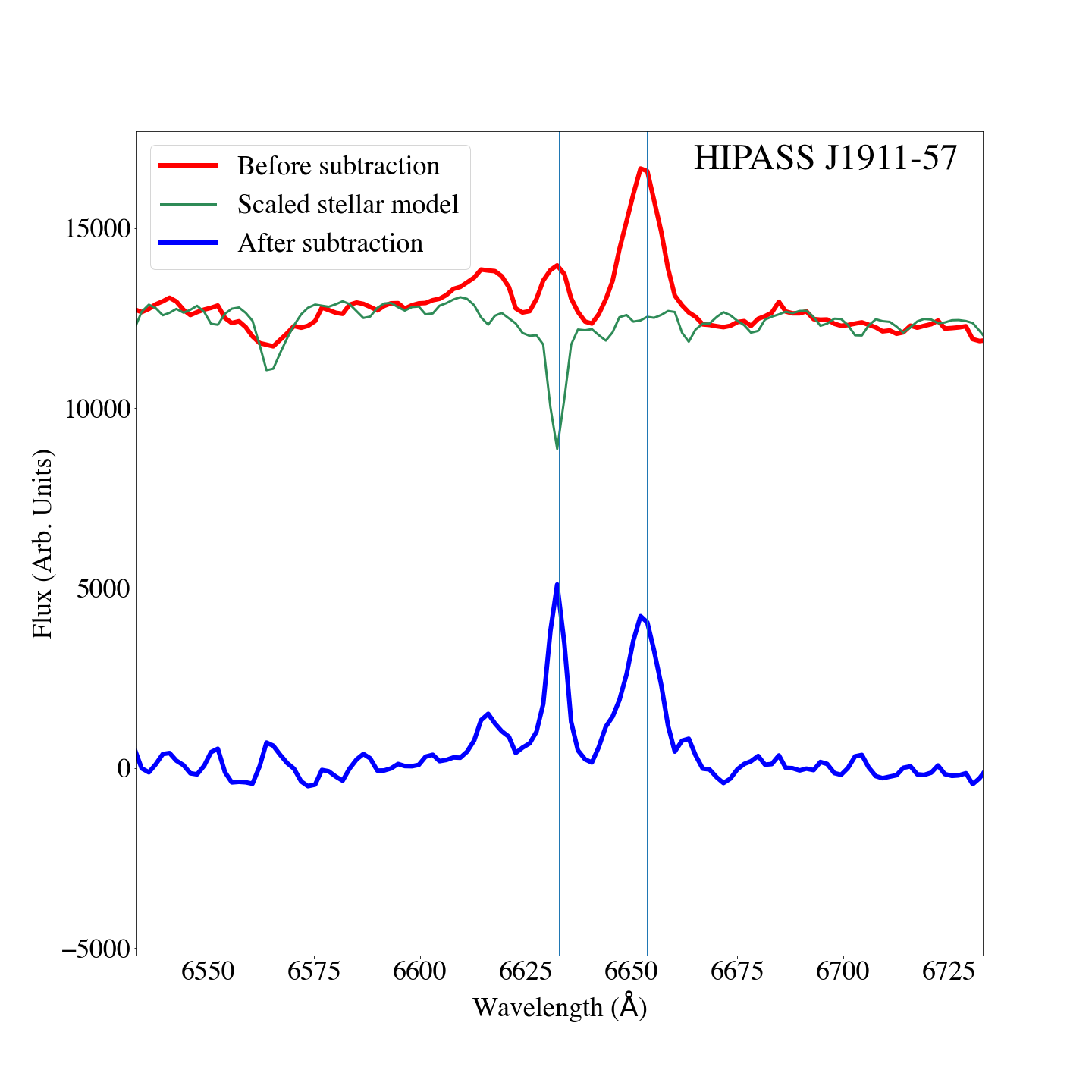}
        \includegraphics[width=0.3\textwidth,angle=0]{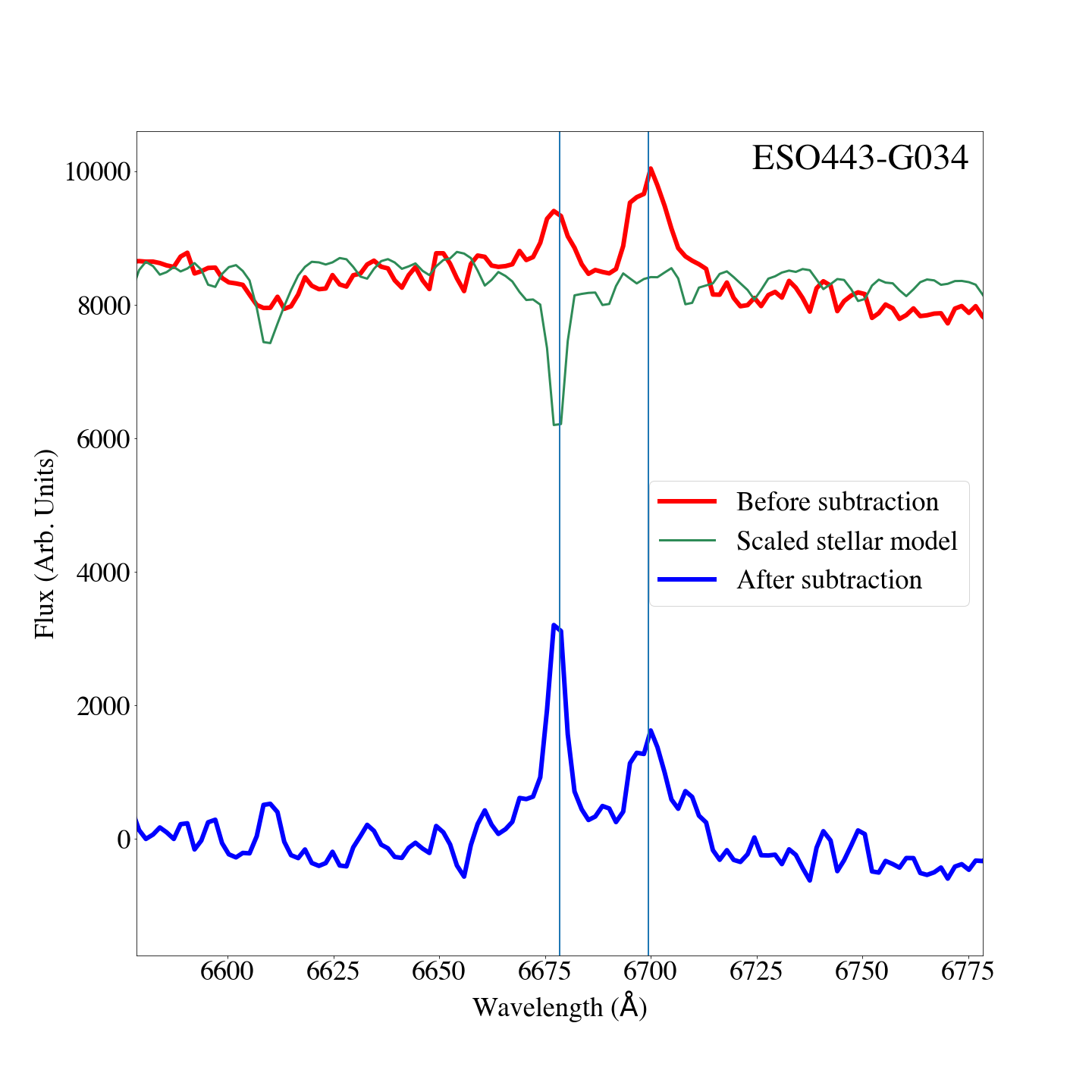}
        \includegraphics[width=0.3\textwidth,angle=0]{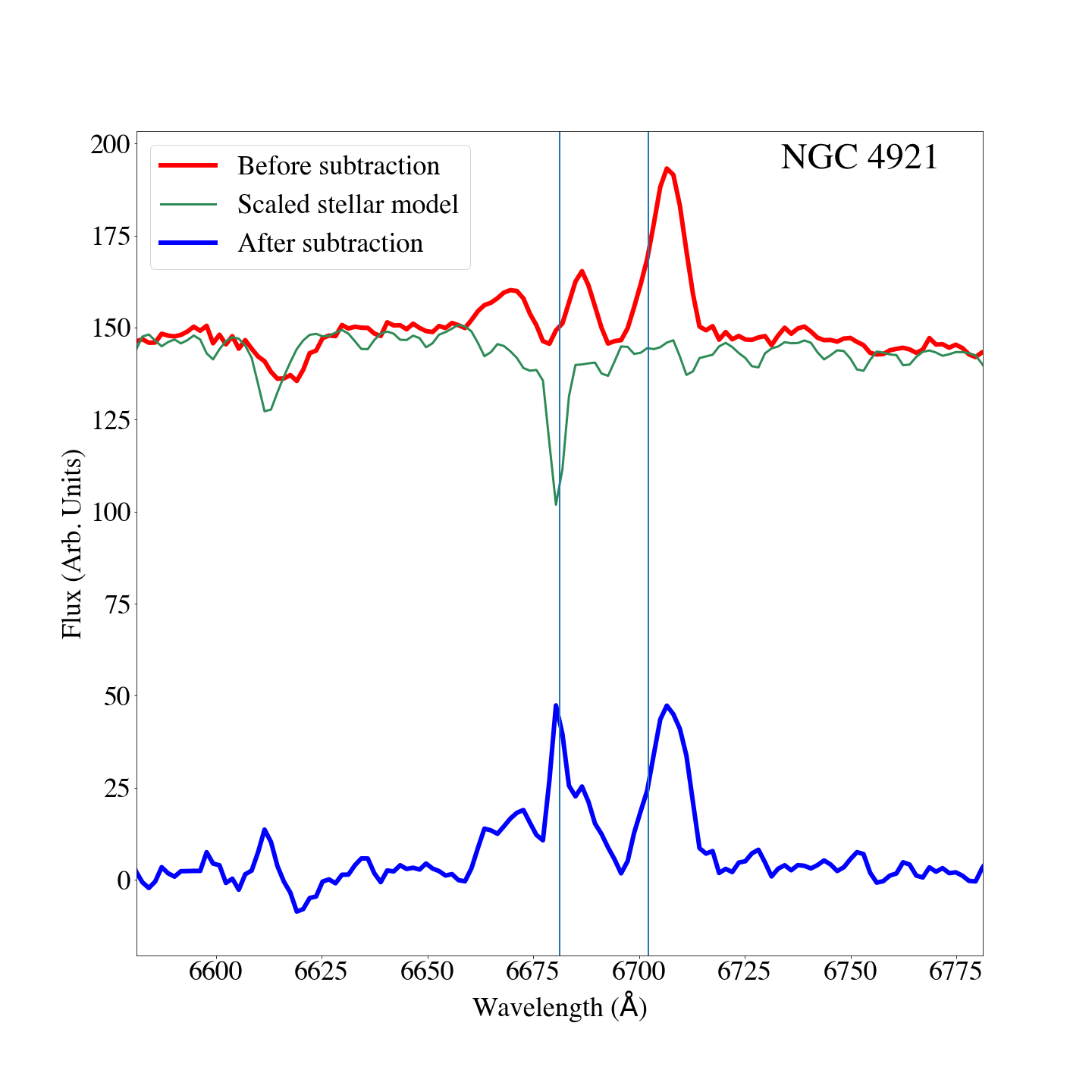}
        \includegraphics[width=0.3\textwidth,angle=0]{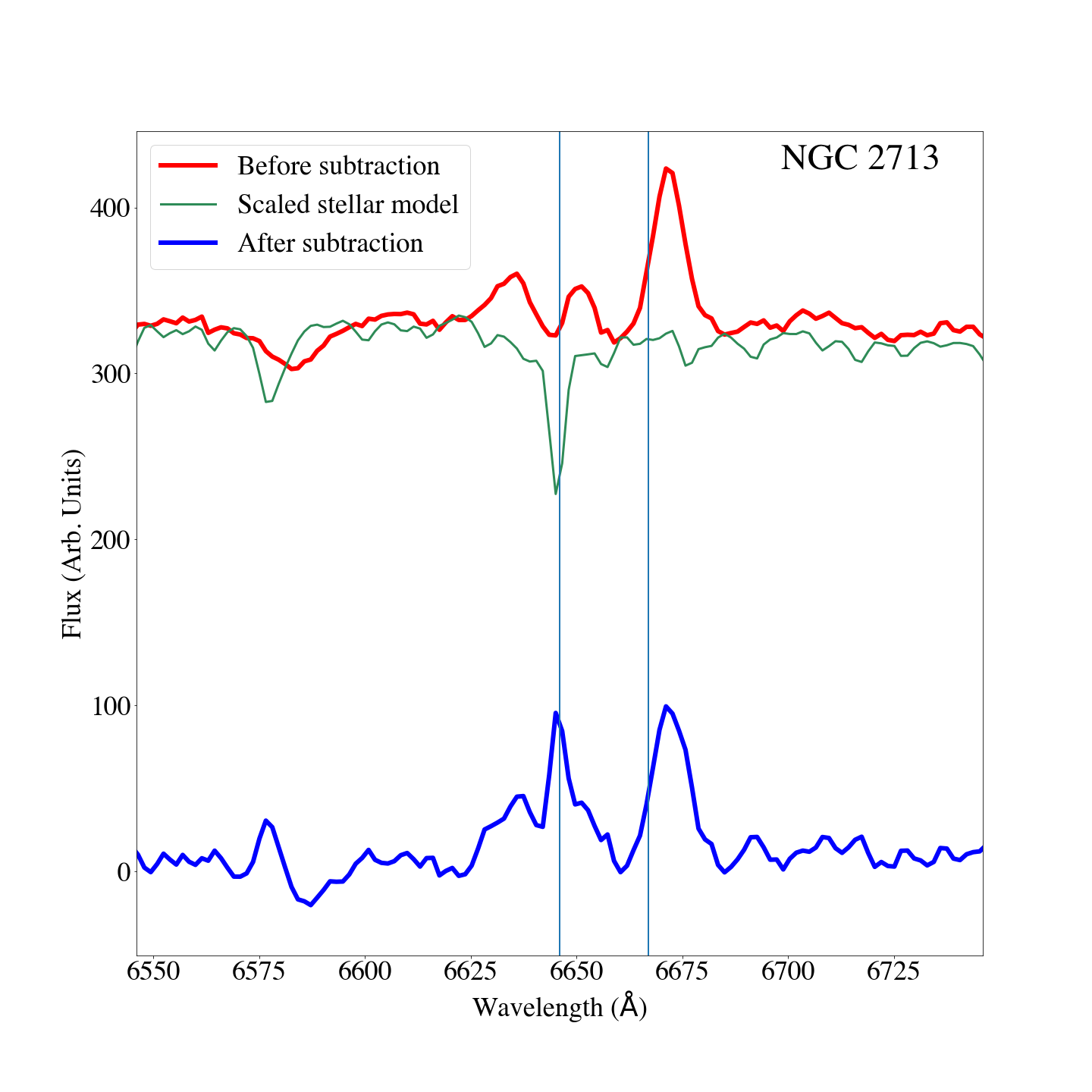}
        \includegraphics[width=0.3\textwidth,angle=0]{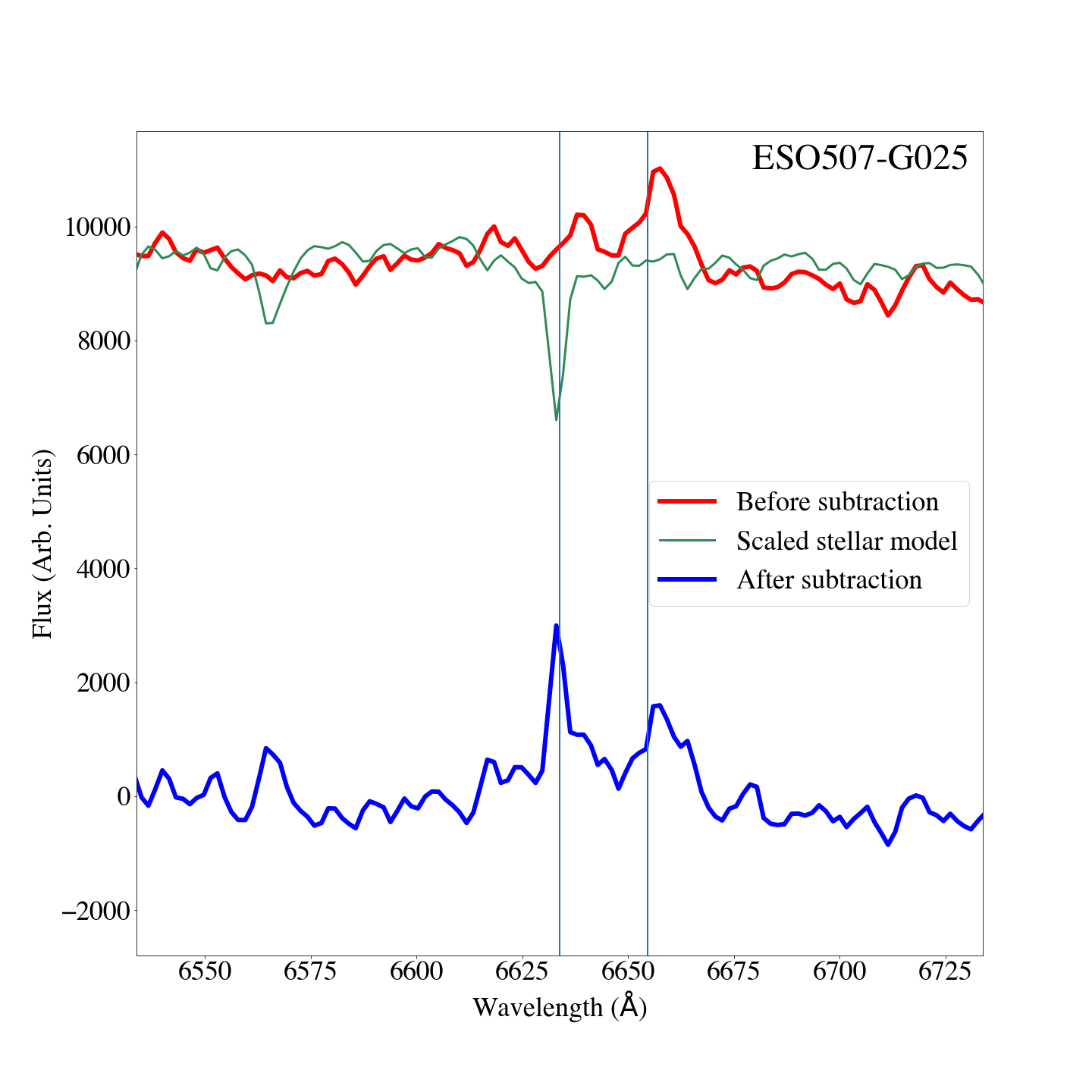}
        \caption{A representative selection of the stellar population subtracted spectra from our sample of LINERs, comprised primarily of spectra from 6dFGS \citep{6df2006, Jones2009}, as well as additional spectra from SDSS \citep{SDSSDR12}, 2MRS Fast Survey \citep{Huchra2012} and the \citet{Ho1995} sample of nearby active galaxies.}
        \label{fig:spectrasub}
    \end{figure*}

\label{lastpage}
\end{document}